\documentclass[aps,prd,nofootinbib,twocolumn,superscriptaddress,preprintnumbers,balancelastpage,longbibliography,nobibnotes]{revtex4-1}

\usepackage{amsmath,amssymb,mathtools,bm}
\usepackage{graphicx, color, hepunits}
\usepackage[dvipsnames]{xcolor}
\usepackage{float}
\usepackage{filecontents}
 \usepackage{hyperref} 
\hypersetup{
    colorlinks=true,       
    linkcolor=blue,        
    citecolor=blue,        
    filecolor=magenta,     
    urlcolor=blue          
}
\usepackage[utf8]{inputenc}
\usepackage[english]{babel}
\let\vec\mathbf

\newcommand{\gagg}{g_{a \gamma \gamma}}

\newcommand{\es}[2] {\begin{equation} \label{#1} \begin{split} #2 \end{split} \end{equation}}

\begin{document}

\title{Green Bank and Effelsberg Radio Telescope Searches for Axion Dark Matter Conversion in Neutron Star Magnetospheres}

\author{Joshua W. Foster}
\email{fosterjw@umich.edu}
\affiliation{Leinweber Center for Theoretical Physics, Department of Physics, University of Michigan, Ann Arbor, MI 48109 U.S.A.}
\author{Yonatan Kahn}
\affiliation{University of Illinois at Urbana-Champaign, Urbana, IL 61801, U.S.A.}
\author{Oscar Macias}
\affiliation{Kavli Institute for the Physics and Mathematics of the Universe (WPI), University of Tokyo, Kashiwa, Chiba 277-8583, Japan}
\affiliation{GRAPPA Institute, Institute of Physics, University of Amsterdam, 1098 XH Amsterdam, The Netherlands}

\author{Zhiquan Sun}
\affiliation{Leinweber Center for Theoretical Physics, Department of Physics, University of Michigan, Ann Arbor, MI 48109 U.S.A.}
\author{Ralph P. Eatough}
\affiliation{National Astronomical Observatories, Chinese Academy of Sciences, 20A Datun Road, Chaoyang District, Beijing 100101, P. R. China}
\affiliation{Max-Planck-Institut fur Radioastronomie, Auf dem Hugel 69, D-53121 Bonn, Germany}
\author{Vladislav I. Kondratiev}
\affiliation{ASTRON, the Netherlands Institute for Radio Astronomy, Oude Hoogeveensedijk 4, 7991 PD Dwingeloo, The Netherlands}
\affiliation{Astro Space Centre, Lebedev Physical Institute, Russian Academy of Sciences, Profsoyuznaya Str. 84/32, Moscow 117997, Russia}
\author{Wendy M. Peters}
\affiliation{Naval Research Laboratory, Remote Sensing Division, Code 7213, Washington, DC 20375-5320, USA}

\author{Christoph Weniger}
\email{C.Weniger@uva.nl}
\affiliation{GRAPPA Institute, Institute of Physics, University of Amsterdam, 1098 XH Amsterdam, The Netherlands}

\author{Benjamin R. Safdi}
\email{bsafdi@umich.edu}
\affiliation{Leinweber Center for Theoretical Physics, Department of Physics, University of Michigan, Ann Arbor, MI 48109 U.S.A.}
\preprint{LCTP-20-04}
\date{\today}

\begin{abstract}
Axion dark matter (DM) may convert to radio-frequency electromagnetic radiation in the strong magnetic fields around neutron stars.  The radio signature of such a process would be an ultra-narrow spectral peak at a frequency determined by the mass of the axion particle.
We analyze data we collected from the Robert C. Byrd Green Bank Telescope in the L-band and the Effelsberg 100-m Telescope in the L-Band and S-band from a number of sources expected to produce bright signals of axion-photon conversion, including the Galactic Center of the Milky Way and the nearby isolated neutron stars RX J0720.4$-$3125 and RX J0806.4$-$4123.  We find no evidence for axion DM and are able to set some of the strongest constraints to-date on the existence of axion DM in the highly-motivated mass range between $\sim$5-11 $\mu$eV.

\end{abstract}
\maketitle
Recently, it was proposed that radio telescope observations of neutron stars (NSs) can probe axion dark matter (DM)~\cite{Pshirkov:2007st,Huang:2018lxq,Hook:2018iia,Safdi:2018oeu,Battye:2019aco}.  In the magnetosphere surrounding a NS, axion DM may resonantly convert into radio-frequency photons at locations where the plasma frequency of the magnetosphere equals the axion mass, with conversion probabilities determined in part by the strength of the magnetic field surrounding the NS. The central frequency of the hypothetical radio signal from an individual NS is set by the mass of the axion, red-shifted by the line-of-sight velocity of the NS.  
The predicted axion-induced radio signal would appear as a nearly monochromatic peak in the otherwise smoothly-varying radio spectrum from the NS and its nearby environment.  The frequency of this peak is universal for all sources and is determined by the currently unknown mass of the axion particle.    
 
In~\cite{Hook:2018iia,Safdi:2018oeu,Battye:2019aco,Leroy:2019ghm} it was shown that high-frequency-resolution observations with radio telescopes such as the Robert C. Byrd Green Bank Telescope (GBT) and the Effelsberg 100-m Telescope towards nearby isolated NSs (INSs) and towards regions of high NS and DM density, such as the Galactic Center (GC) of the Milky Way, would be sensitive to vast regions of previously unexplored axion parameter space.  In this work, we perform such searches with the GBT and the Effelsberg radio telescope.

The quantum chromodynamics (QCD) axion is a well-motivated DM candidate because in addition to explaining the observed abundance of DM~\cite{Preskill:1982cy,Abbott:1982af,Dine:1982ah} it may also resolve the strong {\it CP} problem of the neutron electron dipole moment~\cite{Peccei:1977ur,Peccei:1977hh,Weinberg:1977ma,Wilczek:1977pj} (see \cite{Irastorza:2018dyq} for a detailed review).
The QCD axion may make up the observed abundance of DM over a wide range of masses~\cite{Beringer:1900zz}, but a natural mass range is 5--25 $\mu$eV.
In this work we target axion masses in the range $m_a \in (4.5 , 10.5)$ $\mu$eV, corresponding to radio frequencies \mbox{$f = m_a / (2 \pi) \in (1.1, 2.7)$} GHz.

The conversion of axion DM to radio photons arises from the Lagrangian ${\mathcal L} = g_{a\gamma\gamma} \, a\, {\bf E} \cdot {\bf B}$,
where ${\bf E}$ (${\bf B}$) are electric (magnetic) fields, $a$ is the axion field, and $g_{a\gamma\gamma}$ is a coupling constant with units of inverse energy. For the QCD axion, $g_{a\gamma\gamma}$ is proportional to $m_a$, but models of more general axion-like particles can have $g_{a\gamma\gamma}$ and $m_a$ as independent parameters. The mass range that we target here with radio telescope searches is also the subject of significant longstanding laboratory search efforts for the coupling $\gagg$. The Rochester-Brookhaven-Fermilab (RBF)~\cite{DePanfilis:1987dk,Wuensch:1989sa} and University of Florida (UF)~\cite{Hagmann:1990tj} axion haloscope experiments set competitive constraints on axion DM in the mass range covered by this analysis, though our results exclude new parameter space beyond what was probed by those experiments.  More recently, the ADMX experiment has reached sensitivity to the QCD axion at
$\sim$2--3.5 $\mu$eV
~\cite{PhysRevLett.104.041301,Du:2018uak,Braine:2019fqb}, and the HAYSTAC experiment has set strong constraints on axion DM in the mass range $m_a \sim 23 - 24 \ \mu$eV~\cite{Brubaker:2017rna}.
 \newline
 
 {\it  Data acquisition.---}We collected data in the L-band ($1.15 - 1.73$ GHz) with the GBT and in the L-Band ($1.28-1.46$ GHz) and S-band ($2.4-2.7$ GHz) using the Effelsberg radio telescope to search for axion DM signatures from a variety of different sources.  We describe the data taking procedures from the two telescopes in turn.
 
 \noindent
 {\it GBT observations.} The GBT observations were performed with the VErsatile GBT Astronomical Spectrometer (VEGAS) backend~\cite{vegas}
 on March 10 and 29, 2019 with a notch filter applied from 1.2 to 1.34 GHz, so these frequencies are not included in our analysis (project AGBT19A\_362, PI: Safdi). The nearby INS targets observed by the GBT are summarized in Tab.~\ref{tab:targets}.  Note that we also observed the GC, M31, and M54 with the GBT, but the resulting axion limits are less robust than those from the INSs and from the Effelsberg GC observations and so are presented in the Supplementary Material (SM).  (The GBT GC observations lead to weaker limits than the Effelsberg GC observations because the GBT observations were taken with lower frequency resolution.) 
 All observations used the ``Spectral Line'' observing type and with one beam  
 covering an area on the sky  $\sim \pi({\rm FWHM}/2)^2$, where FWHM is the full width at half maximum of the telescope response, which is $8.4'$ at 1.5 GHz for the GBT. 

 The INS observations used five VEGAS spectrometers in mode 9 across the L-band, leading to the frequency resolution $\delta f_{\rm obs}$ reported in Tab.~\ref{tab:targets}.  
 For our fiducial analyses the data is further down-binned to resolutions $\delta f_{\rm fid}$ given in Tab.~\ref{tab:targets}. Data were collected in both polarizations, though in this analysis we only analyze the polarization-averaged flux. (See~\cite{Hook:2018iia} for possible polarization signatures.) The observations performed position switching so that for a given observational target, half the data collection time was on-source (``ON'') and half was spent observing blank-sky locations at similar elevations (``OFF'') in order to establish a reference baseline for the analysis.  The ON exposure times $t_{\rm exp}$ are listed in Tab.~\ref{tab:targets}. 
 The OFF locations were chosen to be $1.25^\circ$ away from the target of interest.
 The position-switching was carried out at five-minute intervals for each of the targets, leading to four separate observations of ON and OFF positions. 
 
  Over the observing period, data were saved in independent short exposures for ON and OFF observations of RX J0720.4$-$3125 and  RX J0806.4$-$4123. 
  In each successive exposure, a calibration noise diode was alternated between on and off with a switching period of 0.2097 seconds. The timing resolution allows for the identification of transient effects and data filtering, which is discussed further below and in the SM.  The calibration source 3C48 was observed for approximately two minutes to flux-calibrate the INS observations.  Additionally, we observed the star-forming region W3(OH) for approximately five minutes to verify that our analysis framework is able to successfully identify the OH maser lines.

 \noindent
 {\it Effelsberg observations.}  We also carried out L-Band and S-Band observations with the Effelsberg 100-m radio telescope towards the GC (project 77-17, 64-18, PI: Desvignes). The observations were taken with the PSRIX backend \cite{2016MNRAS.458..868L} $-$ performing baseband sampling $-$ in mid-June 2018 and early-February 2019 using the prime (secondary) focus receiver P217mm (S110mm) for the L- and S-band, respectively.  In both cases we recorded orthogonal polarizations, which were later averaged offline for further analysis. Note that the FWHM of the Effelsberg beam is $9.78'$ ($4.58'$) at 1.408 GHz (2.64 GHz). Observations were carried out towards the magnetar SGR~J1745$-$2900, which is $\sim$$2.4''$ away from the GC, and the planetary nebula ${\rm NGC}\,7027$ for subsequent use in the flux calibration procedure. For the measurements towards the GC we used a position switching mode, with ON-source integration times of 61.9 min and 40.0 min for S-band and L-band, respectively, and respective OFF-source integration times of 22.8 min and 37.0 min (see Tab.~\ref{tab:targets}).  
 The ON observation was performed first, followed by a single OFF observation taken $16.4^\circ$ away from the GC.
  \newline
 
  \begin{table}[t!]
\begin{tabular}{|c||c|c|c|c|c|}
\hline
Target        		&  $t_{\rm exp}$ [min]  	& $\delta f_{\rm obs}$ [kHz] & $\delta f_{\rm fid}$ [kHz] &     type			\\ \hline
RX J0806.4$-$4123 	& $20.0$ 			& 0.8 & $8.4$		& \, INS \,       		\\ \hline
RX J0720.4$-$3125 	& $20.0 $     	& 0.8  & $8.4$        		& \, INS  \,       	\\ \hline
GC (Eff., S-Band) 	& $61.9$    	& 3.81 & $11.44$        		& \, pop. \,   		\\ \hline
GC (Eff., L-Band) 	& $40.0$    & $2.44$	& $7.32$        		& \, pop. \,   		\\ \hline

\end{tabular}

\caption{ {The targets observed by the GBT and Effelsberg for evidence of axion DM. ``Pop.'' refers to populations of NSs, while ``INS'' refers to a single isolated NS.  The bin widths $\delta f_{\rm obs}$ correspond to those of the original observation, but we down-bin the data before performing the axion line search to the resolution given by $\delta f_{\rm fid}$ to account for the finite width of the signal.  The INS (GC) observations were performed with the GBT (Effelsberg radio telescope). The GBT INS observations cover the frequency range 1.15 to 1.73 GHz, with a gap from 1.2 to 1.35 GHz, and the L-band (S-band) Effeslberg observation covers 1.28 to 1.46 GHz (2.4 to 2.7 GHz).
Note that the $t_{\rm exp}$ are the ON exposure times.}
\label{tab:targets}
}
\end{table}
 
 {\it Analysis.---}We reduced and calibrated the GBT data following a modified implementation of the \texttt{GBTIDL} data reduction pipeline \cite{garwood_marganian_braatz_maddalena_2005}, extended to include a time-series data filtering performed independently at each channel and a channel-dependent system temperature calibration. The full procedure results in measurements of flux densities $\{d_i\}$ at frequencies $\{f_i\}$, with $i$ labeling the frequency channel. Because the stacked, calibrated data has been constructed by averaging many ($>10^3$) independent antenna measurements together, the $\{d_i\}$ are approximately normally distributed. 

For Effelsberg, high-resolution frequency spectra (131072 spectral channels) were generated from the raw `baseband' data using the \texttt{DSPSR}\footnote{\url{http://dspsr.sourceforge.net}} software tools~\cite{dspsr}. 
We used the full-integrated spectra in our analysis, with a calibration procedure described in the SM.
Before analyzing the data we first down-bin in frequency space to bins of width $\sim$$8$ kHz (see~Tab.~\ref{tab:targets}) to account for the finite width of the signal, such that the majority of the signal should appear in a single frequency bin. 
As discussed further in the SM and first suggested in~\cite{Battye:2019aco}, reflection and refraction of the outgoing electromagnetic waves in the rotating plasma induces a frequency broadening at the level $\delta f / f \sim 5 \times 10^{-6}$ or less from the INSs.  
We note that even though the Effelsberg observations are searching for emission from a population of NSs, the data are at sufficiently high frequency resolution that we may search simply for the brightest converting NS from that population.

To inspect the data for excess flux at frequency channel $i$, we construct the likelihood 
\begin{equation}
\mathcal{L}_i( \vec{d}| A, \mathbf{a}) = \prod_{k} {1 \over \sqrt{2 \pi \sigma_k^2}} \exp \left[ - \frac{(d_k - \mu(f_k | \mathbf{a}) - A \delta_{ik})^2}{2 \sigma_k^2}\right]\,,
\label{eq:Likelihood}
\end{equation}
where $A$ is the excess flux density in the central frequency channel.  Note that the index $k$ labels the analysis-level frequency channel, and the product runs over the frequency bins included in the analysis window.
We model the background in the narrow sliding frequency window with a frequency-dependent mean flux density $\mu(f | \mathbf{a})$ and a single variance parameter $\sigma^2$, such that the variance in each frequency channel is given by  $\sigma_i^2 =  \sigma^2 / \alpha_i$ for an acceptance fraction $\alpha_i$ of data at frequency channel $i$ after the data filtering. Note that $\alpha_i = 1$ for all Effelsberg frequency channels as we do not apply the time-filtering procedure to that data. The nuisance parameter vector ${\bf a}$ characterizes the frequency dependence of the mean; in practice we take $\mu$ to be a quadratic function of $f$ so that ${\bf a}$ has three independent parameters, though our final results are not sensitive to this choice (see the SM).

\begin{figure}
  \includegraphics[width = .49\textwidth]{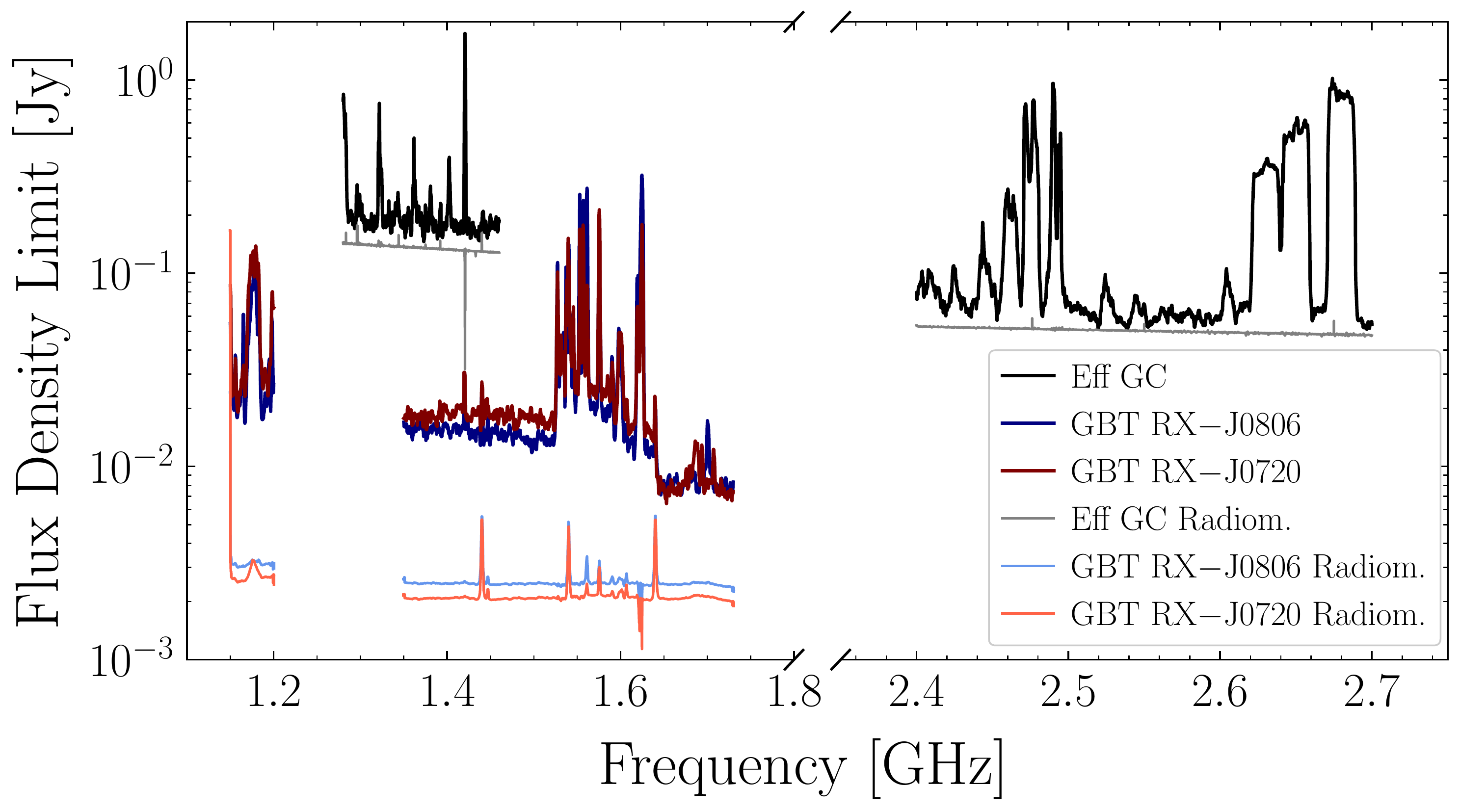} 
  \caption{
  \label{fig:flux_constraints} The 95\% upper limits on the signal flux for the indicated sources from the GBT and Effelsberg observations.  These upper limits apply to monochromatic signals at the widths $\delta f_{\rm fid}$ given in Tab.~\ref{tab:targets}. These curves have been down-sampled for visualization purposes. We compare these limits with the 95\% upper limits expected from the ideal radiometer equation under the assumption that the only source of statistical uncertainty is thermal noise at the total system temperature.
  }
\end{figure}

In our fiducial analysis we include within the sliding analysis window the $10$ frequency bins to the left and to the right of the central frequency channel,
excluding the two bins on either side of the signal bin in case of signal leakage into those bins, if {\it e.g.} the axion mass does not line up with the bin center. Note that to account for this possibility we also perform the analyses with all frequency bins shifted by approximately half a bin spacing.
The variance parameter $\sigma^2$ is fixed by fitting the background-only model to the frequency sidebands with the central frequency channel masked out.
We construct the profile likelihood $\mathcal{L}_i( \vec{d} | A)$ by maximizing $\mathcal{L}_i( \vec{d} | A, \mathbf{a})$ over the nuisance parameters ${\bf a}$ at each fixed value of $A$, and we use the profile likelihood to construct the one-sided 95\% upper limit on the flux density as shown in Fig.~\ref{fig:flux_constraints} (see, {\it e.g.},~\cite{Cowan:2010js}).  In particular, we consider positive and negative values of $A$ and we take the 95\% upper limit to be the value of $A > \hat A$ such that \mbox{$2 [ \ln \mathcal{L}_i( \vec{d} | A) - \ln \mathcal{L}_i( \vec{d} | \hat A)] \approx - 2.71$}, where $\hat A$ is the signal parameter that maximizes the profile likelihood.  We then further power-constrain our limits to avoid setting limits that are stronger than expected due to downward statistical fluctuations~\cite{Cowan:2011an}.  We accomplish this by
recording the actual limit as the maximum of the 16$^{\rm th}$ percentile of the distribution of expected limits under the null hypothesis, as computed using the Asimov procedure~\cite{Cowan:2010js}, and the limit observed on the actual data.
Our test-statistic (TS) for comparing signal and null hypotheses for evidence of an axion is the  log-likelihood ratio $\mathrm{TS}_i \equiv 2 \times [ \ln \mathcal{L}_i(\vec{d} | \hat A ) - \ln \mathcal{L}_i(\vec{d} |0)]$, for $\hat A > 0$, and ${\rm TS}_i = 0$ if $\hat A < 0$. 

We additionally analyze the stacked but uncalibrated OFF spectra.
This is valuable because the OFF data are subtracted and divided from the ON data to remove the instrumental baselines, but this may cause features in the OFF spectra to be imprinted on the calibrated flux densities.
Therefore, statistically significant excesses that appear in both the calibrated source flux density spectra and the OFF system temperatures can be vetoed as they are inconsistent with, or at least do not require, an axion interpretation. In our analysis, we veto any excess in the calibrated ON data which appears with a $97.5^\mathrm{th}$ percentile discovery TS in the OFF data. Note that we determine the TS percentiles by using the full distribution of observed TSs. 

The 95\% upper limits on the flux densities, defined relative to the single-channel frequency bin widths $\delta f_{\rm fid}$ given in Tab.~\ref{tab:targets}, are shown in Fig.~\ref{fig:flux_constraints}.
We compare the upper limits to the expected limits from the ideal radiometer equation, which assumes that all of the noise is thermal at the system temperature.  The true limits are slightly weaker likely because of sources of systematic uncertainty, such as uncertainties in the background model and instrumental  uncertainties not fully captured by the calibration procedure. 

We search for evidence of an axion signal by using the discovery TSs.  We apply a discovery threshold of ${\rm TS} > 100$, which was defined before performing the analysis and not modified afterwards.  From Monte Carlo (MC) simulations of the null hypothesis we find that this TS threshold corresponds to approximately 5$\sigma$ local significance (see the SM for details). 
After applying the analysis procedure described above we find no axion signal candidates at or beyond the detection significance in any of the observations, and the distributions of observed TSs are consistent with the null hypothesis.  Note that HI emission frequencies are excluded automatically in our analysis by the OFF veto criterion. 
\newline

{\it Results.---}To translate the flux-density limits from Fig.~\ref{fig:flux_constraints} into limits on the axion-photon coupling, we closely follow the theoretical modeling presented in~\cite{Hook:2018iia,Safdi:2018oeu} for computing the axion-induced radio fluxes from these specific sources.

The radiated power for a single INS depends on $g_{a \gamma\gamma}$, the polar magnetic field strength $B_0$ (assuming a dipole field configuration), the NS mass (which we fix at $1$ $M_\odot$, since this value does not significantly affect the flux), the NS spin period $P$, the axion mass $m_a$, the DM density $\rho_\infty$ in the neighborhood of the NS but asymptotically far away from its gravitational potential, and the velocity dispersion $v_0$ of the ambient DM.
For the local INSs we take $v_0 = 200$ km$/$s and $\rho_{\infty}  = 0.4$ GeV$/$cm$^3$~\cite{Bovy:2012tw,Read:2014qva,Schutz:2017tfp}. For the GC analysis we assume the DM follows an Navarro-Frenk-White (NFW)~\cite{Navarro:1995iw,Navarro:1996gj} density profile near the GC, normalized to give the local DM density above and with a scale radius of 20 kpc (see, {\it e.g.},~\cite{Safdi:2018oeu}). For RX J0806.4$-$4123 we take $\log_{10}(B_0 / G) =13.40$ and $P = 11.4 \, \mathrm{s}$, while for RX J0720.4$-$3125 we use $\log_{10}(B_0 /G) = 13.53$ and $P = 8.4 \, \mathrm{s}$. We note that these parameters were inferred from spin-down measurements performed in the $X$-ray band~\cite{Kaplan:2005fy,Kaplan:2009ce,Buschmann:2019pfp}. We take RX J0806.4$-$4123 and RX J0720.4$-$3125 to be at distances of 250 pc and 360 pc from Earth, respectively \cite{Kaplan:2009ce}.

Given these parameters, we estimate the radiated power following~\cite{Hook:2018iia,Leroy:2019ghm}.  However, we note that a fully self consistent calculation of the axion-induced radiation has yet to be performed.  Ref.~\cite{Leroy:2019ghm} corrected the assumption in~\cite{Hook:2018iia} that the axions travel along radial trajectories, but~\cite{Leroy:2019ghm} did not account for the fact that the outgoing radiation is strongly refracted in the inhomogeneous magnetosphere, as we point out in the SM.  As a dedicated simulation of the axion-induced radiation is beyond the scope of this work, we estimate the power with the following approximation.  We assume that (i) all axions travel along radial trajectories, as in~\cite{Hook:2018iia}, (ii) that all NSs are aligned rotators (magnetically-misaligned rotators give nearly identical results~\cite{Hook:2018iia}), 
and (iii) that the magnetosphere is well-described by the Goldreich-Julian model~\cite{1969ApJ...157..869G} (see~\cite{Safdi:2018oeu} where it is shown that more complicated magnetosphere models give similar results).  Then, following~\cite{Hook:2018iia} we compute the angular power distribution $dP/d\theta$ of radio emission as a function of the angle from the polar axis $\theta$.  However, we assign to each NS a single power value equal to $\int \frac{dP}{d\theta} \, d\theta$, and we assume that the flux is radiated from each NS isotropically.  With the latter assumption we find results that agree to within a factor $\sim$2 with those in~\cite{Leroy:2019ghm}, which correctly accounted for the isotropic axion phase space.  We chose this simpler formalism, however, because it is likely that the more complicated computation in~\cite{Leroy:2019ghm} must be modified due to the refraction of outgoing radio photons, which could result in an anisotropic signal.  
Given an improved theoretical predictions in the future, our results may be reinterpreted using the \href{https://github.com/joshwfoster/RadioAxionSearch}{Supplementary Data}~\cite{SM}.    

The width of the signal in frequency space is determined in part by the asymptotic energy dispersion of the DM, which is set by $v_0$.  This induces a $\delta f / f \lesssim 10^{-6}$ contribution to the width from the INSs.  However, as discussed more in the SM and in~\cite{Battye:2019aco}, the signals are Doppler-broadened when refracting or reflecting from the rotating plasma, inducing a frequency broadening closer to $\delta f / f \sim 5 \times 10^{-6}$ and justifying the bin widths taken in Tab.~\ref{tab:targets}. 

Since we do not actually know which specific NSs are being targeted in the Effeslberg GC analysis (and similarly in the GBT population analyses discussed in the SM),
we model the population of NSs (number density, spatial distribution, magnetic field, and spin period) within the GC region as a whole, closely following~\cite{Safdi:2018oeu}.  In particular, two models for the NS magnetic field and period distributions were developed in that work, based on fits to existing pulsar data.  We conservatively choose the model which yields weaker constraints as our fiducial model. In practice, our fiducial NS population model (Model II in~\cite{Safdi:2018oeu}) assumes that magnetic fields quickly decay after the NSs cross the pulsar death-line, while the optimistic model (Model I in~\cite{Safdi:2018oeu}) assumes that the magnetic fields decay more slowly.  We also follow~\cite{Safdi:2018oeu} when modeling the spatial distribution of NSs within the Galactic bulge and disk.  
For the Effelsberg analysis, we perform $\mathcal{O}(10^3)$ MC simulations of the NS population model and profile over the simulation results when calculating the expected flux and associated 95\% limit.
 
 \begin{figure}[t!]
  \includegraphics[width = .49\textwidth]{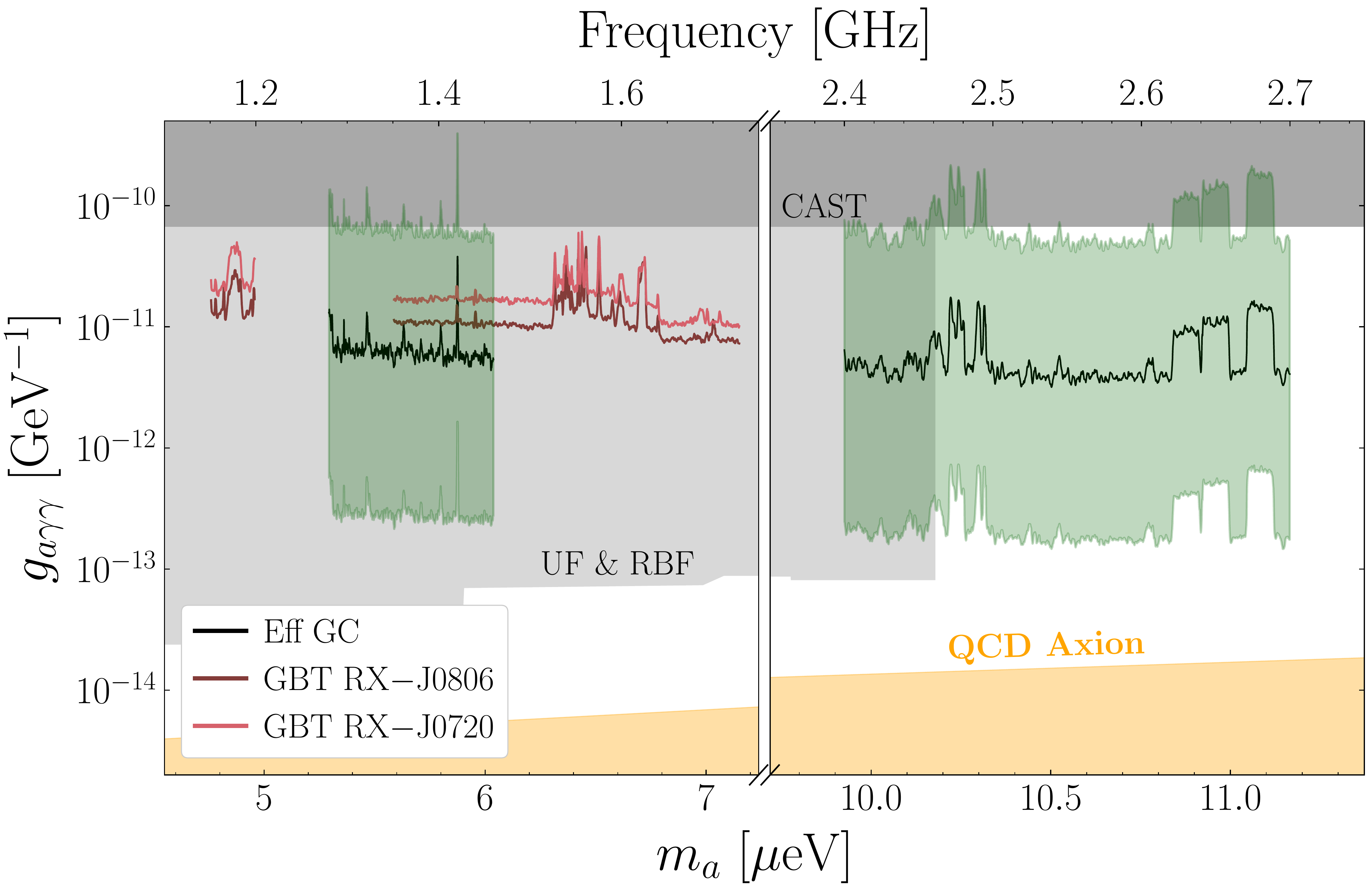} 
  \caption{
  \label{fig:money} 
  The one-sided 95\% upper limits on $\gagg$ as a function of the axion mass $m_a$ from this work are shown as colored lines (GBT INS observations) and black lines (Effelsberg GC observations). Previous limits from the CAST helioscope and the UF and RBF haloscopes are shown in shaded grey. The range of couplings expected for the QCD axion is shaded in orange. Note that the fiducial GC limits assume an NFW DM profile
  and the conservative NS population model (Model II) from~\cite{Safdi:2018oeu}. The green band depicts theoretical uncertainties on the $\gagg$ limit associated with the GC analysis for the Effelsberg data.  The top of the band assumes an NFW DM density profile with a 0.6 kpc core, while the bottom of the band uses the alternate NS population model in~\cite{Safdi:2018oeu} (Model I). 
  }
\end{figure}

Given the fiducial models we have described, we obtain the limits on $\gagg$ shown in Fig.~\ref{fig:money}. The orange band represents the predicted $\gagg$ for the QCD axion, and the grey shaded regions represent existing constraints from other experiments. We obtain limits that are stronger than those from CAST \cite{Anastassopoulos:2017ftl} and comparable to constraints from the UF~\cite{Hagmann:1990tj} and RBF~\cite{DePanfilis:1987dk,Wuensch:1989sa} haloscopes, while the S-band Effelsberg constraints exclude previously unexplored parameter space. The green shaded band in Fig.~\ref{fig:money} represents two dominant sources of uncertainty for the GC analysis. The top of the band is derived by assuming that the DM profile follows a cored density profile with a core radius of 0.6 kpc; this radius is chosen based on recent hydrodynamic simulations which suggest that the DM density may be modified in the inner $\sim$0.6 kpc where the baryons dominate the gravitational potential, though these same simulations suggest an enhancement of the central DM density may also be possible ~\cite{Hopkins:2017ycn}. The lower boundary of the band assumes the fiducial NFW DM profile but takes the alternate NS population model (Model I) from~\cite{Safdi:2018oeu}.
\newline

{\it Discussion.---}In this work we performed the first dedicated radio telescope search for signatures of axion DM from axion-photon conversion in NS magnetospheres.  We found no evidence for axion DM and set some of the strongest constraints to date on the axion DM scenario.
These results show that radio searches for axion DM are a promising path forward, analogous to indirect detection for WIMP DM searches, which should proceed in parallel with laboratory experiments for discovering or excluding axion DM.  Additional flux sensitivity is needed in order to reach the QCD axion band at the frequencies targeted in this work.  This sensitivity may be available with the upcoming Square Kilometer Array-mid \cite{Bull:2018lat} or may already be achievable with the FAST radio telescope~\cite{2011IJMPD..20..989N}, since at constant system temperature the sensitivity to $g_{a\gamma\gamma}$ scales inversely with the square root of the effective area~\cite{Safdi:2018oeu}.

Our work strongly motivates searching with the GBT or Effelsberg radio telescope for evidence of axion DM at higher frequencies, closer to 6 GHz, to probe the axion mass window around $m_a \approx 25$ $\mu$eV.  There is mounting evidence that points towards 25 $\mu$eV as a likely mass for the axion~\cite{Buschmann:2019icd,Klaer:2017ond}, and the axion-photon coupling may also be enhanced~\cite{Agrawal:2017cmd} and thus within reach of GBT and Effelsberg searches.
This work also motivates additional effort in modeling the population evolution of NS magnetic fields and spin periods, as these are the largest sources of uncertainty in our population analyses, as well as further efforts to understand the distribution of DM in the inner Galaxy.  
More work on the axion-induced signal itself from individual INSs would be also useful, as a full calculation of the axion-induced radio signal does not yet exist; such results could lead to reinterpretations of the limits presented in this Letter using the \href{https://github.com/joshwfoster/RadioAxionSearch}{Supplementary Data}~\cite{SM}. 

{\it Acknowledgments.---}{\it The Green Bank Observatory is a facility of the National Science Foundation operated under cooperative agreement by Associated Universities, Inc. This work was based on observations with the $100\,{\rm m}$ telescope of the Max-Planck-Institut f\"{u}r Radioastronomie at Effelsberg. We thank Amber Bonsall for assistance with GBT data-taking, and Anson Hook for collaboration in the early stages of this work.  We also thank 
Karl Van Bibber, Christopher Dessert, Gregory Desvignes, Tom Edwards, Reyco Henning, Jonathan Ouellet, Nicholas Rodd, Katelin Schutz, Chenoa Tremblay, Mauro Valli, and Kathryn Zurek for helpful comments and discussions. This work was supported in part by the DOE Early Career Grant DESC0019225 and through computational resources and services provided by Advanced Research Computing at the University of Michigan, Ann Arbor. JF was supported in part by the Munich Institute for Astro- and Particle Physics (MIAPP) which is funded by the Deutsche Forschungsgemeinschaft (DFG, German Research Foundation) under Germany's Excellence Strategy – EXC-2094 – 390783311.  OM acknowledges support by JSPS KAKENHI Grant Numbers JP17H04836, JP18H04340, JP18H04578 and by World Premier International Research Center Initiative (WPI Initiative), MEXT, Japan. RE acknowledges financial support by the European Research Council for the ERC Synergy Grant BlackHoleCam under contract no. 610058. Basic research in radio astronomy at the Naval Research Laboratory is supported by 6.1 Base funding.  Digitized versions of main figures along with raw and processed data products may be found in the \href{https://github.com/joshwfoster/RadioAxionSearch}{Supplementary Data}~\cite{SM}.}

\bibliography{axion}

\begin{thebibliography}{56}%
\makeatletter
\providecommand \@ifxundefined [1]{%
 \@ifx{#1\undefined}
}%
\providecommand \@ifnum [1]{%
 \ifnum #1\expandafter \@firstoftwo
 \else \expandafter \@secondoftwo
 \fi
}%
\providecommand \@ifx [1]{%
 \ifx #1\expandafter \@firstoftwo
 \else \expandafter \@secondoftwo
 \fi
}%
\providecommand \natexlab [1]{#1}%
\providecommand \enquote  [1]{``#1''}%
\providecommand \bibnamefont  [1]{#1}%
\providecommand \bibfnamefont [1]{#1}%
\providecommand \citenamefont [1]{#1}%
\providecommand \href@noop [0]{\@secondoftwo}%
\providecommand \href [0]{\begingroup \@sanitize@url \@href}%
\providecommand \@href[1]{\@@startlink{#1}\@@href}%
\providecommand \@@href[1]{\endgroup#1\@@endlink}%
\providecommand \@sanitize@url [0]{\catcode `\\12\catcode `\$12\catcode
  `\&12\catcode `\#12\catcode `\^12\catcode `\_12\catcode `\%12\relax}%
\providecommand \@@startlink[1]{}%
\providecommand \@@endlink[0]{}%
\providecommand \url  [0]{\begingroup\@sanitize@url \@url }%
\providecommand \@url [1]{\endgroup\@href {#1}{\urlprefix }}%
\providecommand \urlprefix  [0]{URL }%
\providecommand \Eprint [0]{\href }%
\providecommand \doibase [0]{http://dx.doi.org/}%
\providecommand \selectlanguage [0]{\@gobble}%
\providecommand \bibinfo  [0]{\@secondoftwo}%
\providecommand \bibfield  [0]{\@secondoftwo}%
\providecommand \translation [1]{[#1]}%
\providecommand \BibitemOpen [0]{}%
\providecommand \bibitemStop [0]{}%
\providecommand \bibitemNoStop [0]{.\EOS\space}%
\providecommand \EOS [0]{\spacefactor3000\relax}%
\providecommand \BibitemShut  [1]{\csname bibitem#1\endcsname}%
\let\auto@bib@innerbib\@empty
\bibitem [{\citenamefont {Pshirkov}(2009)}]{Pshirkov:2007st}%
  \BibitemOpen
  \bibfield  {author} {\bibinfo {author} {\bibfnamefont {M.~S.}\ \bibnamefont
  {Pshirkov}},\ }\bibfield  {title} {\enquote {\bibinfo {title} {{Conversion of
  Dark matter axions to photons in magnetospheres of neutron stars}},}\ }\href
  {\doibase 10.1134/S1063776109030030} {\bibfield  {journal} {\bibinfo
  {journal} {J. Exp. Theor. Phys.}\ }\textbf {\bibinfo {volume} {108}},\
  \bibinfo {pages} {384--388} (\bibinfo {year} {2009})},\ \Eprint
  {http://arxiv.org/abs/0711.1264} {arXiv:0711.1264 [astro-ph]} \BibitemShut
  {NoStop}%
\bibitem [{\citenamefont {Huang}\ \emph {et~al.}(2018)\citenamefont {Huang},
  \citenamefont {Kadota}, \citenamefont {Sekiguchi},\ and\ \citenamefont
  {Tashiro}}]{Huang:2018lxq}%
  \BibitemOpen
  \bibfield  {author} {\bibinfo {author} {\bibfnamefont {Fa~Peng}\ \bibnamefont
  {Huang}}, \bibinfo {author} {\bibfnamefont {Kenji}\ \bibnamefont {Kadota}},
  \bibinfo {author} {\bibfnamefont {Toyokazu}\ \bibnamefont {Sekiguchi}}, \
  and\ \bibinfo {author} {\bibfnamefont {Hiroyuki}\ \bibnamefont {Tashiro}},\
  }\bibfield  {title} {\enquote {\bibinfo {title} {{Radio telescope search for
  the resonant conversion of cold dark matter axions from the magnetized
  astrophysical sources}},}\ }\href {\doibase 10.1103/PhysRevD.97.123001}
  {\bibfield  {journal} {\bibinfo  {journal} {Phys. Rev.}\ }\textbf {\bibinfo
  {volume} {D97}},\ \bibinfo {pages} {123001} (\bibinfo {year} {2018})},\
  \Eprint {http://arxiv.org/abs/1803.08230} {arXiv:1803.08230 [hep-ph]}
  \BibitemShut {NoStop}%
\bibitem [{\citenamefont {Hook}\ \emph {et~al.}(2018)\citenamefont {Hook},
  \citenamefont {Kahn}, \citenamefont {Safdi},\ and\ \citenamefont
  {Sun}}]{Hook:2018iia}%
  \BibitemOpen
  \bibfield  {author} {\bibinfo {author} {\bibfnamefont {Anson}\ \bibnamefont
  {Hook}}, \bibinfo {author} {\bibfnamefont {Yonatan}\ \bibnamefont {Kahn}},
  \bibinfo {author} {\bibfnamefont {Benjamin~R.}\ \bibnamefont {Safdi}}, \ and\
  \bibinfo {author} {\bibfnamefont {Zhiquan}\ \bibnamefont {Sun}},\ }\bibfield
  {title} {\enquote {\bibinfo {title} {{Radio Signals from Axion Dark Matter
  Conversion in Neutron Star Magnetospheres}},}\ }\href {\doibase
  10.1103/PhysRevLett.121.241102} {\bibfield  {journal} {\bibinfo  {journal}
  {Phys. Rev. Lett.}\ }\textbf {\bibinfo {volume} {121}},\ \bibinfo {pages}
  {241102} (\bibinfo {year} {2018})},\ \Eprint
  {http://arxiv.org/abs/1804.03145} {arXiv:1804.03145 [hep-ph]} \BibitemShut
  {NoStop}%
\bibitem [{\citenamefont {Safdi}\ \emph {et~al.}(2019)\citenamefont {Safdi},
  \citenamefont {Sun},\ and\ \citenamefont {Chen}}]{Safdi:2018oeu}%
  \BibitemOpen
  \bibfield  {author} {\bibinfo {author} {\bibfnamefont {Benjamin~R.}\
  \bibnamefont {Safdi}}, \bibinfo {author} {\bibfnamefont {Zhiquan}\
  \bibnamefont {Sun}}, \ and\ \bibinfo {author} {\bibfnamefont {Alexander~Y.}\
  \bibnamefont {Chen}},\ }\bibfield  {title} {\enquote {\bibinfo {title}
  {{Detecting Axion Dark Matter with Radio Lines from Neutron Star
  Populations}},}\ }\href {\doibase 10.1103/PhysRevD.99.123021} {\bibfield
  {journal} {\bibinfo  {journal} {Phys. Rev.}\ }\textbf {\bibinfo {volume}
  {D99}},\ \bibinfo {pages} {123021} (\bibinfo {year} {2019})},\ \Eprint
  {http://arxiv.org/abs/1811.01020} {arXiv:1811.01020 [astro-ph.CO]}
  \BibitemShut {NoStop}%
\bibitem [{\citenamefont {Battye}\ \emph {et~al.}(2019)\citenamefont {Battye},
  \citenamefont {Garbrecht}, \citenamefont {McDonald}, \citenamefont {Pace},\
  and\ \citenamefont {Srinivasan}}]{Battye:2019aco}%
  \BibitemOpen
  \bibfield  {author} {\bibinfo {author} {\bibfnamefont {Richard~A.}\
  \bibnamefont {Battye}}, \bibinfo {author} {\bibfnamefont {Bjoern}\
  \bibnamefont {Garbrecht}}, \bibinfo {author} {\bibfnamefont {Jamie~I.}\
  \bibnamefont {McDonald}}, \bibinfo {author} {\bibfnamefont {Francesco}\
  \bibnamefont {Pace}}, \ and\ \bibinfo {author} {\bibfnamefont {Sankarshana}\
  \bibnamefont {Srinivasan}},\ }\bibfield  {title} {\enquote {\bibinfo {title}
  {{Dark matter axion detection in the radio/mm-waveband}},}\ }\href@noop {} {\
   (\bibinfo {year} {2019})},\ \Eprint {http://arxiv.org/abs/1910.11907}
  {arXiv:1910.11907 [astro-ph.CO]} \BibitemShut {NoStop}%
\bibitem [{\citenamefont {Leroy}\ \emph {et~al.}(2019)\citenamefont {Leroy},
  \citenamefont {Chianese}, \citenamefont {Edwards},\ and\ \citenamefont
  {Weniger}}]{Leroy:2019ghm}%
  \BibitemOpen
  \bibfield  {author} {\bibinfo {author} {\bibfnamefont {Mikaël}\ \bibnamefont
  {Leroy}}, \bibinfo {author} {\bibfnamefont {Marco}\ \bibnamefont {Chianese}},
  \bibinfo {author} {\bibfnamefont {Thomas D.~P.}\ \bibnamefont {Edwards}}, \
  and\ \bibinfo {author} {\bibfnamefont {Christoph}\ \bibnamefont {Weniger}},\
  }\bibfield  {title} {\enquote {\bibinfo {title} {{Radio Signal of Axion
  Photon Conversion in Neutron Stars: A Ray Tracing Analysis}},}\ }\href@noop
  {} {\  (\bibinfo {year} {2019})},\ \Eprint {http://arxiv.org/abs/1912.08815}
  {arXiv:1912.08815 [hep-ph]} \BibitemShut {NoStop}%
\bibitem [{\citenamefont {Preskill}\ \emph {et~al.}(1983)\citenamefont
  {Preskill}, \citenamefont {Wise},\ and\ \citenamefont
  {Wilczek}}]{Preskill:1982cy}%
  \BibitemOpen
  \bibfield  {author} {\bibinfo {author} {\bibfnamefont {John}\ \bibnamefont
  {Preskill}}, \bibinfo {author} {\bibfnamefont {Mark~B.}\ \bibnamefont
  {Wise}}, \ and\ \bibinfo {author} {\bibfnamefont {Frank}\ \bibnamefont
  {Wilczek}},\ }\bibfield  {title} {\enquote {\bibinfo {title} {{Cosmology of
  the Invisible Axion}},}\ }\href {\doibase 10.1016/0370-2693(83)90637-8}
  {\bibfield  {journal} {\bibinfo  {journal} {Phys. Lett.}\ }\textbf {\bibinfo
  {volume} {B120}},\ \bibinfo {pages} {127--132} (\bibinfo {year}
  {1983})}\BibitemShut {NoStop}%
\bibitem [{\citenamefont {Abbott}\ and\ \citenamefont
  {Sikivie}(1983)}]{Abbott:1982af}%
  \BibitemOpen
  \bibfield  {author} {\bibinfo {author} {\bibfnamefont {L.~F.}\ \bibnamefont
  {Abbott}}\ and\ \bibinfo {author} {\bibfnamefont {P.}~\bibnamefont
  {Sikivie}},\ }\bibfield  {title} {\enquote {\bibinfo {title} {{A Cosmological
  Bound on the Invisible Axion}},}\ }\href {\doibase
  10.1016/0370-2693(83)90638-X} {\bibfield  {journal} {\bibinfo  {journal}
  {Phys. Lett.}\ }\textbf {\bibinfo {volume} {B120}},\ \bibinfo {pages}
  {133--136} (\bibinfo {year} {1983})}\BibitemShut {NoStop}%
\bibitem [{\citenamefont {Dine}\ and\ \citenamefont
  {Fischler}(1983)}]{Dine:1982ah}%
  \BibitemOpen
  \bibfield  {author} {\bibinfo {author} {\bibfnamefont {Michael}\ \bibnamefont
  {Dine}}\ and\ \bibinfo {author} {\bibfnamefont {Willy}\ \bibnamefont
  {Fischler}},\ }\bibfield  {title} {\enquote {\bibinfo {title} {{The Not So
  Harmless Axion}},}\ }\href {\doibase 10.1016/0370-2693(83)90639-1} {\bibfield
   {journal} {\bibinfo  {journal} {Phys. Lett.}\ }\textbf {\bibinfo {volume}
  {B120}},\ \bibinfo {pages} {137--141} (\bibinfo {year} {1983})}\BibitemShut
  {NoStop}%
\bibitem [{\citenamefont {Peccei}\ and\ \citenamefont
  {Quinn}(1977{\natexlab{a}})}]{Peccei:1977ur}%
  \BibitemOpen
  \bibfield  {author} {\bibinfo {author} {\bibfnamefont {R.~D.}\ \bibnamefont
  {Peccei}}\ and\ \bibinfo {author} {\bibfnamefont {Helen~R.}\ \bibnamefont
  {Quinn}},\ }\bibfield  {title} {\enquote {\bibinfo {title} {{Constraints
  Imposed by CP Conservation in the Presence of Instantons}},}\ }\href
  {\doibase 10.1103/PhysRevD.16.1791} {\bibfield  {journal} {\bibinfo
  {journal} {Phys. Rev.}\ }\textbf {\bibinfo {volume} {D16}},\ \bibinfo {pages}
  {1791--1797} (\bibinfo {year} {1977}{\natexlab{a}})}\BibitemShut {NoStop}%
\bibitem [{\citenamefont {Peccei}\ and\ \citenamefont
  {Quinn}(1977{\natexlab{b}})}]{Peccei:1977hh}%
  \BibitemOpen
  \bibfield  {author} {\bibinfo {author} {\bibfnamefont {R.~D.}\ \bibnamefont
  {Peccei}}\ and\ \bibinfo {author} {\bibfnamefont {Helen~R.}\ \bibnamefont
  {Quinn}},\ }\bibfield  {title} {\enquote {\bibinfo {title} {{CP Conservation
  in the Presence of Instantons}},}\ }\href {\doibase
  10.1103/PhysRevLett.38.1440} {\bibfield  {journal} {\bibinfo  {journal}
  {Phys. Rev. Lett.}\ }\textbf {\bibinfo {volume} {38}},\ \bibinfo {pages}
  {1440--1443} (\bibinfo {year} {1977}{\natexlab{b}})}\BibitemShut {NoStop}%
\bibitem [{\citenamefont {Weinberg}(1978)}]{Weinberg:1977ma}%
  \BibitemOpen
  \bibfield  {author} {\bibinfo {author} {\bibfnamefont {Steven}\ \bibnamefont
  {Weinberg}},\ }\bibfield  {title} {\enquote {\bibinfo {title} {{A New Light
  Boson?}}}\ }\href {\doibase 10.1103/PhysRevLett.40.223} {\bibfield  {journal}
  {\bibinfo  {journal} {Phys.Rev.Lett.}\ }\textbf {\bibinfo {volume} {40}},\
  \bibinfo {pages} {223--226} (\bibinfo {year} {1978})}\BibitemShut {NoStop}%
\bibitem [{\citenamefont {Wilczek}(1978)}]{Wilczek:1977pj}%
  \BibitemOpen
  \bibfield  {author} {\bibinfo {author} {\bibfnamefont {Frank}\ \bibnamefont
  {Wilczek}},\ }\bibfield  {title} {\enquote {\bibinfo {title} {{Problem of
  Strong P and T Invariance in the Presence of Instantons}},}\ }\href {\doibase
  10.1103/PhysRevLett.40.279} {\bibfield  {journal} {\bibinfo  {journal}
  {Phys.Rev.Lett.}\ }\textbf {\bibinfo {volume} {40}},\ \bibinfo {pages}
  {279--282} (\bibinfo {year} {1978})}\BibitemShut {NoStop}%
\bibitem [{\citenamefont {Irastorza}\ and\ \citenamefont
  {Redondo}(2018)}]{Irastorza:2018dyq}%
  \BibitemOpen
  \bibfield  {author} {\bibinfo {author} {\bibfnamefont {Igor~G.}\ \bibnamefont
  {Irastorza}}\ and\ \bibinfo {author} {\bibfnamefont {Javier}\ \bibnamefont
  {Redondo}},\ }\bibfield  {title} {\enquote {\bibinfo {title} {{New
  experimental approaches in the search for axion-like particles}},}\ }\href
  {\doibase 10.1016/j.ppnp.2018.05.003} {\bibfield  {journal} {\bibinfo
  {journal} {Prog. Part. Nucl. Phys.}\ }\textbf {\bibinfo {volume} {102}},\
  \bibinfo {pages} {89--159} (\bibinfo {year} {2018})},\ \Eprint
  {http://arxiv.org/abs/1801.08127} {arXiv:1801.08127 [hep-ph]} \BibitemShut
  {NoStop}%
\bibitem [{\citenamefont {Beringer}\ \emph {et~al.}(2012)\citenamefont
  {Beringer} \emph {et~al.}}]{Beringer:1900zz}%
  \BibitemOpen
  \bibfield  {author} {\bibinfo {author} {\bibfnamefont {J.}~\bibnamefont
  {Beringer}} \emph {et~al.} (\bibinfo {collaboration} {Particle Data Group}),\
  }\bibfield  {title} {\enquote {\bibinfo {title} {{Review of Particle Physics
  (RPP)}},}\ }\href {\doibase 10.1103/PhysRevD.86.010001} {\bibfield  {journal}
  {\bibinfo  {journal} {Phys.Rev.}\ }\textbf {\bibinfo {volume} {D86}},\
  \bibinfo {pages} {010001} (\bibinfo {year} {2012})}\BibitemShut {NoStop}%
\bibitem [{\citenamefont {De~Panfilis}\ \emph {et~al.}(1987)\citenamefont
  {De~Panfilis}, \citenamefont {Melissinos}, \citenamefont {Moskowitz},
  \citenamefont {Rogers}, \citenamefont {Semertzidis}, \citenamefont {Wuensch},
  \citenamefont {Halama}, \citenamefont {Prodell}, \citenamefont {Fowler},\
  and\ \citenamefont {Nezrick}}]{DePanfilis:1987dk}%
  \BibitemOpen
  \bibfield  {author} {\bibinfo {author} {\bibfnamefont {S.}~\bibnamefont
  {De~Panfilis}}, \bibinfo {author} {\bibfnamefont {A.~C.}\ \bibnamefont
  {Melissinos}}, \bibinfo {author} {\bibfnamefont {B.~E.}\ \bibnamefont
  {Moskowitz}}, \bibinfo {author} {\bibfnamefont {J.~T.}\ \bibnamefont
  {Rogers}}, \bibinfo {author} {\bibfnamefont {Y.~K.}\ \bibnamefont
  {Semertzidis}}, \bibinfo {author} {\bibfnamefont {Walter}\ \bibnamefont
  {Wuensch}}, \bibinfo {author} {\bibfnamefont {H.~J.}\ \bibnamefont {Halama}},
  \bibinfo {author} {\bibfnamefont {A.~G.}\ \bibnamefont {Prodell}}, \bibinfo
  {author} {\bibfnamefont {W.~B.}\ \bibnamefont {Fowler}}, \ and\ \bibinfo
  {author} {\bibfnamefont {F.~A.}\ \bibnamefont {Nezrick}},\ }\bibfield
  {title} {\enquote {\bibinfo {title} {{Limits on the Abundance and Coupling of
  Cosmic Axions at 4.5-Microev < m(a) < 5.0-Microev}},}\ }\href {\doibase
  10.1103/PhysRevLett.59.839} {\bibfield  {journal} {\bibinfo  {journal} {Phys.
  Rev. Lett.}\ }\textbf {\bibinfo {volume} {59}},\ \bibinfo {pages} {839}
  (\bibinfo {year} {1987})}\BibitemShut {NoStop}%
\bibitem [{\citenamefont {Wuensch}\ \emph {et~al.}(1989)\citenamefont
  {Wuensch}, \citenamefont {De~Panfilis-Wuensch}, \citenamefont {Semertzidis},
  \citenamefont {Rogers}, \citenamefont {Melissinos}, \citenamefont {Halama},
  \citenamefont {Moskowitz}, \citenamefont {Prodell}, \citenamefont {Fowler},\
  and\ \citenamefont {Nezrick}}]{Wuensch:1989sa}%
  \BibitemOpen
  \bibfield  {author} {\bibinfo {author} {\bibfnamefont {Walter}\ \bibnamefont
  {Wuensch}}, \bibinfo {author} {\bibfnamefont {S.}~\bibnamefont
  {De~Panfilis-Wuensch}}, \bibinfo {author} {\bibfnamefont {Y.~K.}\
  \bibnamefont {Semertzidis}}, \bibinfo {author} {\bibfnamefont {J.~T.}\
  \bibnamefont {Rogers}}, \bibinfo {author} {\bibfnamefont {A.~C.}\
  \bibnamefont {Melissinos}}, \bibinfo {author} {\bibfnamefont {H.~J.}\
  \bibnamefont {Halama}}, \bibinfo {author} {\bibfnamefont {B.~E.}\
  \bibnamefont {Moskowitz}}, \bibinfo {author} {\bibfnamefont {A.~G.}\
  \bibnamefont {Prodell}}, \bibinfo {author} {\bibfnamefont {W.~B.}\
  \bibnamefont {Fowler}}, \ and\ \bibinfo {author} {\bibfnamefont {F.~A.}\
  \bibnamefont {Nezrick}},\ }\bibfield  {title} {\enquote {\bibinfo {title}
  {{Results of a Laboratory Search for Cosmic Axions and Other Weakly Coupled
  Light Particles}},}\ }\href {\doibase 10.1103/PhysRevD.40.3153} {\bibfield
  {journal} {\bibinfo  {journal} {Phys. Rev.}\ }\textbf {\bibinfo {volume}
  {D40}},\ \bibinfo {pages} {3153} (\bibinfo {year} {1989})}\BibitemShut
  {NoStop}%
\bibitem [{\citenamefont {Hagmann}\ \emph {et~al.}(1990)\citenamefont
  {Hagmann}, \citenamefont {Sikivie}, \citenamefont {Sullivan},\ and\
  \citenamefont {Tanner}}]{Hagmann:1990tj}%
  \BibitemOpen
  \bibfield  {author} {\bibinfo {author} {\bibfnamefont {C.}~\bibnamefont
  {Hagmann}}, \bibinfo {author} {\bibfnamefont {P.}~\bibnamefont {Sikivie}},
  \bibinfo {author} {\bibfnamefont {N.~S.}\ \bibnamefont {Sullivan}}, \ and\
  \bibinfo {author} {\bibfnamefont {D.~B.}\ \bibnamefont {Tanner}},\ }\bibfield
   {title} {\enquote {\bibinfo {title} {{Results from a search for cosmic
  axions}},}\ }\href {\doibase 10.1103/PhysRevD.42.1297} {\bibfield  {journal}
  {\bibinfo  {journal} {Phys. Rev.}\ }\textbf {\bibinfo {volume} {D42}},\
  \bibinfo {pages} {1297--1300} (\bibinfo {year} {1990})}\BibitemShut {NoStop}%
\bibitem [{\citenamefont {Asztalos}\ \emph {et~al.}(2010)\citenamefont
  {Asztalos}, \citenamefont {Carosi}, \citenamefont {Hagmann}, \citenamefont
  {Kinion}, \citenamefont {van Bibber}, \citenamefont {Hotz}, \citenamefont
  {Rosenberg}, \citenamefont {Rybka}, \citenamefont {Hoskins}, \citenamefont
  {Hwang}, \citenamefont {Sikivie}, \citenamefont {Tanner}, \citenamefont
  {Bradley},\ and\ \citenamefont {Clarke}}]{PhysRevLett.104.041301}%
  \BibitemOpen
  \bibfield  {author} {\bibinfo {author} {\bibfnamefont {S.~J.}\ \bibnamefont
  {Asztalos}}, \bibinfo {author} {\bibfnamefont {G.}~\bibnamefont {Carosi}},
  \bibinfo {author} {\bibfnamefont {C.}~\bibnamefont {Hagmann}}, \bibinfo
  {author} {\bibfnamefont {D.}~\bibnamefont {Kinion}}, \bibinfo {author}
  {\bibfnamefont {K.}~\bibnamefont {van Bibber}}, \bibinfo {author}
  {\bibfnamefont {M.}~\bibnamefont {Hotz}}, \bibinfo {author} {\bibfnamefont
  {L.~J}\ \bibnamefont {Rosenberg}}, \bibinfo {author} {\bibfnamefont
  {G.}~\bibnamefont {Rybka}}, \bibinfo {author} {\bibfnamefont
  {J.}~\bibnamefont {Hoskins}}, \bibinfo {author} {\bibfnamefont
  {J.}~\bibnamefont {Hwang}}, \bibinfo {author} {\bibfnamefont
  {P.}~\bibnamefont {Sikivie}}, \bibinfo {author} {\bibfnamefont {D.~B.}\
  \bibnamefont {Tanner}}, \bibinfo {author} {\bibfnamefont {R.}~\bibnamefont
  {Bradley}}, \ and\ \bibinfo {author} {\bibfnamefont {J.}~\bibnamefont
  {Clarke}},\ }\bibfield  {title} {\enquote {\bibinfo {title} {{SQUID-Based
  Microwave Cavity Search for Dark-Matter Axions}},}\ }\href {\doibase
  10.1103/PhysRevLett.104.041301} {\bibfield  {journal} {\bibinfo  {journal}
  {Phys. Rev. Lett.}\ }\textbf {\bibinfo {volume} {104}},\ \bibinfo {pages}
  {041301} (\bibinfo {year} {2010})}\BibitemShut {NoStop}%
\bibitem [{\citenamefont {Du}\ \emph {et~al.}(2018)\citenamefont {Du} \emph
  {et~al.}}]{Du:2018uak}%
  \BibitemOpen
  \bibfield  {author} {\bibinfo {author} {\bibfnamefont {N.}~\bibnamefont {Du}}
  \emph {et~al.} (\bibinfo {collaboration} {ADMX}),\ }\bibfield  {title}
  {\enquote {\bibinfo {title} {{A Search for Invisible Axion Dark Matter with
  the Axion Dark Matter Experiment}},}\ }\href {\doibase
  10.1103/PhysRevLett.120.151301} {\bibfield  {journal} {\bibinfo  {journal}
  {Phys. Rev. Lett.}\ }\textbf {\bibinfo {volume} {120}},\ \bibinfo {pages}
  {151301} (\bibinfo {year} {2018})},\ \Eprint
  {http://arxiv.org/abs/1804.05750} {arXiv:1804.05750 [hep-ex]} \BibitemShut
  {NoStop}%
\bibitem [{\citenamefont {Braine}\ \emph {et~al.}(2020)\citenamefont {Braine}
  \emph {et~al.}}]{Braine:2019fqb}%
  \BibitemOpen
  \bibfield  {author} {\bibinfo {author} {\bibfnamefont {T.}~\bibnamefont
  {Braine}} \emph {et~al.} (\bibinfo {collaboration} {ADMX}),\ }\bibfield
  {title} {\enquote {\bibinfo {title} {{Extended Search for the Invisible Axion
  with the Axion Dark Matter Experiment}},}\ }\href {\doibase
  10.1103/PhysRevLett.124.101303} {\bibfield  {journal} {\bibinfo  {journal}
  {Phys. Rev. Lett.}\ }\textbf {\bibinfo {volume} {124}},\ \bibinfo {pages}
  {101303} (\bibinfo {year} {2020})},\ \Eprint
  {http://arxiv.org/abs/1910.08638} {arXiv:1910.08638 [hep-ex]} \BibitemShut
  {NoStop}%
\bibitem [{\citenamefont {Brubaker}\ \emph {et~al.}(2017)\citenamefont
  {Brubaker}, \citenamefont {Zhong}, \citenamefont {Lamoreaux}, \citenamefont
  {Lehnert},\ and\ \citenamefont {van Bibber}}]{Brubaker:2017rna}%
  \BibitemOpen
  \bibfield  {author} {\bibinfo {author} {\bibfnamefont {B.~M.}\ \bibnamefont
  {Brubaker}}, \bibinfo {author} {\bibfnamefont {L.}~\bibnamefont {Zhong}},
  \bibinfo {author} {\bibfnamefont {S.~K.}\ \bibnamefont {Lamoreaux}}, \bibinfo
  {author} {\bibfnamefont {K.~W.}\ \bibnamefont {Lehnert}}, \ and\ \bibinfo
  {author} {\bibfnamefont {K.~A.}\ \bibnamefont {van Bibber}},\ }\bibfield
  {title} {\enquote {\bibinfo {title} {{HAYSTAC axion search analysis
  procedure}},}\ }\href {\doibase 10.1103/PhysRevD.96.123008} {\bibfield
  {journal} {\bibinfo  {journal} {Phys. Rev.}\ }\textbf {\bibinfo {volume}
  {D96}},\ \bibinfo {pages} {123008} (\bibinfo {year} {2017})},\ \Eprint
  {http://arxiv.org/abs/1706.08388} {arXiv:1706.08388 [astro-ph.IM]}
  \BibitemShut {NoStop}%
\bibitem [{\citenamefont {{Prestage}}\ \emph {et~al.}(2015)\citenamefont
  {{Prestage}}, \citenamefont {{Bloss}}, \citenamefont {{Brandt}},
  \citenamefont {{Chen}}, \citenamefont {{Creager}}, \citenamefont
  {{Demorest}}, \citenamefont {{Ford}}, \citenamefont {{Jones}}, \citenamefont
  {{Kepley}}, \citenamefont {{Kobelski}}, \citenamefont {{Marganian}},
  \citenamefont {{Mello}}, \citenamefont {{McMahon}}, \citenamefont
  {{McCullough}}, \citenamefont {{Ray}}, \citenamefont {{Roshi}}, \citenamefont
  {{Werthimer}},\ and\ \citenamefont {{Whitehead}}}]{vegas}%
  \BibitemOpen
  \bibfield  {author} {\bibinfo {author} {\bibfnamefont {Richard~M.}\
  \bibnamefont {{Prestage}}}, \bibinfo {author} {\bibfnamefont {Marty}\
  \bibnamefont {{Bloss}}}, \bibinfo {author} {\bibfnamefont {Joe}\ \bibnamefont
  {{Brandt}}}, \bibinfo {author} {\bibfnamefont {Hong}\ \bibnamefont {{Chen}}},
  \bibinfo {author} {\bibfnamefont {Ray}\ \bibnamefont {{Creager}}}, \bibinfo
  {author} {\bibfnamefont {Paul}\ \bibnamefont {{Demorest}}}, \bibinfo {author}
  {\bibfnamefont {John}\ \bibnamefont {{Ford}}}, \bibinfo {author}
  {\bibfnamefont {Glenn}\ \bibnamefont {{Jones}}}, \bibinfo {author}
  {\bibfnamefont {Amanda}\ \bibnamefont {{Kepley}}}, \bibinfo {author}
  {\bibfnamefont {Adam}\ \bibnamefont {{Kobelski}}}, \bibinfo {author}
  {\bibfnamefont {Paul}\ \bibnamefont {{Marganian}}}, \bibinfo {author}
  {\bibfnamefont {Melinda}\ \bibnamefont {{Mello}}}, \bibinfo {author}
  {\bibfnamefont {David}\ \bibnamefont {{McMahon}}}, \bibinfo {author}
  {\bibfnamefont {Randy}\ \bibnamefont {{McCullough}}}, \bibinfo {author}
  {\bibfnamefont {Jason}\ \bibnamefont {{Ray}}}, \bibinfo {author}
  {\bibfnamefont {D.~Anish}\ \bibnamefont {{Roshi}}}, \bibinfo {author}
  {\bibfnamefont {Dan}\ \bibnamefont {{Werthimer}}}, \ and\ \bibinfo {author}
  {\bibfnamefont {Mark}\ \bibnamefont {{Whitehead}}},\ }\bibfield  {title}
  {\enquote {\bibinfo {title} {{The versatile GBT astronomical spectrometer
  (VEGAS): Current status and future plans}},}\ }in\ \href {\doibase
  10.1109/USNC-URSI.2015.7303578} {\emph {\bibinfo {booktitle} {2015 URSI-USNC
  Radio Science Meeting}}}\ (\bibinfo {year} {2015})\ p.~\bibinfo {pages}
  {4}\BibitemShut {NoStop}%
\bibitem [{\citenamefont {{Lazarus}}\ \emph {et~al.}(2016)\citenamefont
  {{Lazarus}}, \citenamefont {{Karuppusamy}}, \citenamefont {{Graikou}},
  \citenamefont {{Caballero}}, \citenamefont {{Champion}}, \citenamefont
  {{Lee}}, \citenamefont {{Verbiest}},\ and\ \citenamefont
  {{Kramer}}}]{2016MNRAS.458..868L}%
  \BibitemOpen
  \bibfield  {author} {\bibinfo {author} {\bibfnamefont {P.}~\bibnamefont
  {{Lazarus}}}, \bibinfo {author} {\bibfnamefont {R.}~\bibnamefont
  {{Karuppusamy}}}, \bibinfo {author} {\bibfnamefont {E.}~\bibnamefont
  {{Graikou}}}, \bibinfo {author} {\bibfnamefont {R.~N.}\ \bibnamefont
  {{Caballero}}}, \bibinfo {author} {\bibfnamefont {D.~J.}\ \bibnamefont
  {{Champion}}}, \bibinfo {author} {\bibfnamefont {K.~J.}\ \bibnamefont
  {{Lee}}}, \bibinfo {author} {\bibfnamefont {J.~P.~W.}\ \bibnamefont
  {{Verbiest}}}, \ and\ \bibinfo {author} {\bibfnamefont {M.}~\bibnamefont
  {{Kramer}}},\ }\bibfield  {title} {\enquote {\bibinfo {title} {{Prospects for
  high-precision pulsar timing with the new Effelsberg PSRIX backend}},}\
  }\href {\doibase 10.1093/mnras/stw189} {\bibfield  {journal} {\bibinfo
  {journal} {MNRAS}\ }\textbf {\bibinfo {volume} {458}},\ \bibinfo {pages}
  {868--880} (\bibinfo {year} {2016})},\ \Eprint
  {http://arxiv.org/abs/1601.06194} {arXiv:1601.06194 [astro-ph.IM]}
  \BibitemShut {NoStop}%
\bibitem [{\citenamefont {Garwood}\ \emph {et~al.}(2005)\citenamefont
  {Garwood}, \citenamefont {Marganian}, \citenamefont {Braatz},\ and\
  \citenamefont {Maddalena}}]{garwood_marganian_braatz_maddalena_2005}%
  \BibitemOpen
  \bibfield  {author} {\bibinfo {author} {\bibfnamefont {Bob}\ \bibnamefont
  {Garwood}}, \bibinfo {author} {\bibfnamefont {Paul}\ \bibnamefont
  {Marganian}}, \bibinfo {author} {\bibfnamefont {Jim}\ \bibnamefont {Braatz}},
  \ and\ \bibinfo {author} {\bibfnamefont {Ron}\ \bibnamefont {Maddalena}},\
  }\href {http://gbtidl.nrao.edu/} {\enquote {\bibinfo {title} {\texttt{GBTIDL}
  data analysis software},}\ } (\bibinfo {year} {2005})\BibitemShut {NoStop}%
\bibitem [{\citenamefont {{van Straten}}\ and\ \citenamefont
  {{Bailes}}(2011)}]{dspsr}%
  \BibitemOpen
  \bibfield  {author} {\bibinfo {author} {\bibfnamefont {W.}~\bibnamefont {{van
  Straten}}}\ and\ \bibinfo {author} {\bibfnamefont {M.}~\bibnamefont
  {{Bailes}}},\ }\bibfield  {title} {\enquote {\bibinfo {title} {{DSPSR:
  Digital Signal Processing Software for Pulsar Astronomy}},}\ }\href {\doibase
  10.1071/AS10021} {\bibfield  {journal} {\bibinfo  {journal} {Publications of
  the Astron. Soc. of Australia}\ }\textbf {\bibinfo {volume} {28}},\ \bibinfo
  {pages} {1--14} (\bibinfo {year} {2011})},\ \Eprint
  {http://arxiv.org/abs/1008.3973} {arXiv:1008.3973 [astro-ph.IM]} \BibitemShut
  {NoStop}%
\bibitem [{\citenamefont {Cowan}\ \emph
  {et~al.}(2011{\natexlab{a}})\citenamefont {Cowan}, \citenamefont {Cranmer},
  \citenamefont {Gross},\ and\ \citenamefont {Vitells}}]{Cowan:2010js}%
  \BibitemOpen
  \bibfield  {author} {\bibinfo {author} {\bibfnamefont {Glen}\ \bibnamefont
  {Cowan}}, \bibinfo {author} {\bibfnamefont {Kyle}\ \bibnamefont {Cranmer}},
  \bibinfo {author} {\bibfnamefont {Eilam}\ \bibnamefont {Gross}}, \ and\
  \bibinfo {author} {\bibfnamefont {Ofer}\ \bibnamefont {Vitells}},\ }\bibfield
   {title} {\enquote {\bibinfo {title} {{Asymptotic formulae for
  likelihood-based tests of new physics}},}\ }\href {\doibase
  10.1140/epjc/s10052-011-1554-0, 10.1140/epjc/s10052-013-2501-z} {\bibfield
  {journal} {\bibinfo  {journal} {Eur. Phys. J.}\ }\textbf {\bibinfo {volume}
  {C71}},\ \bibinfo {pages} {1554} (\bibinfo {year} {2011}{\natexlab{a}})},\
  \bibinfo {note} {[Erratum: Eur. Phys. J.C73,2501(2013)]},\ \Eprint
  {http://arxiv.org/abs/1007.1727} {arXiv:1007.1727 [physics.data-an]}
  \BibitemShut {NoStop}%
\bibitem [{\citenamefont {Cowan}\ \emph
  {et~al.}(2011{\natexlab{b}})\citenamefont {Cowan}, \citenamefont {Cranmer},
  \citenamefont {Gross},\ and\ \citenamefont {Vitells}}]{Cowan:2011an}%
  \BibitemOpen
  \bibfield  {author} {\bibinfo {author} {\bibfnamefont {Glen}\ \bibnamefont
  {Cowan}}, \bibinfo {author} {\bibfnamefont {Kyle}\ \bibnamefont {Cranmer}},
  \bibinfo {author} {\bibfnamefont {Eilam}\ \bibnamefont {Gross}}, \ and\
  \bibinfo {author} {\bibfnamefont {Ofer}\ \bibnamefont {Vitells}},\ }\bibfield
   {title} {\enquote {\bibinfo {title} {{Power-Constrained Limits}},}\
  }\href@noop {} {\  (\bibinfo {year} {2011}{\natexlab{b}})},\ \Eprint
  {http://arxiv.org/abs/1105.3166} {arXiv:1105.3166 [physics.data-an]}
  \BibitemShut {NoStop}%
\bibitem [{\citenamefont {Bovy}\ and\ \citenamefont
  {Tremaine}(2012)}]{Bovy:2012tw}%
  \BibitemOpen
  \bibfield  {author} {\bibinfo {author} {\bibfnamefont {Jo}~\bibnamefont
  {Bovy}}\ and\ \bibinfo {author} {\bibfnamefont {Scott}\ \bibnamefont
  {Tremaine}},\ }\bibfield  {title} {\enquote {\bibinfo {title} {{On the local
  dark matter density}},}\ }\href {\doibase 10.1088/0004-637X/756/1/89}
  {\bibfield  {journal} {\bibinfo  {journal} {Astrophys. J.}\ }\textbf
  {\bibinfo {volume} {756}},\ \bibinfo {pages} {89} (\bibinfo {year} {2012})},\
  \Eprint {http://arxiv.org/abs/1205.4033} {arXiv:1205.4033 [astro-ph.GA]}
  \BibitemShut {NoStop}%
\bibitem [{\citenamefont {Read}(2014)}]{Read:2014qva}%
  \BibitemOpen
  \bibfield  {author} {\bibinfo {author} {\bibfnamefont {J.~I.}\ \bibnamefont
  {Read}},\ }\bibfield  {title} {\enquote {\bibinfo {title} {{The Local Dark
  Matter Density}},}\ }\href {\doibase 10.1088/0954-3899/41/6/063101}
  {\bibfield  {journal} {\bibinfo  {journal} {J. Phys.}\ }\textbf {\bibinfo
  {volume} {G41}},\ \bibinfo {pages} {063101} (\bibinfo {year} {2014})},\
  \Eprint {http://arxiv.org/abs/1404.1938} {arXiv:1404.1938 [astro-ph.GA]}
  \BibitemShut {NoStop}%
\bibitem [{\citenamefont {Schutz}\ \emph {et~al.}(2018)\citenamefont {Schutz},
  \citenamefont {Lin}, \citenamefont {Safdi},\ and\ \citenamefont
  {Wu}}]{Schutz:2017tfp}%
  \BibitemOpen
  \bibfield  {author} {\bibinfo {author} {\bibfnamefont {Katelin}\ \bibnamefont
  {Schutz}}, \bibinfo {author} {\bibfnamefont {Tongyan}\ \bibnamefont {Lin}},
  \bibinfo {author} {\bibfnamefont {Benjamin~R.}\ \bibnamefont {Safdi}}, \ and\
  \bibinfo {author} {\bibfnamefont {Chih-Liang}\ \bibnamefont {Wu}},\
  }\bibfield  {title} {\enquote {\bibinfo {title} {{Constraining a Thin Dark
  Matter Disk with Gaia}},}\ }\href {\doibase 10.1103/PhysRevLett.121.081101}
  {\bibfield  {journal} {\bibinfo  {journal} {Phys. Rev. Lett.}\ }\textbf
  {\bibinfo {volume} {121}},\ \bibinfo {pages} {081101} (\bibinfo {year}
  {2018})},\ \Eprint {http://arxiv.org/abs/1711.03103} {arXiv:1711.03103
  [astro-ph.GA]} \BibitemShut {NoStop}%
\bibitem [{\citenamefont {Navarro}\ \emph {et~al.}(1996)\citenamefont
  {Navarro}, \citenamefont {Frenk},\ and\ \citenamefont
  {White}}]{Navarro:1995iw}%
  \BibitemOpen
  \bibfield  {author} {\bibinfo {author} {\bibfnamefont {Julio~F.}\
  \bibnamefont {Navarro}}, \bibinfo {author} {\bibfnamefont {Carlos~S.}\
  \bibnamefont {Frenk}}, \ and\ \bibinfo {author} {\bibfnamefont {Simon D.~M.}\
  \bibnamefont {White}},\ }\bibfield  {title} {\enquote {\bibinfo {title} {{The
  Structure of cold dark matter halos}},}\ }\href {\doibase 10.1086/177173}
  {\bibfield  {journal} {\bibinfo  {journal} {Astrophys. J.}\ }\textbf
  {\bibinfo {volume} {462}},\ \bibinfo {pages} {563--575} (\bibinfo {year}
  {1996})},\ \Eprint {http://arxiv.org/abs/astro-ph/9508025} {astro-ph/9508025}
  \BibitemShut {NoStop}%
\bibitem [{\citenamefont {Navarro}\ \emph {et~al.}(1997)\citenamefont
  {Navarro}, \citenamefont {Frenk},\ and\ \citenamefont
  {White}}]{Navarro:1996gj}%
  \BibitemOpen
  \bibfield  {author} {\bibinfo {author} {\bibfnamefont {Julio~F.}\
  \bibnamefont {Navarro}}, \bibinfo {author} {\bibfnamefont {Carlos~S.}\
  \bibnamefont {Frenk}}, \ and\ \bibinfo {author} {\bibfnamefont {Simon D.~M.}\
  \bibnamefont {White}},\ }\bibfield  {title} {\enquote {\bibinfo {title} {{A
  Universal density profile from hierarchical clustering}},}\ }\href {\doibase
  10.1086/304888} {\bibfield  {journal} {\bibinfo  {journal} {Astrophys. J.}\
  }\textbf {\bibinfo {volume} {490}},\ \bibinfo {pages} {493--508} (\bibinfo
  {year} {1997})},\ \Eprint {http://arxiv.org/abs/astro-ph/9611107}
  {arXiv:astro-ph/9611107 [astro-ph]} \BibitemShut {NoStop}%
\bibitem [{\citenamefont {Kaplan}\ and\ \citenamefont {van
  Kerkwijk}(2005)}]{Kaplan:2005fy}%
  \BibitemOpen
  \bibfield  {author} {\bibinfo {author} {\bibfnamefont {David~L.}\
  \bibnamefont {Kaplan}}\ and\ \bibinfo {author} {\bibfnamefont {M.~H.}\
  \bibnamefont {van Kerkwijk}},\ }\bibfield  {title} {\enquote {\bibinfo
  {title} {{A Coherent timing solution for the nearby isolated neutron star RX
  J0720.4-3125}},}\ }\href {\doibase 10.1086/432536} {\bibfield  {journal}
  {\bibinfo  {journal} {Astrophys. J.}\ }\textbf {\bibinfo {volume} {628}},\
  \bibinfo {pages} {L45--L48} (\bibinfo {year} {2005})},\ \Eprint
  {http://arxiv.org/abs/astro-ph/0506419} {arXiv:astro-ph/0506419 [astro-ph]}
  \BibitemShut {NoStop}%
\bibitem [{\citenamefont {Kaplan}\ and\ \citenamefont {van
  Kerkwijk}(2009)}]{Kaplan:2009ce}%
  \BibitemOpen
  \bibfield  {author} {\bibinfo {author} {\bibfnamefont {D.~L.}\ \bibnamefont
  {Kaplan}}\ and\ \bibinfo {author} {\bibfnamefont {M.~H.}\ \bibnamefont {van
  Kerkwijk}},\ }\bibfield  {title} {\enquote {\bibinfo {title} {{Constraining
  the Spin-down of the Nearby Isolated Neutron Star RX J0806.4-4123, and
  Implications for the Population of Nearby Neutron Stars}},}\ }\href {\doibase
  10.1088/0004-637X/705/1/798} {\bibfield  {journal} {\bibinfo  {journal}
  {Astrophys. J.}\ }\textbf {\bibinfo {volume} {705}},\ \bibinfo {pages}
  {798--808} (\bibinfo {year} {2009})},\ \Eprint
  {http://arxiv.org/abs/0909.5218} {arXiv:0909.5218 [astro-ph.HE]} \BibitemShut
  {NoStop}%
\bibitem [{\citenamefont {Buschmann}\ \emph
  {et~al.}(2019{\natexlab{a}})\citenamefont {Buschmann}, \citenamefont {Co},
  \citenamefont {Dessert},\ and\ \citenamefont {Safdi}}]{Buschmann:2019pfp}%
  \BibitemOpen
  \bibfield  {author} {\bibinfo {author} {\bibfnamefont {Malte}\ \bibnamefont
  {Buschmann}}, \bibinfo {author} {\bibfnamefont {Raymond~T.}\ \bibnamefont
  {Co}}, \bibinfo {author} {\bibfnamefont {Christopher}\ \bibnamefont
  {Dessert}}, \ and\ \bibinfo {author} {\bibfnamefont {Benjamin~R.}\
  \bibnamefont {Safdi}},\ }\bibfield  {title} {\enquote {\bibinfo {title}
  {{X-ray Search for Axions from Nearby Isolated Neutron Stars}},}\ }\href@noop
  {} {\  (\bibinfo {year} {2019}{\natexlab{a}})},\ \Eprint
  {http://arxiv.org/abs/1910.04164} {arXiv:1910.04164 [hep-ph]} \BibitemShut
  {NoStop}%
\bibitem [{\citenamefont {{Goldreich}}\ and\ \citenamefont
  {{Julian}}(1969)}]{1969ApJ...157..869G}%
  \BibitemOpen
  \bibfield  {author} {\bibinfo {author} {\bibfnamefont {P.}~\bibnamefont
  {{Goldreich}}}\ and\ \bibinfo {author} {\bibfnamefont {W.~H.}\ \bibnamefont
  {{Julian}}},\ }\bibfield  {title} {\enquote {\bibinfo {title} {{Pulsar
  Electrodynamics}},}\ }\href {\doibase 10.1086/150119} {\bibfield  {journal}
  {\bibinfo  {journal} {\apj}\ }\textbf {\bibinfo {volume} {157}},\ \bibinfo
  {pages} {869} (\bibinfo {year} {1969})}\BibitemShut {NoStop}%
\bibitem [{\citenamefont {et~al.}(2020)}]{SM}%
  \BibitemOpen
  \bibfield  {author} {\bibinfo {author} {\bibfnamefont {J.~Foster}\
  \bibnamefont {et~al.}},\ }\href@noop {} {\emph {\bibinfo {title}
  {Supplementary Data}}} (\bibinfo {year} {2020}),\ \bibinfo {note}
  {\url{https://github.com/joshwfoster/RadioAxionSearch}}\BibitemShut {NoStop}%
\bibitem [{\citenamefont {Anastassopoulos}\ \emph {et~al.}(2017)\citenamefont
  {Anastassopoulos} \emph {et~al.}}]{Anastassopoulos:2017ftl}%
  \BibitemOpen
  \bibfield  {author} {\bibinfo {author} {\bibfnamefont {V.}~\bibnamefont
  {Anastassopoulos}} \emph {et~al.} (\bibinfo {collaboration} {CAST}),\
  }\bibfield  {title} {\enquote {\bibinfo {title} {{New CAST Limit on the
  Axion-Photon Interaction}},}\ }\href {\doibase 10.1038/nphys4109} {\bibfield
  {journal} {\bibinfo  {journal} {Nature Phys.}\ }\textbf {\bibinfo {volume}
  {13}},\ \bibinfo {pages} {584--590} (\bibinfo {year} {2017})},\ \Eprint
  {http://arxiv.org/abs/1705.02290} {arXiv:1705.02290 [hep-ex]} \BibitemShut
  {NoStop}%
\bibitem [{\citenamefont {Hopkins}\ \emph {et~al.}(2018)\citenamefont {Hopkins}
  \emph {et~al.}}]{Hopkins:2017ycn}%
  \BibitemOpen
  \bibfield  {author} {\bibinfo {author} {\bibfnamefont {Philip~F}\
  \bibnamefont {Hopkins}} \emph {et~al.},\ }\bibfield  {title} {\enquote
  {\bibinfo {title} {{FIRE-2 Simulations: Physics versus Numerics in Galaxy
  Formation}},}\ }\href {\doibase 10.1093/mnras/sty1690} {\bibfield  {journal}
  {\bibinfo  {journal} {Mon. Not. Roy. Astron. Soc.}\ }\textbf {\bibinfo
  {volume} {480}},\ \bibinfo {pages} {800--863} (\bibinfo {year} {2018})},\
  \Eprint {http://arxiv.org/abs/1702.06148} {arXiv:1702.06148 [astro-ph.GA]}
  \BibitemShut {NoStop}%
\bibitem [{\citenamefont {Weltman}\ \emph {et~al.}(2020)\citenamefont {Weltman}
  \emph {et~al.}}]{Bull:2018lat}%
  \BibitemOpen
  \bibfield  {author} {\bibinfo {author} {\bibfnamefont {A.}~\bibnamefont
  {Weltman}} \emph {et~al.},\ }\bibfield  {title} {\enquote {\bibinfo {title}
  {{Fundamental Physics with the Square Kilometre Array}},}\ }\href {\doibase
  10.1017/pasa.2019.42} {\bibfield  {journal} {\bibinfo  {journal} {Publ.
  Astron. Soc. Austral.}\ }\textbf {\bibinfo {volume} {37}},\ \bibinfo {pages}
  {e002} (\bibinfo {year} {2020})},\ \Eprint {http://arxiv.org/abs/1810.02680}
  {arXiv:1810.02680 [astro-ph.CO]} \BibitemShut {NoStop}%
\bibitem [{\citenamefont {{Nan}}\ \emph {et~al.}(2011)\citenamefont {{Nan}},
  \citenamefont {{Li}}, \citenamefont {{Jin}}, \citenamefont {{Wang}},
  \citenamefont {{Zhu}}, \citenamefont {{Zhu}}, \citenamefont {{Zhang}},
  \citenamefont {{Yue}},\ and\ \citenamefont {{Qian}}}]{2011IJMPD..20..989N}%
  \BibitemOpen
  \bibfield  {author} {\bibinfo {author} {\bibfnamefont {Rendong}\ \bibnamefont
  {{Nan}}}, \bibinfo {author} {\bibfnamefont {Di}~\bibnamefont {{Li}}},
  \bibinfo {author} {\bibfnamefont {Chengjin}\ \bibnamefont {{Jin}}}, \bibinfo
  {author} {\bibfnamefont {Qiming}\ \bibnamefont {{Wang}}}, \bibinfo {author}
  {\bibfnamefont {Lichun}\ \bibnamefont {{Zhu}}}, \bibinfo {author}
  {\bibfnamefont {Wenbai}\ \bibnamefont {{Zhu}}}, \bibinfo {author}
  {\bibfnamefont {Haiyan}\ \bibnamefont {{Zhang}}}, \bibinfo {author}
  {\bibfnamefont {Youling}\ \bibnamefont {{Yue}}}, \ and\ \bibinfo {author}
  {\bibfnamefont {Lei}\ \bibnamefont {{Qian}}},\ }\bibfield  {title} {\enquote
  {\bibinfo {title} {{The Five-Hundred Aperture Spherical Radio Telescope
  (fast) Project}},}\ }\href {\doibase 10.1142/S0218271811019335} {\bibfield
  {journal} {\bibinfo  {journal} {International Journal of Modern Physics D}\
  }\textbf {\bibinfo {volume} {20}},\ \bibinfo {pages} {989--1024} (\bibinfo
  {year} {2011})},\ \Eprint {http://arxiv.org/abs/1105.3794} {arXiv:1105.3794
  [astro-ph.IM]} \BibitemShut {NoStop}%
\bibitem [{\citenamefont {Buschmann}\ \emph
  {et~al.}(2019{\natexlab{b}})\citenamefont {Buschmann}, \citenamefont
  {Foster},\ and\ \citenamefont {Safdi}}]{Buschmann:2019icd}%
  \BibitemOpen
  \bibfield  {author} {\bibinfo {author} {\bibfnamefont {Malte}\ \bibnamefont
  {Buschmann}}, \bibinfo {author} {\bibfnamefont {Joshua~W.}\ \bibnamefont
  {Foster}}, \ and\ \bibinfo {author} {\bibfnamefont {Benjamin~R.}\
  \bibnamefont {Safdi}},\ }\bibfield  {title} {\enquote {\bibinfo {title}
  {{Early-Universe Simulations of the Cosmological Axion}},}\ }\href@noop {} {\
   (\bibinfo {year} {2019}{\natexlab{b}})},\ \Eprint
  {http://arxiv.org/abs/1906.00967} {arXiv:1906.00967 [astro-ph.CO]}
  \BibitemShut {NoStop}%
\bibitem [{\citenamefont {Klaer}\ and\ \citenamefont
  {Moore}(2017)}]{Klaer:2017ond}%
  \BibitemOpen
  \bibfield  {author} {\bibinfo {author} {\bibfnamefont {Vincent~B..}\
  \bibnamefont {Klaer}}\ and\ \bibinfo {author} {\bibfnamefont {Guy~D.}\
  \bibnamefont {Moore}},\ }\bibfield  {title} {\enquote {\bibinfo {title} {{The
  dark-matter axion mass}},}\ }\href {\doibase 10.1088/1475-7516/2017/11/049}
  {\bibfield  {journal} {\bibinfo  {journal} {JCAP}\ }\textbf {\bibinfo
  {volume} {1711}},\ \bibinfo {pages} {049} (\bibinfo {year} {2017})},\ \Eprint
  {http://arxiv.org/abs/1708.07521} {arXiv:1708.07521 [hep-ph]} \BibitemShut
  {NoStop}%
\bibitem [{\citenamefont {Agrawal}\ \emph {et~al.}(2018)\citenamefont
  {Agrawal}, \citenamefont {Fan}, \citenamefont {Reece},\ and\ \citenamefont
  {Wang}}]{Agrawal:2017cmd}%
  \BibitemOpen
  \bibfield  {author} {\bibinfo {author} {\bibfnamefont {Prateek}\ \bibnamefont
  {Agrawal}}, \bibinfo {author} {\bibfnamefont {JiJi}\ \bibnamefont {Fan}},
  \bibinfo {author} {\bibfnamefont {Matthew}\ \bibnamefont {Reece}}, \ and\
  \bibinfo {author} {\bibfnamefont {Lian-Tao}\ \bibnamefont {Wang}},\
  }\bibfield  {title} {\enquote {\bibinfo {title} {{Experimental Targets for
  Photon Couplings of the QCD Axion}},}\ }\href {\doibase
  10.1007/JHEP02(2018)006} {\bibfield  {journal} {\bibinfo  {journal} {JHEP}\
  }\textbf {\bibinfo {volume} {02}},\ \bibinfo {pages} {006} (\bibinfo {year}
  {2018})},\ \Eprint {http://arxiv.org/abs/1709.06085} {arXiv:1709.06085
  [hep-ph]} \BibitemShut {NoStop}%
\bibitem [{\citenamefont {{Perley}}\ and\ \citenamefont
  {{Butler}}(2013)}]{Perley_2013}%
  \BibitemOpen
  \bibfield  {author} {\bibinfo {author} {\bibfnamefont {R.~A.}\ \bibnamefont
  {{Perley}}}\ and\ \bibinfo {author} {\bibfnamefont {B.~J.}\ \bibnamefont
  {{Butler}}},\ }\bibfield  {title} {\enquote {\bibinfo {title} {{Integrated
  Polarization Properties of 3C48, 3C138, 3C147, and 3C286}},}\ }\href
  {\doibase 10.1088/0067-0049/206/2/16} {\bibfield  {journal} {\bibinfo
  {journal} {The Astrophysical Journal Supplement Series}\ }\textbf {\bibinfo
  {volume} {206}},\ \bibinfo {eid} {16} (\bibinfo {year} {2013})},\ \Eprint
  {http://arxiv.org/abs/1302.6662} {arXiv:1302.6662 [astro-ph.IM]} \BibitemShut
  {NoStop}%
\bibitem [{\citenamefont {Perley}\ and\ \citenamefont
  {Butler}(2017)}]{Perley_2017}%
  \BibitemOpen
  \bibfield  {author} {\bibinfo {author} {\bibfnamefont {R.~A.}\ \bibnamefont
  {Perley}}\ and\ \bibinfo {author} {\bibfnamefont {B.~J.}\ \bibnamefont
  {Butler}},\ }\bibfield  {title} {\enquote {\bibinfo {title} {An accurate flux
  density scale from 50 {MHz} to 50 {GHz}},}\ }\href {\doibase
  10.3847/1538-4365/aa6df9} {\bibfield  {journal} {\bibinfo  {journal} {The
  Astrophysical Journal Supplement Series}\ }\textbf {\bibinfo {volume}
  {230}},\ \bibinfo {pages} {7} (\bibinfo {year} {2017})}\BibitemShut {NoStop}%
\bibitem [{\citenamefont {{Hotan}}\ \emph {et~al.}(2004)\citenamefont
  {{Hotan}}, \citenamefont {{van Straten}},\ and\ \citenamefont
  {{Manchester}}}]{psrchive}%
  \BibitemOpen
  \bibfield  {author} {\bibinfo {author} {\bibfnamefont {A.~W.}\ \bibnamefont
  {{Hotan}}}, \bibinfo {author} {\bibfnamefont {W.}~\bibnamefont {{van
  Straten}}}, \ and\ \bibinfo {author} {\bibfnamefont {R.~N.}\ \bibnamefont
  {{Manchester}}},\ }\bibfield  {title} {\enquote {\bibinfo {title} {{PSRCHIVE
  and PSRFITS: An Open Approach to Radio Pulsar Data Storage and Analysis}},}\
  }\href {\doibase 10.1071/AS04022} {\bibfield  {journal} {\bibinfo  {journal}
  {Publications of the Astron. Soc. of Australia}\ }\textbf {\bibinfo {volume}
  {21}},\ \bibinfo {pages} {302--309} (\bibinfo {year} {2004})},\ \Eprint
  {http://arxiv.org/abs/arXiv:astro-ph/0404549} {arXiv:astro-ph/0404549}
  \BibitemShut {NoStop}%
\bibitem [{\citenamefont {{Crocker}}\ \emph {et~al.}(2011)\citenamefont
  {{Crocker}}, \citenamefont {{Jones}}, \citenamefont {{Aharonian}},
  \citenamefont {{Law}}, \citenamefont {{Melia}}, \citenamefont {{Oka}},\ and\
  \citenamefont {{Ott}}}]{2011MNRAS.413..763C}%
  \BibitemOpen
  \bibfield  {author} {\bibinfo {author} {\bibfnamefont {R.~M.}\ \bibnamefont
  {{Crocker}}}, \bibinfo {author} {\bibfnamefont {D.~I.}\ \bibnamefont
  {{Jones}}}, \bibinfo {author} {\bibfnamefont {F.}~\bibnamefont
  {{Aharonian}}}, \bibinfo {author} {\bibfnamefont {C.~J.}\ \bibnamefont
  {{Law}}}, \bibinfo {author} {\bibfnamefont {F.}~\bibnamefont {{Melia}}},
  \bibinfo {author} {\bibfnamefont {T.}~\bibnamefont {{Oka}}}, \ and\ \bibinfo
  {author} {\bibfnamefont {J.}~\bibnamefont {{Ott}}},\ }\bibfield  {title}
  {\enquote {\bibinfo {title} {{Wild at Heart: the particle astrophysics of the
  Galactic Centre}},}\ }\href {\doibase 10.1111/j.1365-2966.2010.18170.x}
  {\bibfield  {journal} {\bibinfo  {journal} {MNRAS}\ }\textbf {\bibinfo
  {volume} {413}},\ \bibinfo {pages} {763--788} (\bibinfo {year} {2011})},\
  \Eprint {http://arxiv.org/abs/1011.0206} {arXiv:1011.0206 [astro-ph.GA]}
  \BibitemShut {NoStop}%
\bibitem [{\citenamefont {{Reich}}\ \emph {et~al.}(1990)\citenamefont
  {{Reich}}, \citenamefont {{Reich}},\ and\ \citenamefont
  {{Fuerst}}}]{1990A&AS...83..539R}%
  \BibitemOpen
  \bibfield  {author} {\bibinfo {author} {\bibfnamefont {W.}~\bibnamefont
  {{Reich}}}, \bibinfo {author} {\bibfnamefont {P.}~\bibnamefont {{Reich}}}, \
  and\ \bibinfo {author} {\bibfnamefont {E.}~\bibnamefont {{Fuerst}}},\
  }\bibfield  {title} {\enquote {\bibinfo {title} {{The Effelsberg 21 CM radio
  continuum survey of theGalacticplane between L = 357 deg. and L = 95.5
  deg.}}}\ }\href@noop {} {\bibfield  {journal} {\bibinfo  {journal} {Astronomy
  and Astrophysics, Supplement}\ }\textbf {\bibinfo {volume} {83}},\ \bibinfo
  {pages} {539} (\bibinfo {year} {1990})}\BibitemShut {NoStop}%
\bibitem [{\citenamefont {Wilks}(1938)}]{Wilks:1938dza}%
  \BibitemOpen
  \bibfield  {author} {\bibinfo {author} {\bibfnamefont {S.~S.}\ \bibnamefont
  {Wilks}},\ }\bibfield  {title} {\enquote {\bibinfo {title} {{The Large-Sample
  Distribution of the Likelihood Ratio for Testing Composite Hypotheses}},}\
  }\href {\doibase 10.1214/aoms/1177732360} {\bibfield  {journal} {\bibinfo
  {journal} {Annals Math. Statist.}\ }\textbf {\bibinfo {volume} {9}},\
  \bibinfo {pages} {60--62} (\bibinfo {year} {1938})}\BibitemShut {NoStop}%
\bibitem [{\citenamefont {Gray}(2012)}]{gray_2012}%
  \BibitemOpen
  \bibfield  {author} {\bibinfo {author} {\bibfnamefont {Malcolm}\ \bibnamefont
  {Gray}},\ }\enquote {\bibinfo {title} {Maser molecules},}\ in\ \href
  {\doibase 10.1017/CBO9780511977534.006} {\emph {\bibinfo {booktitle} {Maser
  Sources in Astrophysics}}},\ \bibinfo {series and number} {Cambridge
  Astrophysics}\ (\bibinfo  {publisher} {Cambridge University Press},\ \bibinfo
  {year} {2012})\ p.\ \bibinfo {pages} {156–185}\BibitemShut {NoStop}%
\bibitem [{\citenamefont {{Norris}}\ and\ \citenamefont
  {{Booth}}(1981)}]{1981MNRAS.195..213N}%
  \BibitemOpen
  \bibfield  {author} {\bibinfo {author} {\bibfnamefont {R.~P.}\ \bibnamefont
  {{Norris}}}\ and\ \bibinfo {author} {\bibfnamefont {R.~S.}\ \bibnamefont
  {{Booth}}},\ }\bibfield  {title} {\enquote {\bibinfo {title} {{Observations
  of OH masers in W3 OH.}}}\ }\href {\doibase 10.1093/mnras/195.2.213}
  {\bibfield  {journal} {\bibinfo  {journal} {MNRAS}\ }\textbf {\bibinfo
  {volume} {195}},\ \bibinfo {pages} {213--226} (\bibinfo {year}
  {1981})}\BibitemShut {NoStop}%
\bibitem [{\citenamefont {{Qiao}}\ \emph {et~al.}(2014)\citenamefont {{Qiao}},
  \citenamefont {{Li}}, \citenamefont {{Shen}}, \citenamefont {{Chen}},\ and\
  \citenamefont {{Zheng}}}]{2014MNRAS.441.3137Q}%
  \BibitemOpen
  \bibfield  {author} {\bibinfo {author} {\bibfnamefont {Haihua}\ \bibnamefont
  {{Qiao}}}, \bibinfo {author} {\bibfnamefont {Juan}\ \bibnamefont {{Li}}},
  \bibinfo {author} {\bibfnamefont {Zhiqiang}\ \bibnamefont {{Shen}}}, \bibinfo
  {author} {\bibfnamefont {Xi}~\bibnamefont {{Chen}}}, \ and\ \bibinfo {author}
  {\bibfnamefont {Xingwu}\ \bibnamefont {{Zheng}}},\ }\bibfield  {title}
  {\enquote {\bibinfo {title} {{The catalogues and mid-infrared environment of
  interstellar OH masers}},}\ }\href {\doibase 10.1093/mnras/stu776} {\bibfield
   {journal} {\bibinfo  {journal} {MNRAS}\ }\textbf {\bibinfo {volume} {441}},\
  \bibinfo {pages} {3137--3147} (\bibinfo {year} {2014})}\BibitemShut {NoStop}%
\bibitem [{\citenamefont {{Caswell}}(1998)}]{1998MNRAS.297..215C}%
  \BibitemOpen
  \bibfield  {author} {\bibinfo {author} {\bibfnamefont {J.~L.}\ \bibnamefont
  {{Caswell}}},\ }\bibfield  {title} {\enquote {\bibinfo {title} {{Positions of
  hydroxyl masers at 1665 and 1667 MHz}},}\ }\href {\doibase
  10.1046/j.1365-8711.1998.01468.x} {\bibfield  {journal} {\bibinfo  {journal}
  {MNRAS}\ }\textbf {\bibinfo {volume} {297}},\ \bibinfo {pages} {215--235}
  (\bibinfo {year} {1998})}\BibitemShut {NoStop}%
\bibitem [{\citenamefont {McCall}\ and\ \citenamefont
  {Censor}(2007)}]{doi:10.1119/1.2772281}%
  \BibitemOpen
  \bibfield  {author} {\bibinfo {author} {\bibfnamefont {Martin}\ \bibnamefont
  {McCall}}\ and\ \bibinfo {author} {\bibfnamefont {Dan}\ \bibnamefont
  {Censor}},\ }\bibfield  {title} {\enquote {\bibinfo {title} {Relativity and
  mathematical tools: Waves in moving media},}\ }\href {\doibase
  10.1119/1.2772281} {\bibfield  {journal} {\bibinfo  {journal} {American
  Journal of Physics}\ }\textbf {\bibinfo {volume} {75}},\ \bibinfo {pages}
  {1134--1140} (\bibinfo {year} {2007})},\ \Eprint
  {http://arxiv.org/abs/https://doi.org/10.1119/1.2772281}
  {https://doi.org/10.1119/1.2772281} \BibitemShut {NoStop}%
\end{thebibliography}%

\clearpage

\onecolumngrid
\begin{center}
  \textbf{\large Supplementary Material for Green Bank and Effelsberg Radio Telescopes Searches for Axion Dark Matter Conversion in Neutron Star Magnetospheres}\\[.2cm]
  \vspace{0.05in}
  {Joshua W. Foster, \ Yonatan Kahn, \ Oscar Macias,  \ Zhiquan Sun, \ Ralph P. Eatough, \ Vladislav I. Kondratiev, \ Wendy M. Peters,  \ Christoph Weniger, \ and \ Benjamin R. Safdi}
\end{center}

\onecolumngrid
\setcounter{equation}{0}
\setcounter{figure}{0}
\setcounter{table}{0}
\setcounter{section}{0}
\setcounter{page}{1}
\makeatletter
\renewcommand{\theequation}{S\arabic{equation}}
\renewcommand{\thefigure}{S\arabic{figure}}
\renewcommand{\thetable}{S\arabic{table}}

This Supplementary Material contains additional results and explanations of our methods that clarify and support the results presented in the main Letter. 

\section{GBT Results for the Galactic Center, M31, and M54}

In this section we present results for analyses of GBT data for the following observation targets: the GC, the Andromeda galaxy (M31), and the globular cluster M54.  The M54 and M31 are complementary targets to the nearby INS targets and the GC.  Like the GC observations, the M54 and M31 observations also search for axion-photon conversion from a population of NSs in DM-dense environments.  M31 has the advantage of being a large and nearby galaxy, similar to our own Milky Way.  M54, on the other hand, is observed to lie at the center of the Sagittarius dwarf galaxy.  It is also relatively nearby at a distance around 25 kpc from the Sun.  Note that the velocity dispersions of the DM in Sagittarius and the NSs in M54 are $\sim$10 km/s, which increases the gravitational capture cross section for the NSs pulling in ambient DM.  On the other hand, the DM density profile of Sagittarius is especially uncertain given its tidal interactions with the Milky Way (see~\cite{Safdi:2018oeu} for details).   

The parameters of the GBT observations of these targets are summarized in Tab.~\ref{tab:targets-GBT}.  These observations proceeded analogously to the INS observations, with a few minor differences.  
The GC and M31 observations were performed with one VEGAS spectrometer across the L-band (mode 2), while the M54 observation used three VEGAS spectrometers (mode 4).
For the GC observations, ON and OFF locations were separated by $2.5^\circ$ since the signal is expected to be spatially extended in this case.  Otherwise, the ON and OFF locations were separated by $1.25^\circ$.  Note that we also performed an off-center GC observation, in addition to an on-center observation, because of the possibility of a higher signal-to-noise ratio by looking slightly away from the very bright center of the galaxy.  
The angular positions of all of the targets are well known at the accuracy needed for the GBT at these wavelengths, but for the GC we chose Galactic coordinate $(\ell,b) = (359.9443^\circ, -0.04614^\circ)$ for the center observation and $(\ell,b) = (359.9996^\circ, 0.9958^\circ)$ for the off-center observation.
The data were saved in exposures of 0.5 seconds for observations of all other targets in Tab.~\ref{tab:targets-GBT}.

We note that the GBT observations took place in two sessions, and during each session calibrator observations were also performed.  The calibrators were 3C286 for the first session, which included the GC observations and M54, and 3C48 for the second session, which included M31 and the INS targets discussed in the main text.  The calibrators were observed for approximately two minutes in each of the modes used in that session.

\begin{figure}[htb]
  \includegraphics[width = .49\textwidth]{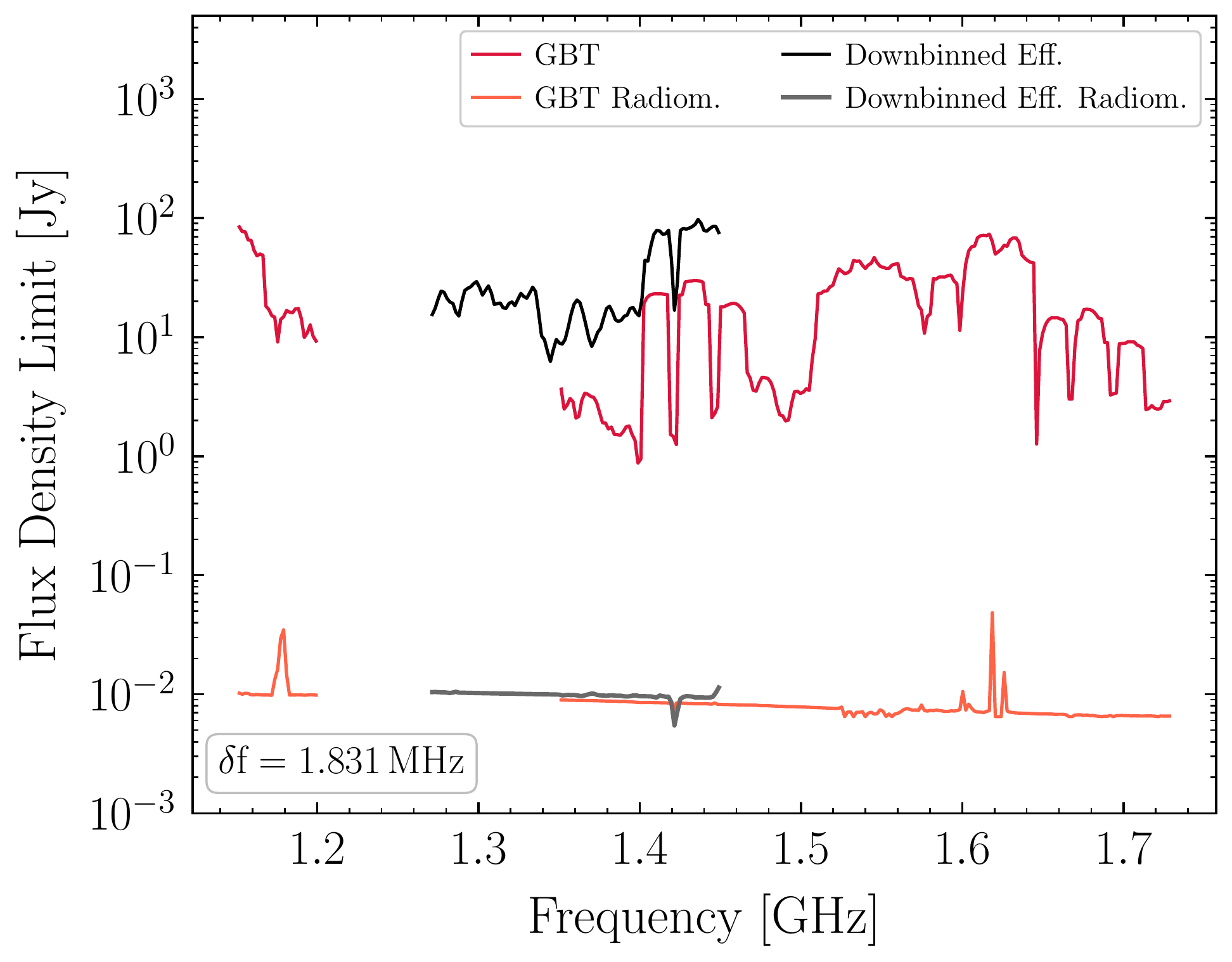} \includegraphics[width =.49\textwidth]{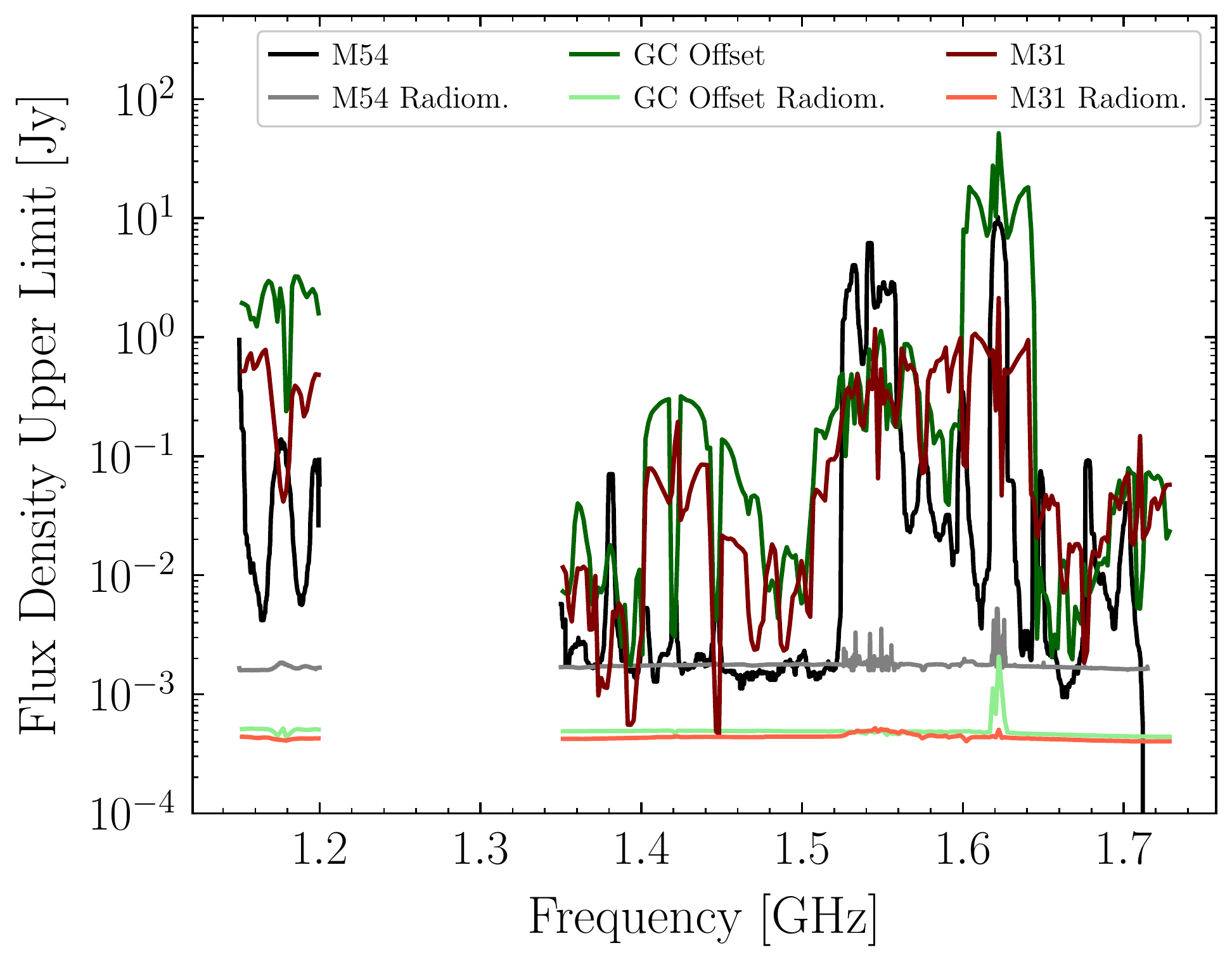} 
  \caption{(\textit{Left}) A comparison of the 95\% upper limits of the flux density spectra measured with our windowed analysis for the GBT and Effelsberg observations of the Galactic Center and radiometer expectations. Data are analyzed at an approximately 1.831 MHz frequency resolution corresponding to the fiducial resolution for the GBT analysis. Although the Effelsberg data is consistent with the radiometer expectations at its original resolution, when down-binned to the GBT resolution, it demonstrates similar incompatibility with the radiometer expectations. (\textit{Right})  The 95\% upper limits on the signal flux for the indicated sources from the GBT observations. These signal flux limits are compared to the expected flux density limit appropriately computed from the radiometer equation. The analysis is performed at the fiducial analysis bandwidth, see Tab.~\ref{tab:targets-GBT}.
  \label{fig:Bandwidth_Systematics}}
\end{figure}

For the population analyses with GBT data we down-bin the data to the level $\delta f/f \sim 10^{-3}$ to account for the fact that we are searching for emission from the whole population of NSs, each of which is Doppler-shifted with respect to the intrinsic frequency~\cite{Safdi:2018oeu}.  The M54 data is down-binned slightly less to account for the smaller velocity dispersion in the globular cluster compared to the GC and M31~\cite{Safdi:2018oeu}. This is a qualitatively different approach to that taken with the Effelsberg data, which is at higher frequency resolution, because with the GBT data from the populations we do not have high enough frequency resolution to search for the brightest converting NS.  Instead, we search for the broader emission from the whole population of NSs.  The frequency bins $\delta f_{\rm fid}$ used in the analyses are given in Tab.~\ref{tab:targets-GBT}.

 \begin{table}[htb]
\begin{tabular}{|c||c|c|c|c|c|}
\hline
Target        		&  $t_{\rm exp}$ [min]  	& $\delta f_{\rm obs}$ [kHz] & $\delta f_{\rm fid}$ [kHz] &     type			\\ \hline
GC 	& $30.0$    	& 92.0 & $1831.0$        		& \, pop. \,       		\\ \hline
GC (off-center)	& $30.0$   & 92.0 	& $1831.0$        		& \, pop. \,  		\\ \hline
M31 	& $30.0$    	& 92.0 & $1831.0$        		& \, pop. \,    		\\ \hline
M54	& $30.0$    &5.7 	& $114.4$        		& \, pop. \,   		\\ \hline
\end{tabular}
\caption{ As in Tab.~\ref{tab:targets} but for the GBT observations of the GC, M31, and M54.}
\label{tab:targets-GBT}
\end{table}

The 95\% one-sided upper limits on the line flux densities from these GBT observations are shown in Fig.~\ref{fig:Bandwidth_Systematics}.  Note that these limits used the frequency-down-binned data, with bin widths $\delta f_{\rm fid}$.  Just as in Fig.~\ref{fig:flux_constraints}, we also show the expectations for each observation from the radiometer equation, which only accounts for statistical uncertainties.

In the left panel of Fig.~\ref{fig:Bandwidth_Systematics}, we show the GBT GC flux density limits.  These limits are significantly higher than the radiometer expectations, indicating that the uncertainties, which are determined in a data-driven fashion by fitting the null model to the sideband data, are predominantly systematic and not statistical. 
This may seem surprising, when considering that in Fig.~\ref{fig:flux_constraints} the Effelsberg constraints were comparable to the radiometer expectations.  However, it is important to remember that the bandwidths $\delta f_{\rm fid}$ for the Effelsberg analyses in Fig.~\ref{fig:flux_constraints} are significantly narrower than those that go into the left panel of Fig.~\ref{fig:Bandwidth_Systematics} for the GBT (7.32 kHz for L-band Effelsberg in Fig.~\ref{fig:flux_constraints} versus 1831 kHz for the GBT in Fig.~\ref{fig:Bandwidth_Systematics}).  Indeed, in Fig.~\ref{fig:Bandwidth_Systematics} we also show what happens if we down-bin the L-band Effelsberg data to 1831 kHz before performing the analysis; in this case, the results are similar to those from the GBT.  We thus see that as the bandwidths are increased, the statistical uncertainties become less important relative to the systematic uncertainties.  This is consistent with the expectation that the statistical uncertainties in the individual frequency bins shrink with more data, which is acquired by increasing the bin widths, while the systematic uncertainties need not change with bin width.

In the right panel of Fig.~\ref{fig:Bandwidth_Systematics} we show the flux density upper limits from the GC offset, M31, and M54 GBT observations, compared to the radiometer equation expectations.  Note that these analyses use the $\delta f_{\rm fid}$ bin widths given in Tab.~\ref{tab:targets-GBT}.  In a narrow  frequency range around around 1.35 - 1.5 GHz, all of the upper limits become comparable to the radiometer expectation, while above 1.5 GHz, all of the upper limits become significantly weaker than expected from statistical uncertainties only.  However, this is precisely what is expected from the known sources of RFI that affect GBT.\footnote{\mbox{See, {\it e.g.}, \url{http://www.gb.nrao.edu/IPG/rfiarchive_files/GBTDataImages/GBTRFI04_08_19_L.html}.}}  Thus, we conclude that the loss of sensitivity above $\sim$1.5 GHz is likely due to sources of RFI that induce variance in the null-hypothesis model. 

The interpretations of the additional GBT observations in terms of the axion model are given later in this SM.

\section{GBT Data Processing}

In this section, we describe the procedure by which we filter the time-series antenna data collected at the GBT and the modified implementation of \texttt{GBTIDL} used to process the time-series data leading to a frequency-dependent stacked antenna temperature \cite{garwood_marganian_braatz_maddalena_2005}. This antenna temperature is then translated to a flux density measurement through comparison to a reference calibration source.

\subsection{Determining the Antenna Temperature with Data Filtering}

\subsubsection{System temperature calibration} For each observation (the procedure that we describe in this section applies to all GBT observations, including those only discussed in the SM), data was collected simultaneously for XX and YY polarizations across a range of frequency channels. For the INSs, data was recorded for $0.2097 \, \mathrm{s}$ exposures, while for all other targets, data was recorded for $0.5 \, \mathrm{s}$ exposures. In alternating exposures, a calibration noise diode with known effective temperature in each polarization is turned on and off, so that the first antenna measurements in each polarization do not include contributions from the noise diode, the second measurements do, the third measurements do not, and so on. Periodically, the observing position was alternated between ON and OFF positions. We will denote the $i^\mathrm{th}$ antenna measurement at the $j^\mathrm{th}$ frequency channel in the $ZZ$ polarization (where $ZZ$ stands for either $XX$ or $YY$) for the ON data at by $d_{i,j}^{ZZ}$ so that the noise diode is on and off for even and odd $i$, respectively. Likewise, we will denote the $i^\mathrm{th}$ antenna measurement at the $j^\mathrm{th}$ frequency in the $ZZ$ polarization for the OFF data by $\tilde d_{i, j}^{ZZ}$. Note that except for a handful of locations due to switching the observation position, the data $d_{i+1, j}^{ZZ}$ ($\tilde d_{i+1,j }^{ZZ}$) is collected immediately after $d_{i, j}^{ZZ}$ ($\tilde d_{i,j}^{ZZ}$). However, for a given ${i, j}$, $d_{i,j}^{ZZ}$ and $\tilde d_{i,j}^{ZZ}$ are \emph{not} collected simultaneously. 

The \textit{system temperature} can be measured at each frequency channel in each polarization using the OFF-position data. The noise diode is alternated on at even $i$ and off at odd $i$, enabling us to obtain the system temperature by 
\begin{equation}
    T_{sys, j}^{ZZ} = T_{cal}^{ZZ} \left[\frac{\sum_{\mathrm{odd} \, i } \tilde d_{i,j }^{ZZ}}{\left(\sum_{\mathrm{even} \, i } \tilde d_{i,j }^{ZZ}\right)- \left(\sum_{\mathrm{odd} \, i } \tilde d_{i,j }^{ZZ}\right)} + \frac{1}{2} \right]\;.
\label{eq:SystemTemperature}
\end{equation}
This frequency-dependent treatment is recommended \cite{garwood_marganian_braatz_maddalena_2005} but not implemented by the standard \texttt{GBTIDL} procedure, in which the sums are extended to be additionally performed over the inner 80\% of all frequency channels so that the observation is characterized by a single system temperature. We found the frequency-dependent procedure to result in a cleaner calibration. Next, using the system temperature as determined by the OFF-position data, we can determine the antenna temperature associated with the ON-position data by
\begin{equation}
    T_{a, j}^{ZZ} = T_{sys, j}^{ZZ} \times \frac{\sum_i d_{i,j }^{ZZ} - \tilde d_{i,j }^{ZZ}}{\sum_i \tilde d_{i,j }^{ZZ}}.
\label{eq:AntennaTemperature}
\end{equation}

\subsubsection{Data cleaning} Because the antenna temperature and therefore the flux density spectrum are directly determined by time averages over the ON-position and OFF-position data, improved calibration precision and a cleaner flux density spectrum, less affected by spurious features, can be achieved by filtering out transient noise features that appear in the data. Here, we accomplish this by filtering the data independently at each frequency channel and in each polarization, in both ON and OFF positions, using the Kolmogorov-Smirnov (KS) test. This is a novel procedure we have devised and implemented in order to obtain clean flux density measurements for the high-precision axion search. Alternate non-parametric $k$-sample tests, such as the Anderson-Darling test, may also be appropriate but are more difficult to implement efficiently.  We note also that while these cleaning procedures do help reduce the noise in the data, they are not strictly necessary and we find qualitatively similar results without data cleaning. 

The filtering is performed in the following way. For a particular observation position, frequency channel, and polarization, we have time-series data which we divide into equally-sized intervals of the shortest length possible satisfying the following conditions:
\begin{itemize}
    \item the interval represents at least 15 seconds of continuous exposure;
    \item the interval contains an even number of antenna measurements;
    \item the interval contains at least 60 independent antenna measurements.
\end{itemize}
Denoting the $j^\mathrm{th}$ interval $y_j$, we determine the $y_j$ with the smallest mean and designate this as our reference interval, which we denote by $\hat y$. The reference interval is accepted by default. We then compare each $y_j$ with $\hat y$ under the KS test when each dataset is unaltered and when each dataset has its mean independently subtracted off. Datasets $y_j$ which are not found incompatible at the 5$\sigma$ level with $\hat y$ under both tests are accepted, with the remaining excluded.

The requirement that the intervals represent at least 15 seconds of exposure is motivated by choosing a duration such that our filtering procedure will never select for intervals which do not contain flux density from an axion signal, which would lead to a spurious exclusion of a signal if it were present. The axion signal from a given NS will have a periodicity set by the NS period, and the time-averaged axion signal flux is expected to be constant over time intervals corresponding to many such periods. Therefore, since NS spin periods are expected to be less than 15 seconds, a potential signal will be present in all intervals.  We additionally require at least 60 independent antenna measurements so that there is a sufficiently large sample for the KS test to have discriminating power. Choosing the shortest interval possible satisfying these criteria then maximizes the resolution of our filtering process. 
Moreover, performing the test with and without the channel mean removed is intended to improve the discriminating power of the KS test. Requiring that each interval is even-sized ensures that an equal amount of noise diode-on and noise diode-off data is contained in each interval and therefore in the accepted ensemble.

\begin{figure}
\includegraphics[width = .8\textwidth]{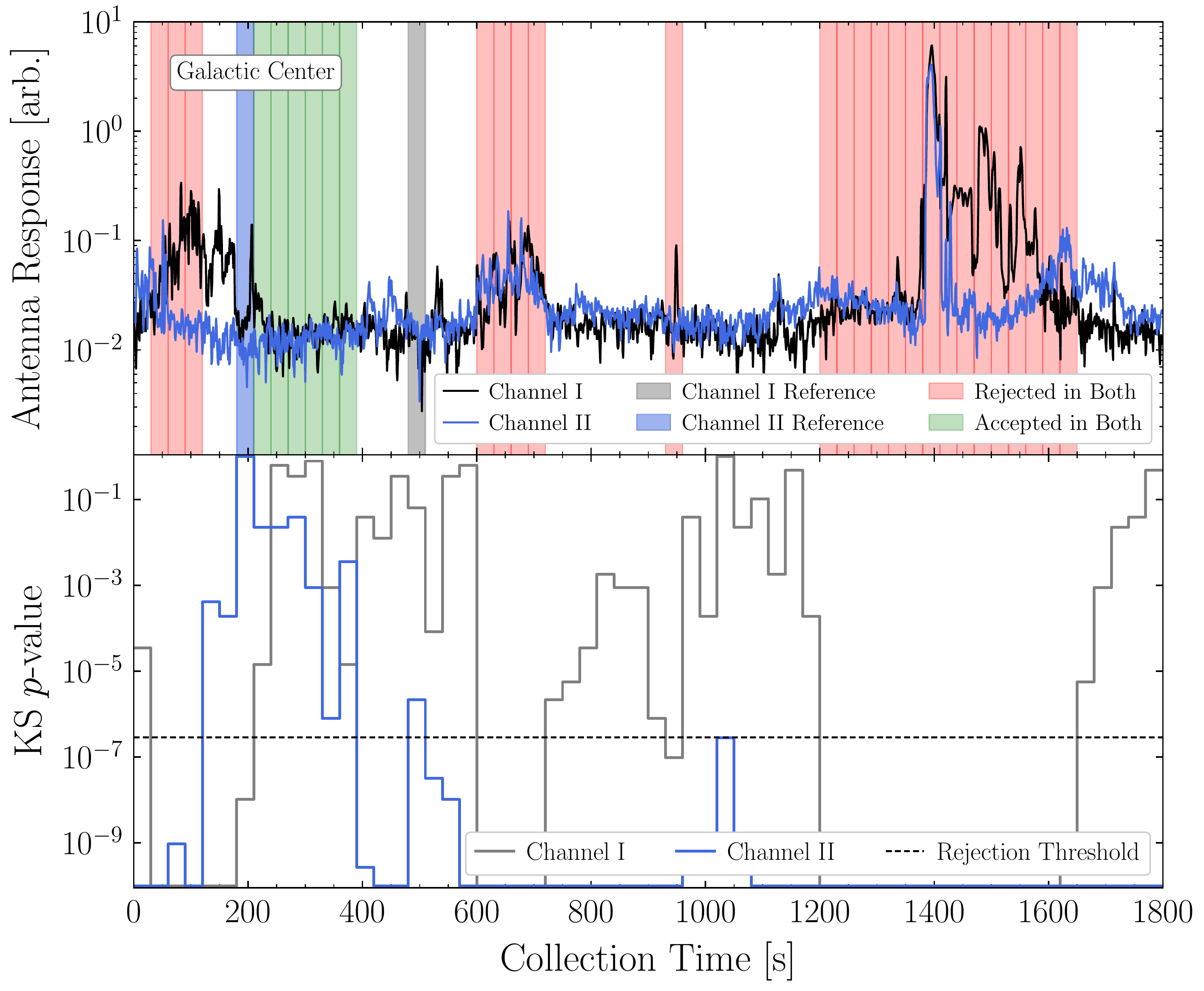} 
\caption{The interval-by-interval acceptance for two adjacent frequency channels for data taken from the GC observation by the GBT. Channels I and II are located at 1.61908569 GHz and  1.61899414 GHz, respectively. Data for each channel are shown in black and blue, respectively, with correspondingly colored highlighted regions identifying the reference interval for each channel. The antenna response is shown in arbitrary units. Time intervals accepted in both channels are highlighted in green, with those rejected in both channels highlighted in red. Intervals which are accepted in only one channel are not highlighted.}
\label{fig:ChannelCompare}
\end{figure}

\begin{figure}
\includegraphics[width = .99 \textwidth]{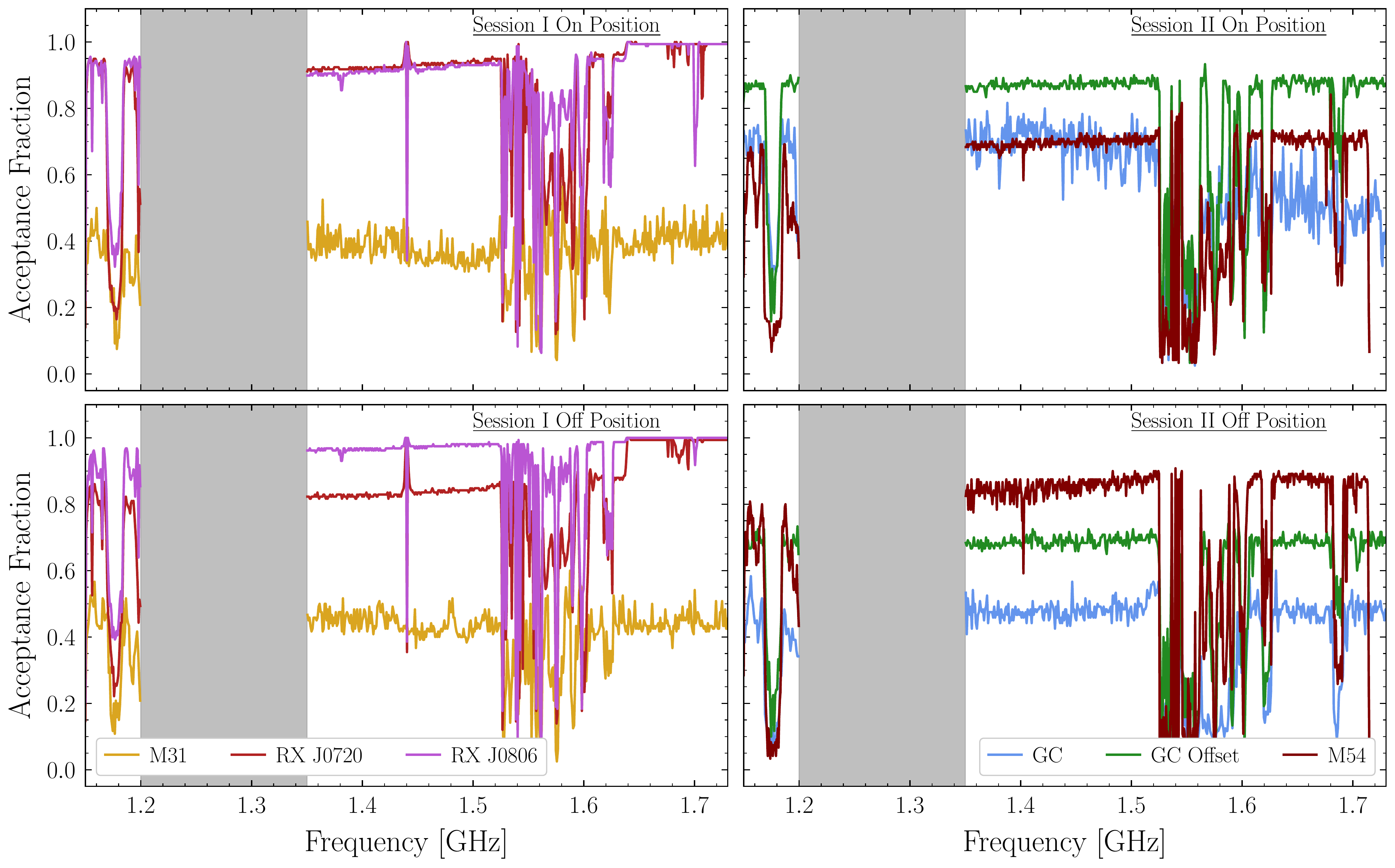} 
\caption{The channel-by-channel acceptance fraction for each ON-position measurements of the observation target in each observing session. Acceptances are averaged over the two polarizations and downsampled by a factor of 50 for presentation purposes. In the top row, we display the acceptances in the ON observing position; in the bottom row, the acceptances in the OFF observing position.
}
\label{fig:Acceptances}
\end{figure}

An example of the data filtering as applied to two adjacent frequencies in ON position data collected at the GC can be seen in Fig.~\ref{fig:ChannelCompare}, where the advantage of independently filtering the data at each channel can be seen by the nontrivial differences in the data and the acceptance results in each channel. The polarization-averaged acceptances for the ON-position data for each observation, sorted by observing session, can be seen in Fig.~\ref{fig:Acceptances}. For the INSs, the acceptance fraction is quite high, while the acceptance fraction tends to be lower from lower frequency resolution sources like M31 and the GC.  This may be due to the reduction in statistical uncertainties when going to larger bandwidths and the increased importance of systematic uncertainties. 

With a successful filtering procedure applied to the data, the calculation of the system temperature is modified so that only data which is accepted in the filtering is included in the calculation of the system temperature, and subsequently the antenna temperature. Once we have determined antenna temperatures in each polarization, we determine a single antenna temperature by averaging over the polarizations, which typically differ at the percent level or below, as
\begin{equation}
     T_{a, j} = \frac{T_{a, j}^{XX} + T_{a, j}^{YY}}{2}\;.
     \label{eq:AvgAntennaTemperature}
\end{equation}
Since the antenna temperature $T_{a,j}$ is proportional to the flux density, we down-bin by averaging in antenna temperature. 

\subsection{Calibrating to a Flux Density}

We calibrate to a flux density through an observation of a flux calibrator. For Session I Observations (M31, RX J0720.4$-$3125, and RX J0806.4$-$4123), we observed the quasar 3C48. For Session II Observations (GC, GC Offset, M54), we observed the  quasar 3C286. Several observations of each calibrator were made so that each target observation can be calibrated to a calibration observation with identical spectrometer configuration and frequency resolution.

Because the observations of the flux calibrators were relatively brief, the data are not amenable to a time-series filtering, so we directly determine an antenna temperature using~\eqref{eq:SystemTemperature},~\eqref{eq:AntennaTemperature}. For a particular calibrator, we have a calibrated flux density scale from \cite{Perley_2013, Perley_2017}. We then determine a frequency-dependent calibration scale for our data, $c_j$, by comparing the observed antenna temperature $T_{a, j}$ from the calibrator to the flux density expected from the source:
\begin{equation}\label{eq:Ta_to_Jy}
    c_j = \frac{e_j}{\mathrm{med}(T_{a, j}, 31)} \,
\end{equation}
Here $e_j$ is the expected flux density at the $j^\mathrm{th}$ frequency channel computed from \cite{Perley_2013, Perley_2017}.  We smooth the antenna temperature at the analysis frequency resolution from the calibrator by a median filter with kernel size of $31$ channels, which is wider than our analysis window, in order avoid calibrating against small-scale features that appear in the data, since the calibration data was taken at a lower time resolution. This frequency-dependent calibration factor then allows us to calibrate our antenna temperature measurement from our target observations.

\section{Effelsberg Data Processing}

We note that the Effelsberg P217mm receiver provides circular polarizations with a receiver system equivalent flux density of $\sim$ $17$ Jy  prior to averaging the two polarizations. The S110mm receiver similarly provides circular polarizations with a receiver system equivalent flux density of $\sim$ $11$ Jy.
Effelsberg data at L- and S-Band were calibrated to a flux density scale via observations of the reference source ${\rm NGC}\,7027$ and using the publicly available software package \texttt{PSRCHIVE}\footnote{\url{http://psrchive.sourceforge.net/}}~\cite{psrchive}.
Unfortunately, an insufficient signal-to-noise ratio for the noise diode signal, in the fine frequency channels used in this work, resulted in strong frequency-dependent artifacts in the resulting calibrated spectra. Because these effects impinged on efforts to identify lines of interest, we assumed an ideal flat frequency response for the telescope receiver (except at the band edges) and defined calibration factors from the integrated spectrum for both ON and OFF target positions; in this case the GC. 

\begin{figure}
\includegraphics[width = .9\textwidth]{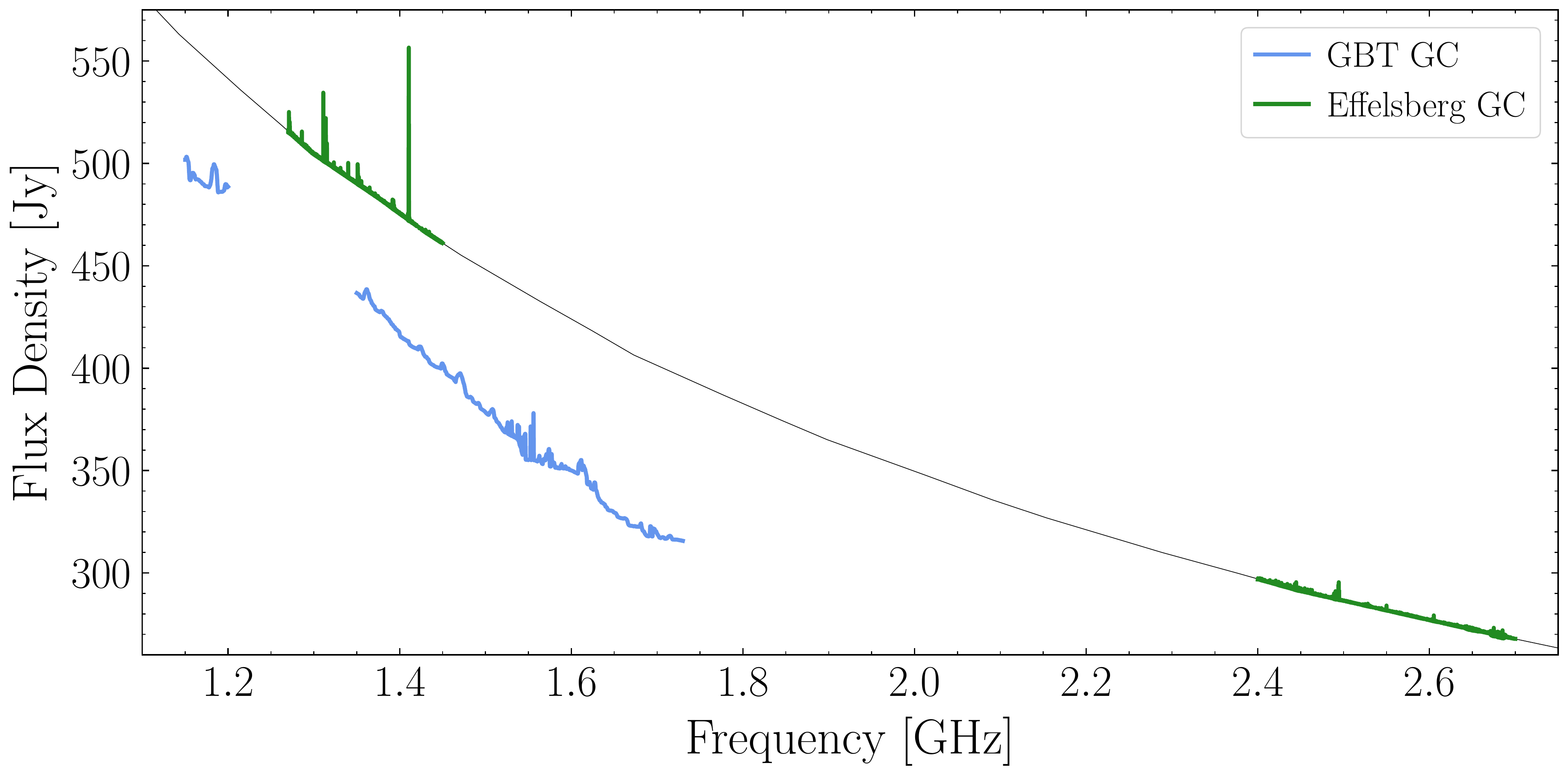} 
\caption{A comparison of the calibrated flux density for the GBT observation of the GC (\textit{blue}) to the Effelsberg observations of the GC in the L-Band and S-Band (\textit{green}).  Note that the Effelsberg data is calibrated to follow the black curve, averaged over large frequency scales. The calibrated L-band Effelsberg data is around 20\% different than the calibrated GBT data, suggesting that errors from the calibration procedure impacting sensitivity to $\gagg$ are only on the order of 10\% and subdominant compared to other sources of uncertainty.}
\label{fig:GC_Comparison}
\end{figure}

However, we find that the maximum flux density in our data under this calibration procedure is in agreement with the expected flux density of the Galactic plane synchotron radiation flux density reported in \cite{2011MNRAS.413..763C}. We therefore calibrate both our L-Band and S-Band data by instead the calibration factor
\begin{equation}
    c_j = \frac{e_j}{\mathrm{med}(d_j, 201)} \,,
\end{equation}
where $e_j$ is the expected flux density at the $j^\mathrm{th}$ frequency channel from \cite{2011MNRAS.413..763C} and $\mathrm{med}(d_j, 201)$ is a median smoothing with kernel size 201 of the antenna data. This calibration assumes there to be no instrumental backgrounds on the antenna, which may not be true, and it also assumes that the expected flux density spectrum from \cite{2011MNRAS.413..763C}, which was constructed by fitting a functional form to data from different telescopes across a range of frequencies, may be applied to the Effelsberg angular beam size, which is also likely not completely true.
A comparison of the calibrated Effelsberg data to the calibrated GBT data for the GC is shown in Fig.~\ref{fig:GC_Comparison}. The absolute magnitude of the Effelsberg flux density is typically $20 \%$ larger in the L-Band than in the corresponding GBT data.  The discrepancy between the GBT and the Effelsberg spectra is likely due to the Effelsberg calibration procedure. By comparison, in \cite{1990A&AS...83..539R}, a 21 cm observation of the GC using the Effelsberg telescope determined a flux density of 450 Jy, which is more compatible with the GBT GC spectra found in this work. 
Because the axion constraints scale like the square-root of the flux density, however, these differences have only a roughly 10\% effect on the constraint strength and so we do not pursue them further. Moreover, we note that systematic errors in radio data are typically on the order of $\mathcal{O}(10\%)$, independent of calibration scheme.

\section{Axion Signal Modeling}
Accurately modeling the axion signal is an important step in our limit-setting procedure. In order to model the radio signal from axion-photon conversion from a population of NSs, we choose a DM density profile and use the NS population models from \cite{Safdi:2018oeu} to determine the distributions of the NS locations, magnetic fields, and spin periods. We then draw from these distributions to construct a MC sample. NSs within the beam width will contribute to a potentially observable signal. At each axion mass, we compute a flux density associated with conversion at each NS within the beam that appears in the data, at a frequency corresponding to the axion mass shifted by a randomly-drawn Doppler factor corresponding to the peculiar velocity of that NS along the line of sight. 

\begin{figure}
\includegraphics[width = .95 \textwidth]{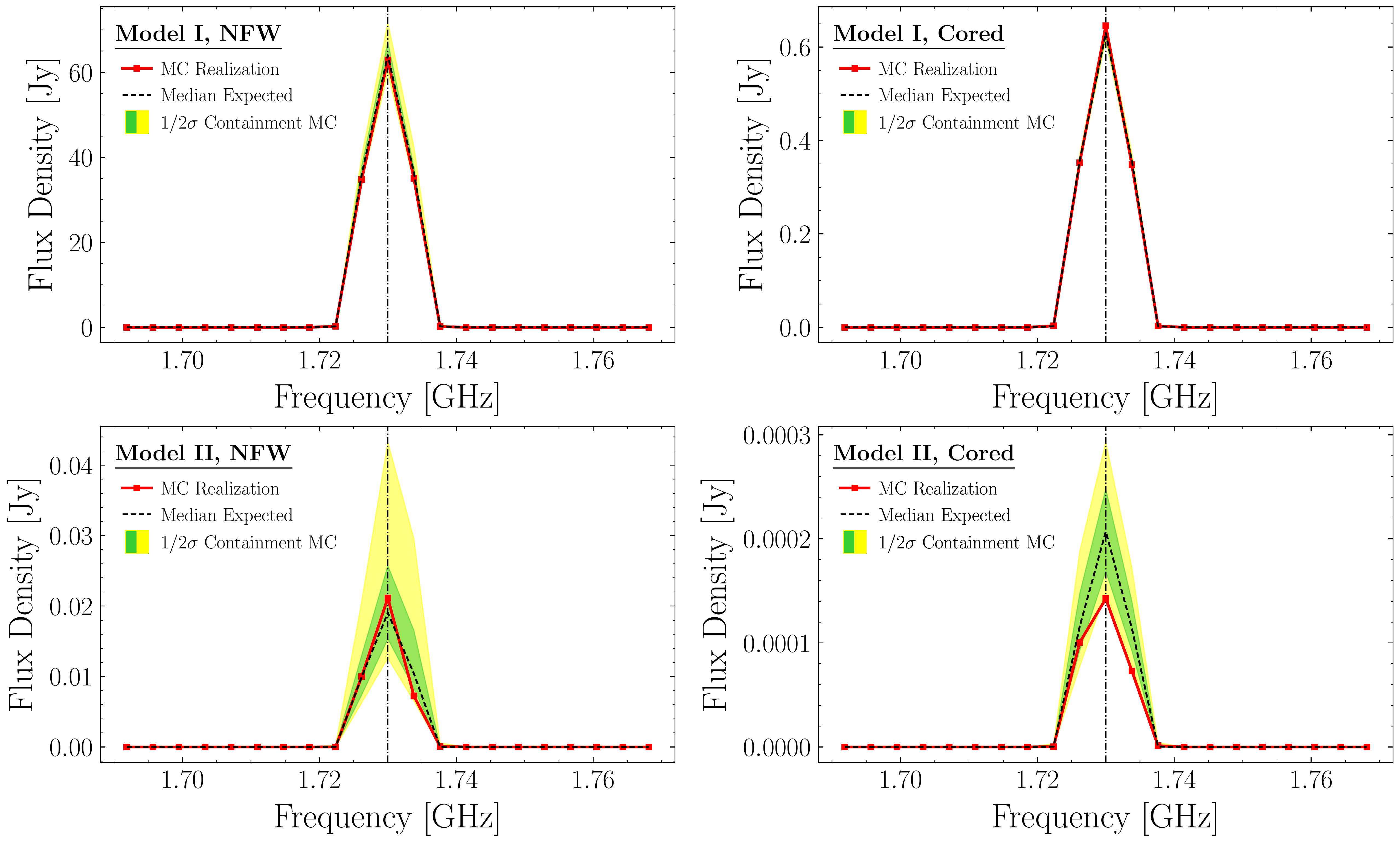}
\caption{(\textit{Top Left}) A noise-free example flux density spectra for an axion of mass $m_a = 3.46 \pi$ GHz with a coupling strength of $g_{a \gamma \gamma} = 10^{-11}\, \mathrm{GeV}^{-1}$ generated for the GBT observation of the GC at $\delta f_{\rm fid} = 1.831$ MHz. We assume Model I for the NS population (the model with more NSs participating in the conversion process) and take the DM to follow an NFW density profile. (\textit{Top Right}) As in the top left, but using a cored DM density profile with a core radius of 600 pc. (\textit{Bottom Left}) As in the top left, but assuming the conservative Model II for the NS population. (\textit{Bottom Right}) As in the top right, but assuming the conservative Model II for the NS population.}
\label{fig:ModelI_GC_Spectra}
\end{figure}

Examples of axion flux density spectra generated for analysis of the GBT GC observations (for population searches) can be seen in Fig.~\ref{fig:ModelI_GC_Spectra}.  Note that these observations are down-binned such that the signals from all NSs appear mostly in the central bin, though there is some leakage to the neighboring two frequency bins.  Fig.~\ref{fig:ModelI_GC_Spectra} shows the expected flux over multiple MC realizations for an axion with mass $m_a = 3.46 \times \pi$ GHz and $g_{a\gamma\gamma} = 10^{-11}$ GeV$^{-1}$.  The dashed black curves show the median expected power, while the green and yellow bands show the 68\% and 95\% containment regions, respectively, for the power.  The red data points illustrate one representative MC realization. Note that in Fig.~\ref{fig:ModelI_GC_Spectra}  we assume  NFW DM or  cored profile for the Milky Way, as indicated. We also alternate between NS Model I, which has more relevant, converting NSs, and our fiducial NS Model II, again as indicated.  Note that the spread is greater for Model II since in this case there are less NSs participating in the conversion process.  

\begin{figure}
\includegraphics[width= .95\textwidth]{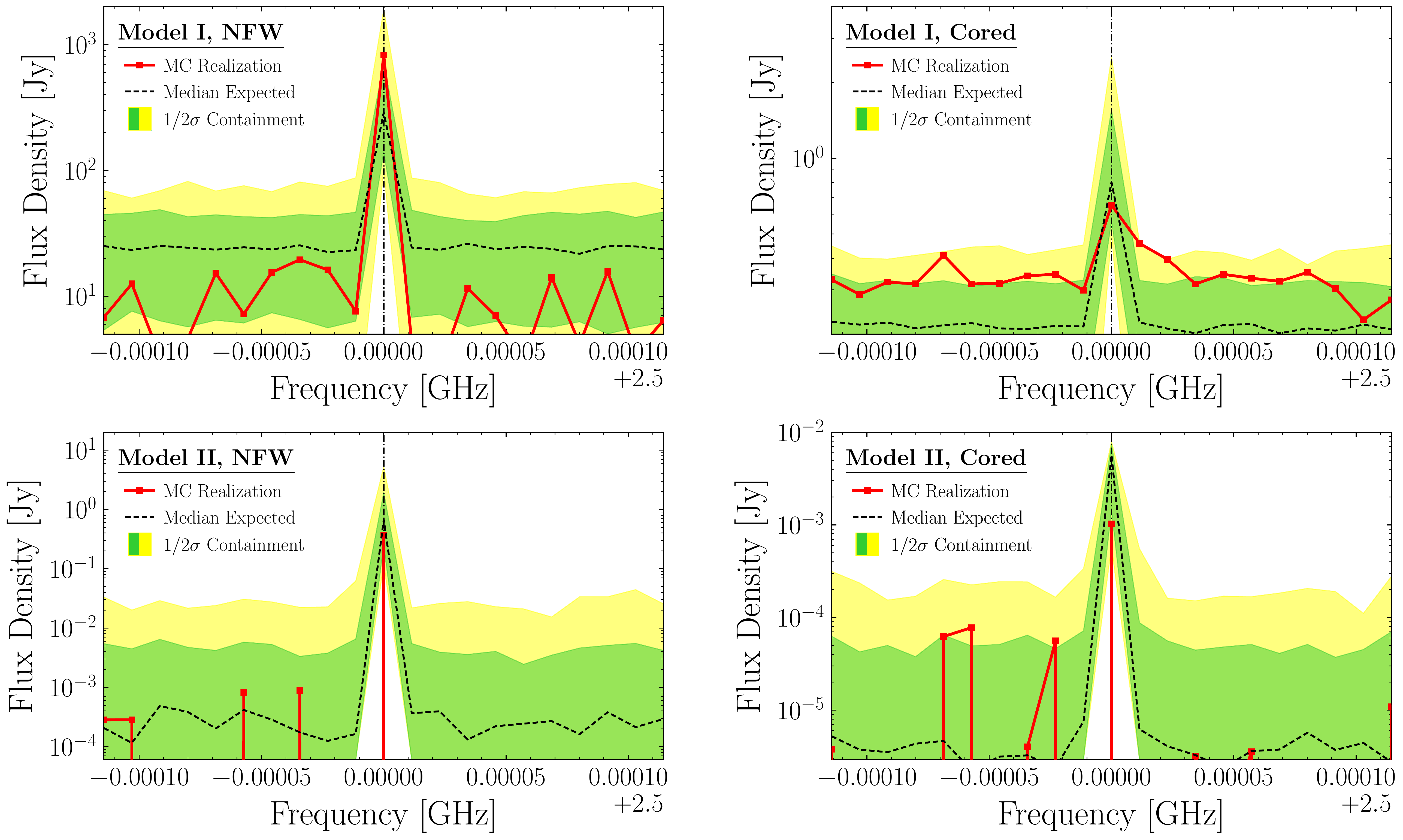} 
\caption{As in Fig.~\ref{fig:ModelI_GC_Spectra} but for the Effelsberg observations of the GC in the S-band, for an axion with $m_a = 2 \pi \times 2.5$ GHz and $g_{a \gamma \gamma} = 10^{-11}\, \mathrm{GeV}^{-1}$.  The panels indicate the assumed DM density profile for the Milky Way (NFW or cored NFW with a core radius of 0.6 kpc) and also the NS population model (Model I or Model II, as described in the text).  Note that in this case we search for the brightest converting NS.  We have shifted each of the MC realizations around in frequency space such that the brightest converting NS appears at $f = 2.5$ GHz.  Note that in the scenario with Model 1 and a cored DM profile, the brightest converting NS is not always that much brighter than the signal flux in the sidebands, from other NSs within the Effelsberg beam, which makes it harder to discover an axion signal in this case.   }
\label{fig:ModelI_Eff_Spectra}
\end{figure}

By contrast, the Effelsberg data is of sufficiently high frequency resolution to enable the search for the brightest converting NS. Example flux density spectra are shown in Fig.~\ref{fig:ModelI_Eff_Spectra}.  The different panels indicate which NS population model is used in the analysis along with the assumed form of the DM density profile (NFW or cored NFW).
Note that in these figures we have shifted the MC in frequency so that the brightest converting NS always shows up at the same frequency (here taken to be 2.5 GHz).  Importantly, we see that the brightest converting NS typically makes the central frequency bin much brighter than the neighboring frequency bins, which are shown at the frequency resolution used in our analysis.  This is especially true when simulating with NS model II, since in this case there are less converting NSs.  The difference between the central and neighboring frequency bins also becomes more apparent with the NFW DM profile, relative to the cored DM profile, since the NFW profile causes those NSs close to the GC to have a bright axion-induced radio flux.  

The situation is slightly more complicated in the NS Model I, cored DM profile case, since in this case the difference between the signal flux in the central frequency bin and the neighboring bins is not as pronounced, as seen in the MC example in top right panel of Fig.~\ref{fig:ModelI_Eff_Spectra}.  In practice this means that using the narrow-frequency bin approach leads to some difficulty in detecting an axion signal in the case of NS Model I and a cored DM profile with this analysis approach.  For example, we may perform multiple MC simulations of the signal only (as shown in red in the top right panel of Fig.~\ref{fig:ModelI_Eff_Spectra}) and then perform our analysis framework on this signal-only data.  This is an idealized analysis, since we are neglecting other sources of noise, but it allows us to quantify the degradation to our detection capability because of signal leakage to the sidebands.  We find, for example, that in this case (for Model I and cored DM profile) that at 95\% confidence the discovery TS should be bigger than $\sim$30.  On the other hand, for our fiducial scenario (Model II and an NFW DM profile) at 95\% confidence the TS is greater than $\sim 10^3$, which indicates that signal leakage into the sidebands is not a concern in this case.  For NS Model I and an NFW DM profile the TS is greater than $\sim$130 at 95\% confidence, which is also greater than our pre-determined discovery threshold.  Thus, the only concern is that in principle an axion signal produced in the S-band for NS Model I and a cored DM profile would not produce a discovery TS above our threshold of 100, at 95\% confidence.  Still, we note that in the S-band there are only two frequency channels with un-vetoed excesses with ${\rm TS} > 30$ (one with ${\rm TS } = 37$ and the other ${\rm TS} = 41$), and in both of these cases the OFF data shows spectral features in the central frequency bin as well, though not at a significant enough level to be vetoed. Thus, we find it unlikely that these signals arise from axions. These excesses, along with others are depicted in Fig.~\ref{fig:ExcessCandidates}. Note that we account for the signal variance in the population models when setting 95\% upper limits on the axion-photon coupling from the radio data (see Sec.~\ref{sec:ax_const}).

\section{Data Analysis}
\label{sec:DataAnalysis}

Here we describe the procedures by which we seek to discover an axion signal and by which we impose upper limits on the axion-photon coupling $g_{a \gamma \gamma}$ in the absence of signal detection. 

\subsection{Profile Likelihood for Antenna Temperature Excesses}

After our data processing and calibration procedure, we have obtained an antenna temperature $T_{a,j}$ and calibration factor $c_j$ over a range of frequency channels around a central frequency $f_j$. Axion-sourced signals are expected to appear as excesses at a narrow frequency bandwidth, so we analyze the antenna temperature downsampled to a resolution at which most (if not all) of the axion signal is expected to appear in a single frequency channel. In principle, we could analyze the calibrated flux density $c_j \times T_{a, j}$ for excesses, but we chose to not to do this so as to avoid injecting spectral features from a calibration spectrum into the calibrated flux density prior to analysis, which could result in the detection of spurious excesses. Analyzing the antenna temperature is acceptable because our smoothed calibration scale $c_j$ should vary slowly as a function of frequency on the scale on which the signal is expected to appear. If the calibration scale were to vary significantly from frequency channel to frequency channel, it would indicate regions in which the data is untrustworthy due to uncontrolled instrumental response.  Indeed, this is the case for the GBT population analyses, which indicates that these analyses are dominated by instrumental systematics, as discussed previously in this SM. 

We search for an excess in a single channel $f_i$ using the Gaussian likelihood
\begin{equation}
\mathcal{L}_i( \vec{T}_a | A, \mathbf{a}) =\prod_{k = i-j}^{i+j} \frac{\exp \left[ - \frac{(T_{a, k} - \mu(f_k | \mathbf{a}) - A \delta_{ik})^2}{2 \sigma_k^2}\right]}{2 \pi \sigma_k^2} \;,
\label{eq:Likelihood}
\end{equation}
where $A$ is signal parameter describing the size of the excess at the $i^\mathrm{th}$ channel and $\mu$ is a quadratic polynomial describing the background determined by the nuisance parameter vector $\mathbf{a}$. The likelihood is calculated including data from the central channel where we seek to identify a putative excess and the $j$ sideband channels to both the left and right. In our fiducial analysis, we take $j = 10$. In each analysis, we exclude the sideband channels immediately adjacent to the central frequency channel in order to not be biased by less than 100\% signal containment. The variance of the data $\sigma_k^2 $ is determined by $\sigma_k^2 = \sigma^2/\alpha_k$ where $\alpha_k$ is the acceptance fraction of the data at the $k^\mathrm{th}$ frequency channel and $\sigma^2$ is a fitted parameter. We fit $\sigma^2$ by fitting the null model ($A = 0$) to the analysis window with the central frequency channel (and hence any putative excess) masked out. We then do not profile over the value of $\sigma^2$ in the analysis. This is done to avoid biasing the calculation of the variance, and therefore the likelihood, in the presence of a large central-channel excess.

Equipped with our full likelihood in terms of both a signal and background model, we determine a profile likelihood purely as a function of the central-channel excess strength $A$ by profiling over the nuisance parameter vector ${\bf a}$: 
\begin{equation}
\mathcal{L}_i(\vec{T}_a |A) = \max_{\mathbf{a}} [\mathcal{L}_i( \vec{T}_a | A, \mathbf{a})]\;.
\end{equation}
We determine the significance of the evidence for a flux density excess described by the best-fit excess parameter $\hat A$ using the test statistic (TS)
\begin{equation}
\mathrm{TS}_i = 2 \times [ \log \mathcal{L}_i(\vec{T}_a | \hat A ) - \log \mathcal{L}_i(\vec{T}_a |0)] \,,
\end{equation}
unless $\hat A < 0$, in which case $\mathrm{TS}_i = 0$. By Wilks' theorem, this test statistic will be asymptotically half-$\chi^2$ distributed \cite{Wilks:1938dza}. In our analysis, our threshold for discovery is set at $\mathrm{TS} = 100$. For details regarding the discovery threshold and TS distribution as informed by Monte Carlo, see Sec.~\ref{sec:Survival}.

\subsection{Excess Vetoing Procedure}

We find multiple excesses in the Antenna temperature data, such as excesses corresponding to 21 cm line emission, that may be vetoed by finding similar excesses in the OFF-position data.
We  analyze the OFF-position data corresponding to each observation, {\it i.e.}, the $\tilde d_{i,j}^{ZZ}$ that were used in~\eqref{eq:AntennaTemperature} to subtract off backgrounds and instrumental baselines. This is important because narrow spectral features present in the OFF-position data indicate locations where backgrounds or antenna responses are systematically mismodeled and narrow spectral features in the calibrated spectrum do not require an axion interpretation.  Note that by mismodeled we do not mean that a real astrophysical line is not necessarily present (such as the 21 cm line), but we mean that our simple quadratic background model is not sufficient to describe all of the astrophysical emission or instrumental backgrounds.     

\begin{figure}
\includegraphics[width = .7\textwidth]{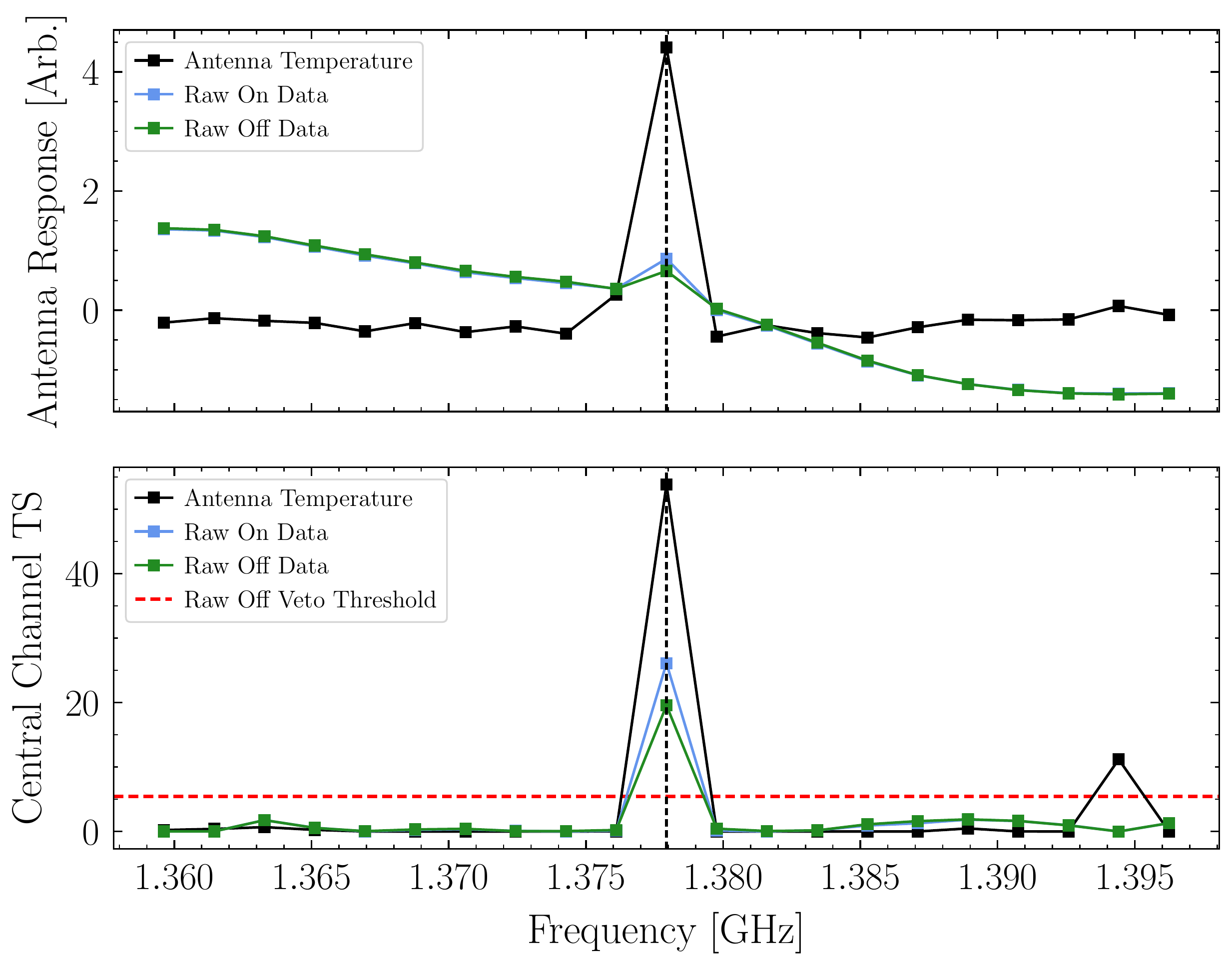} 
\caption{(\textit{Above}) The ON-position antenna temperature and raw antenna data for ON- and OFF-position measurements for the M31 observation. A narrow feature appears at the indicated central frequency channel in each of the datasets, although with larger relative magnitude in the antenna temperature. The fact that the feature appears in all datasets suggests it is not an axion signal. (\textit{Below}) The test statistics for the central channel excess as a function of the central channel for the analysis of the ON-position antenna temperature and the raw antenna data for ON- and OFF-position measurements for the M31 observation. At the location of the narrow feature, the test statistic is quite large for all analyzed datasets, and the excess in the antenna temperature is vetoed as the test statistic in the OFF-position raw antenna data exceeds the veto threshold.}
\label{fig:AllExcessVetoExample}
\end{figure}

We analyze the OFF dataset $\vec{\tilde d}$, which is constructed from the polarization-summed OFF data that has been accepted by our time-series filtering. We analyze this data with a modified version of the likelihood in~\eqref{eq:Likelihood}, now allowing a signal to appear as a flat excess across across one, three, or five frequency channels centered on the central frequency channel. This allows us to identify spurious features slightly wider than the signal we search for in the antenna temperature data that would nonetheless indicate locations of uncontrolled backgrounds and instrumental response. The likelihood is given by
\begin{equation}
\begin{split}
\mathcal{L}_i( \vec{d} | A, \mathbf{a}, w) &= \prod_{k = i-j}^{i-w-1} \frac{\exp \left[ - \frac{(\tilde d_k - \mu(f_k | \mathbf{a}))^2}{2 \sigma_k^2}\right]}{2 \pi \sigma_k^2} \prod_{k = i+w+1}^{i+j} \frac{\exp \left[ - \frac{(\tilde d_k - \mu(f_k | \mathbf{a}))^2}{2 \sigma_k^2}\right]}{2 \pi \sigma_k^2} \\ 
& \times \frac{\exp \left[- \frac{(\sum_{k = i-w}^{i+w} \tilde d_k - \mu(f_k | \mathbf{a}) - A)^2}{2 \sum_{k = i-w}^{i+w} \sigma_k^2}\right]}{2 \pi \sum_{k = i-w}^{i+w} \sigma_k^2} \,.
\end{split}
\label{eq:OffLikelihood}
\end{equation}
This likelihood is at full spectral resolution in the sidebands (the top line in Fig.~\eqref{eq:OffLikelihood} accounts for the left and right sidebands) but at downbinned resolution corresponding to the window width in the signal region (the bottom line in Fig.~\eqref{eq:OffLikelihood}).  Note that we consider three values for $w$ in~\eqref{eq:OffLikelihood}: $w = 0,1,2$, corresponding to signal-region widths of 1, 3, and 5 bins. 
Our variance parameter $\sigma_k^2$ is treated analogously to before, and the TS is taken to be
\begin{equation}
\tilde{\mathrm{TS}}_i(w) = 2 \times [\mathrm{max}_{A, \mathbf{a}}\mathcal{L}_i ( \vec{T}_a | A, \mathbf{a}, w) - \mathrm{max}_{\mathbf{a}}\mathcal{L}_i ( \vec{T}_a |0, \mathbf{a}, w)].
\end{equation}
Note that we calculate the TS for each choice of $w$.
Here we do not zero out TSs associated with negative best-fit signal parameters, as our interest is only in identifying locations of spurious features of any sign. We also do not mask out the frequency channels adjacent to the signal region in this analysis as we are not seeking to identify an axion signal in this data. Large TSs in the antenna temperature data are then vetoed if they are at identical or directly adjacent frequency channels to a channel in the OFF-position data with a TS above the $97.5^\mathrm{th}$ percentile value of the OFF-position TS ensemble.  We perform this test for each of the three choices for $w$. In practice, we find that vetoing on the five-channel wide analysis has negligible effect. An example of vetoed excesses can be see in Fig.~\ref{fig:AllExcessVetoExample}. 

\subsection{Axion Constraints}
\label{sec:ax_const}
Setting constraints on the axion-photon coupling $g_{a \gamma \gamma}$ requires a more involved data analysis procedure than our discovery search. This is because variance in the expected axion signal due to scatter in the NS population prevents a direct conversion of a flux density limit, obtained through Wilks' theorem using our profile likelihood, to a constraint on $g_{a \gamma \gamma}$. 

Instead, we take a profile-likelihood MC approach that allows us to place limits that account for signal variance. To do this, we analyze the real data under the likelihood of~\eqref{eq:Likelihood} to determine the best-fit signal strength parameter $\hat A$ and the best-fit null-model background specified by the best-fit model parameters $\hat{\mathbf{a}}$ and $\sigma^2$. We generate 200 MC realizations of the data under the best-fit null model. For each null model spectra, we construct a unique realization of an axion signal generated under the assumed NS population and DM density profile. The axion signal is injected in the null model spectra with varying strength corresponding to adjusting the value of $g_{a \gamma \gamma}$. We analyze each MC data realization in order to determine the fraction which result in a best-fit signal strength parameter $\hat A_{MC}$ that is less than the best-fit signal strength parameter $\hat A$ in the real data. We can then obtain a frequentist 95\% upper limit on the $g_{a \gamma \gamma}$ by determining the value of $g_{a \gamma \gamma}$ at which only 5\% of the $\hat A_{MC}$ are less than $\hat A$.  The advantage of this procedure is a safe accounting for variance in the signal strength and Doppler shifting that lead to less than perfect signal containment in the central frequency channel. 

\section{Signal Injection Tests}
\begin{figure}
\includegraphics[width = .9\textwidth]{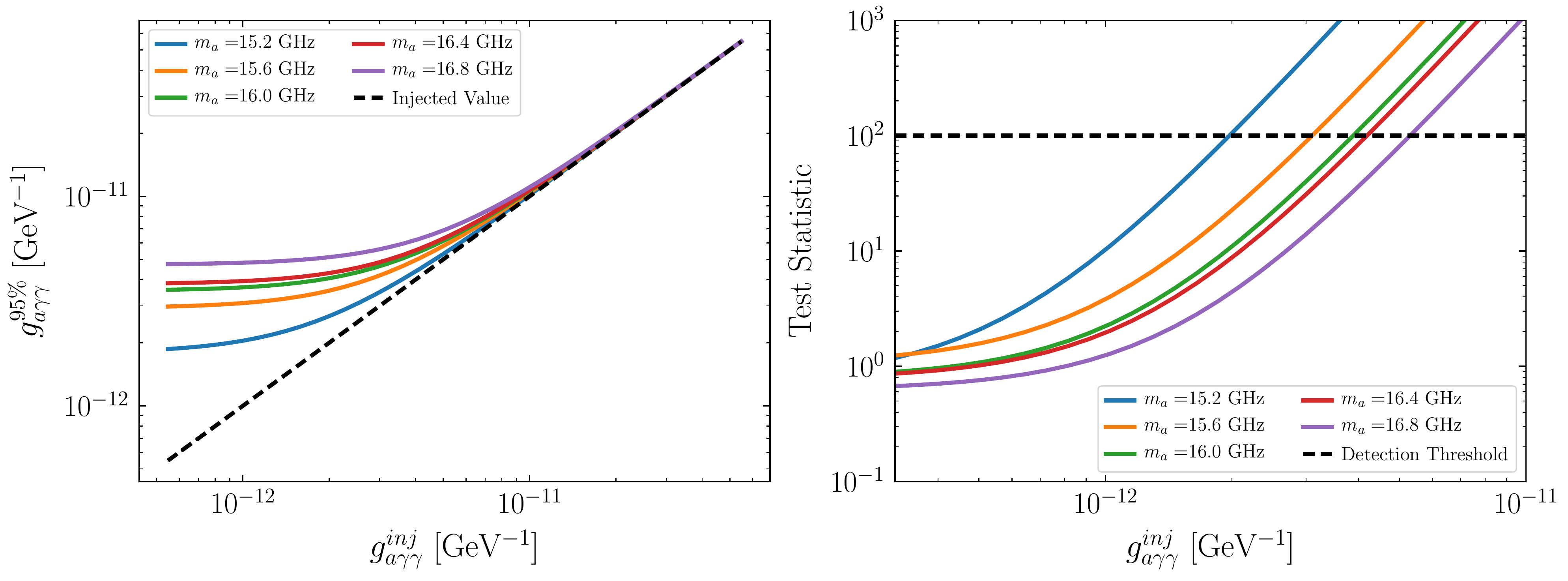} 
\caption{(\textit{Left}) The one-sided 95\% upper limit on the axion-photon coupling as a function of the injected signal strength. The limit lies above the injected signal strength, indicating that we are not excluding an axion signal when present. (\textit{Right}) The test statistic (TS) for discovery as a function of the injected signal strength. 
For sufficiently large signal strengths the TS exceeds our ${\rm TS} = 100$ threshold for an axion signal to be discovered.
}
\label{fig:InjectionTest}
\end{figure}

To demonstrate that our framework is capable of setting appropriate upper limits and discovering an axion were such a signal present in the data, we perform signal injection tests where we inject synthetic axion signals into the actual data.  Here we demonstrate these tests for signal flux density spectra injected at several frequencies in Effelsberg S-Band data. We inject the signal flux density spectra with varying amplitudes corresponding to varying the strength of $g_{a \gamma \gamma}$. The results of the signal injection tests can be seen in Fig.~\ref{fig:InjectionTest}.  In the left panel we show the 95\% one-sided upper limits $g_{a\gamma\gamma}^{\rm 95\%}$ as a function of the injected signal strength $g_{a\gamma\gamma}^{\rm inj}$ for three different and randomly chosen axion masses.  Importantly, we never exclude the injected signal strength, which is indicated by the dashed black line.  In the right panel we show the discovery TS in favor of the axion model as a function of the injected signal strength, with our ${\rm TS} = 100$ discovery threshold indicated in horizontal dashed black.  As expected, the TS increases for increasing signal strengths.

\section{Maser Line Detection Test}

The observation of known spectral line sources enables additional opportunities to test our analysis pipeline. During the March 10$^\mathrm{th}$ observing session, 10 minutes of observing time was used to collect data in the L-Band at the compact HII region W3OH. Data was collected at 92 kHz frequency resolution for five minutes in the ON position, followed by five minutes of data collection at identical frequency resolution in the OFF position. Data was collected over the frequency range 1.15-1.73 GHz with a notch filter applied in the 1.2 to 1.34 GHz region as in all other GBT observations. Narrow maser emission at rest frequencies approximately 1.612 GHz, 1.665 GHz, 1.667 GHz, and 1.72 GHz associated with transitions in hydroxyl molecules have been observed in W3OH \cite{gray_2012, 1981MNRAS.195..213N}, enabling a direct test of the ability of our analysis pipeline to identify bright spectral lines of a similar character to those that might appear due to axion-to-photon conversion in NS magnetospheres.

\begin{figure}
\includegraphics[width = .95\textwidth]{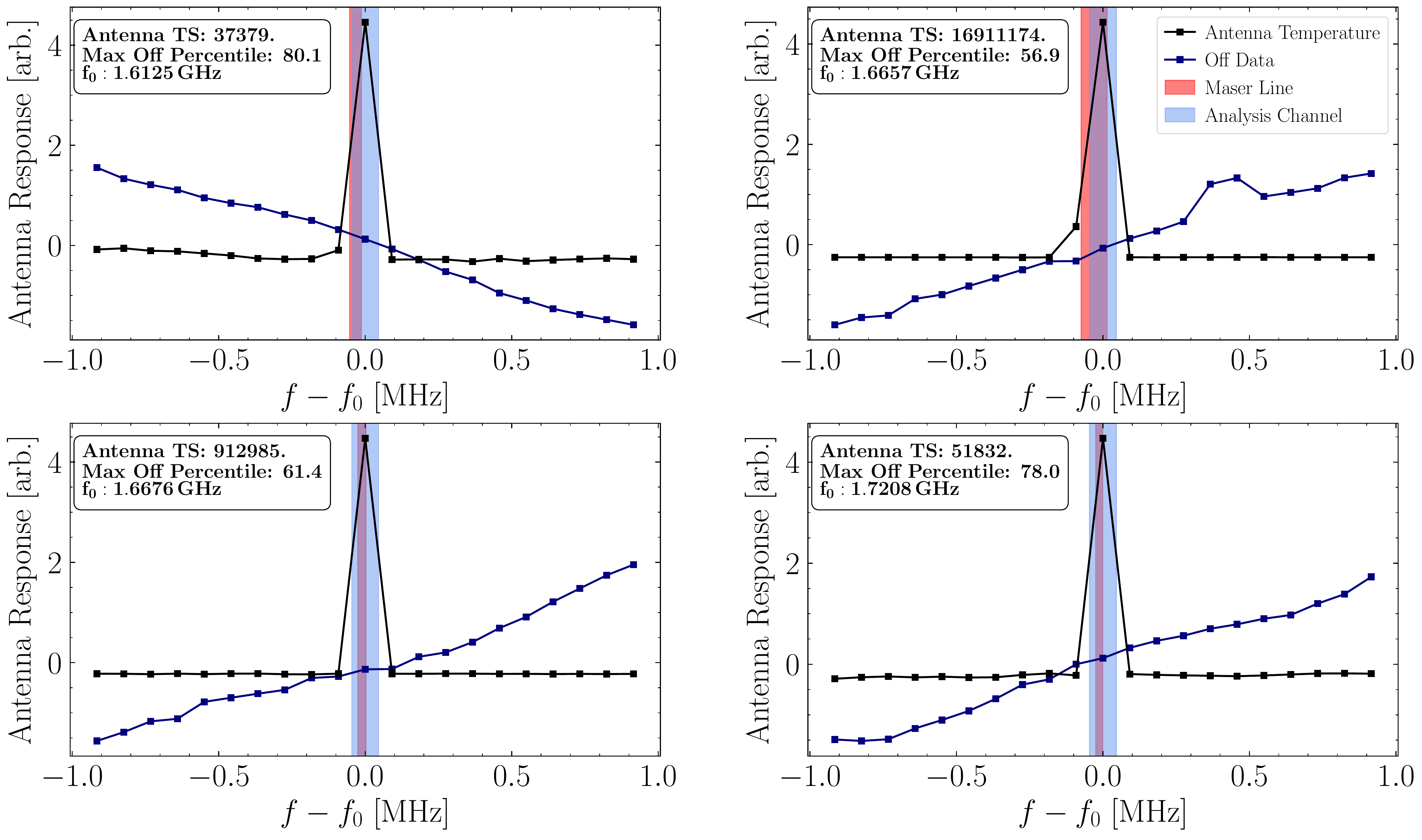} 
\caption{Maser lines as detected in the GBT data. For each maser line, we show the antenna temperature (black) and the raw OFF data (blue), with each independently rescaled so as to fit within the same figure. The expected frequency location and width of the maser line, which is set by the line-of-sight velocity of W3OH, is indicated by the light red band. The width of the central frequency channel in which the maser line is detected is indicated by the light blue band. We additionally provide the TS associated with the maser line detection in the antenna temperature and the maximum percentile of the variable-width OFF position TS for each line. None of the detections are vetoed as none the maximum OFF position TS percentiles exceed the 97.5$^\mathrm{th}$ percentile value that triggers vetoing.}
\label{fig:MaserTest_W3OH}
\end{figure}

We analyze the collected data over the full frequency range at the original 92 kHz frequency resolution with the procedure described in Sec.~\ref{sec:DataAnalysis} in order to search for any spectral lines which may appear. All OH maser lines
are detected at their expected frequency locations consistent with the measured Doppler shifting and line-widths of 1665-MHz OH emission within our observing beam \cite{2014MNRAS.441.3137Q}. Moreover, none of the detected OH maser lines are vetoed by a coincident excess in the OFF data, validating our exclusion criteria. Results are shown in Fig.~\ref{fig:MaserTest_W3OH}. Additional high significance excesses $\mathrm{TS} \geq 100$ are observed at the frequencies 1.35065308, 1.374823, and  1.64664307 GHz. The excess at 1.64664307 GHz is coincident with an excess observed in the L-Band during GBT RFI scans,\footnote{RFI Scan data available at \texttt{https://science.nrao.edu/facilities/gbt/interference-protection/ipg/rfi-scans}.} while the origins and appropriate interpretations of the other two excesses are less certain. Despite the presence of two unidentified lines, these tests effectively demonstrate the ability of our framework to recover spectral lines present in the data. Analogous data were collected for 1665-MHz OH maser 351.775$4-$0.536~\cite{1998MNRAS.297..215C} during the March 29$^\mathrm{th}$ observing session, with similar results regarding maser line detection.

\section{Extended Results}
Here we provide extended results for the analyses presented in the main Letter and for the additional GBT observations described at the beginning of this SM.

\subsection{Survival Functions and Excess Candidates}
\label{sec:Survival}
\begin{figure}
\includegraphics[width = 1.0\textwidth]{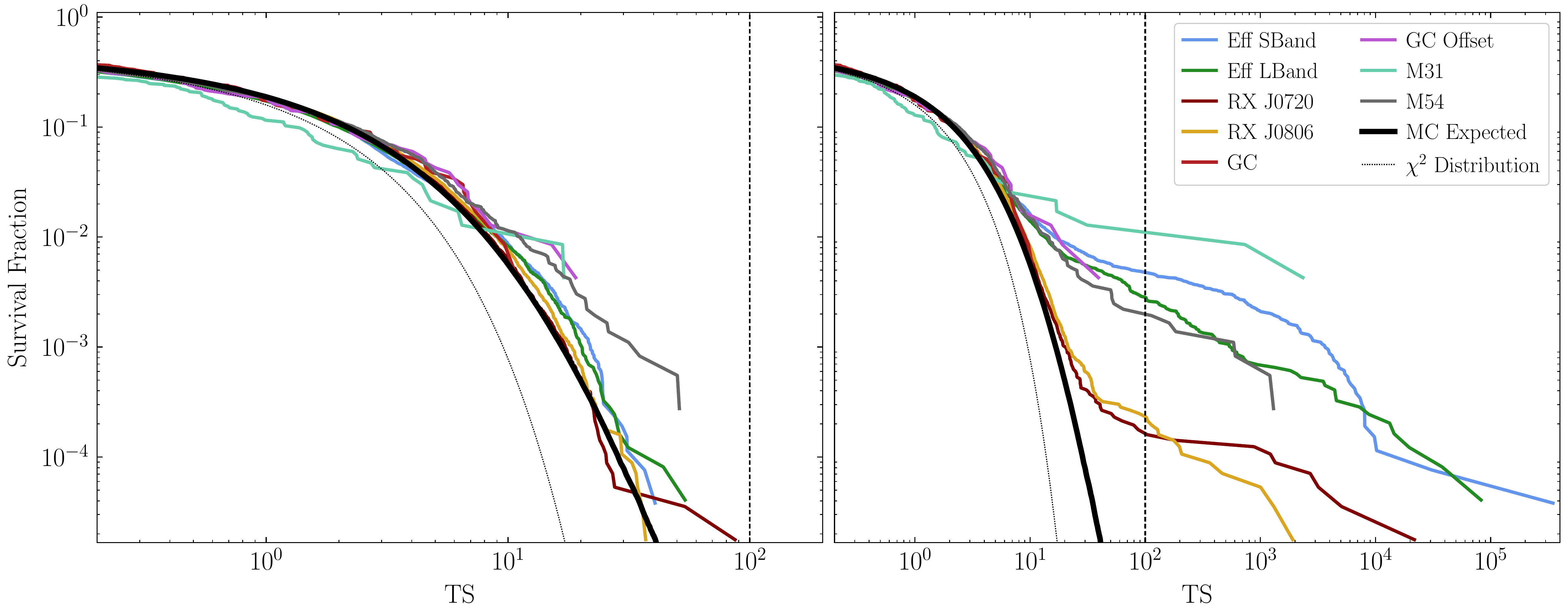} 
\caption{(\textit{Left}) The discovery TS survival function for all of the observations considered in this Letter.  Note that the survival function is defined as the fraction of TSs observed at or above the indicated value.  This figure excludes frequencies that are vetoed from the OFF position observation analyses.  The ``MC Expected" curve shows the expectation under the null hypothesis, as determined by MC simulations.  We note that all observations are from GBT except those labeled ``Eff", which are from the Effelsberg telescope. (\textit{Right}) As in the left panel, but including frequencies that would be vetoed by the OFF data.  Without the OFF vetoes there would be a significant number of frequencies with TSs exceeding the TS detection threshold, which emphasizes the importance of the OFF position vetoing procedure.}
\label{fig:SurvivalFunction}
\end{figure}

For each observation, we compute the survival function associated with the observed TS distribution. In all observations, no significant TSs exceeding our discovery threshold are observed, and the survival functions appear to match the expectation under the null hypothesis, as shown in
Fig.~\ref{fig:SurvivalFunction}.  Note that the survival function figure shows the faction of TSs at or above the indicated TS on the $x$-axis.  

From Wilks' theorem we would expect the distribution of TSs to be asymptotically $\chi^2$-distributed under the null hypothesis.  This asymptotic expectation is indicated in Fig.~\ref{fig:SurvivalFunction}.  However, our expected distribution of TSs does not follow this asymptotic expectation, even under the null hypothesis, because our analysis window contains only a finite  number of frequency bins in the sidebands.   When we perform MC simulations under the null hypothesis (first, we fit the null model to the data to determine the model parameters, and then we simulate data from those best-fit parameters and re-analyze the simulated data) we find that passing that simulated data through our analysis pipeline results in a survival function, under the null hypothesis, that is not $\chi^2$-distributed.  In Fig.~\ref{fig:SurvivalFunction} we show the simulated survival function under the null hypothesis (``MC Expected"), which is constructed by averaging the simulated survival functions across all observations (though we find that all observations produce consistent expectations for the survival function).  We note that if we modify our analysis framework to include more frequency channels in the sidebands, the survival function better approaches the $\chi^2$ distribution.  On the other hand, the fact that under the null hypothesis our survival function is not precisely $\chi^2$-distributed does not mean that our analysis framework is in any way not valid.  It simply means that when assigning TS values significances ({\it e.g.}, $p$-values), we should use the simulated survival function and not the asymptotic expectation from Wilks' theorem.  For example, ${\rm TS} = 100$ corresponds to $10\sigma$ significance under Wilks' theorem.  On the other hand, using the simulated TS distribution under the null hypothesis we find that in fact for our analysis framework ${\rm TS} = 100$ corresponds to approximately $5\sigma$ (local) significance ($p$-value of approximately $6 \times 10^{-7}$).

In the left panel of Fig.~\ref{fig:SurvivalFunction} we have already applied the vetoes from analyses of the OFF data.  That is, frequencies that are vetoed from the OFF data analyses are not shown in Fig.~\ref{fig:SurvivalFunction}.  In this case, the observed TS distributions are found to closely match the MC expecations under the null hypothesis.   In the right panel of Fig.~\ref{fig:SurvivalFunction} we show what would happen if we did not apply the OFF data vetoes.  In this case, there are many significant detections.  Some of these detections correspond to real astrophysical lines, such as the 21 cm line, that also appear in the OFF data, while other lines may be due to RFI or instrumental backgrounds. 

Note that we veto excesses at 27 of 56362 analyzed channels in the RX J0720.4$-$3125 data, 37 of 56362 analyzed channels in the RX J0806.4$-$4123 data, 24 of 26214 analyzed channels in the Effelsberg S-Band data, and 22 of 24576 analyzed channels in the Effelsberg L-Band data.

\begin{figure}
\includegraphics[width = .99\textwidth]{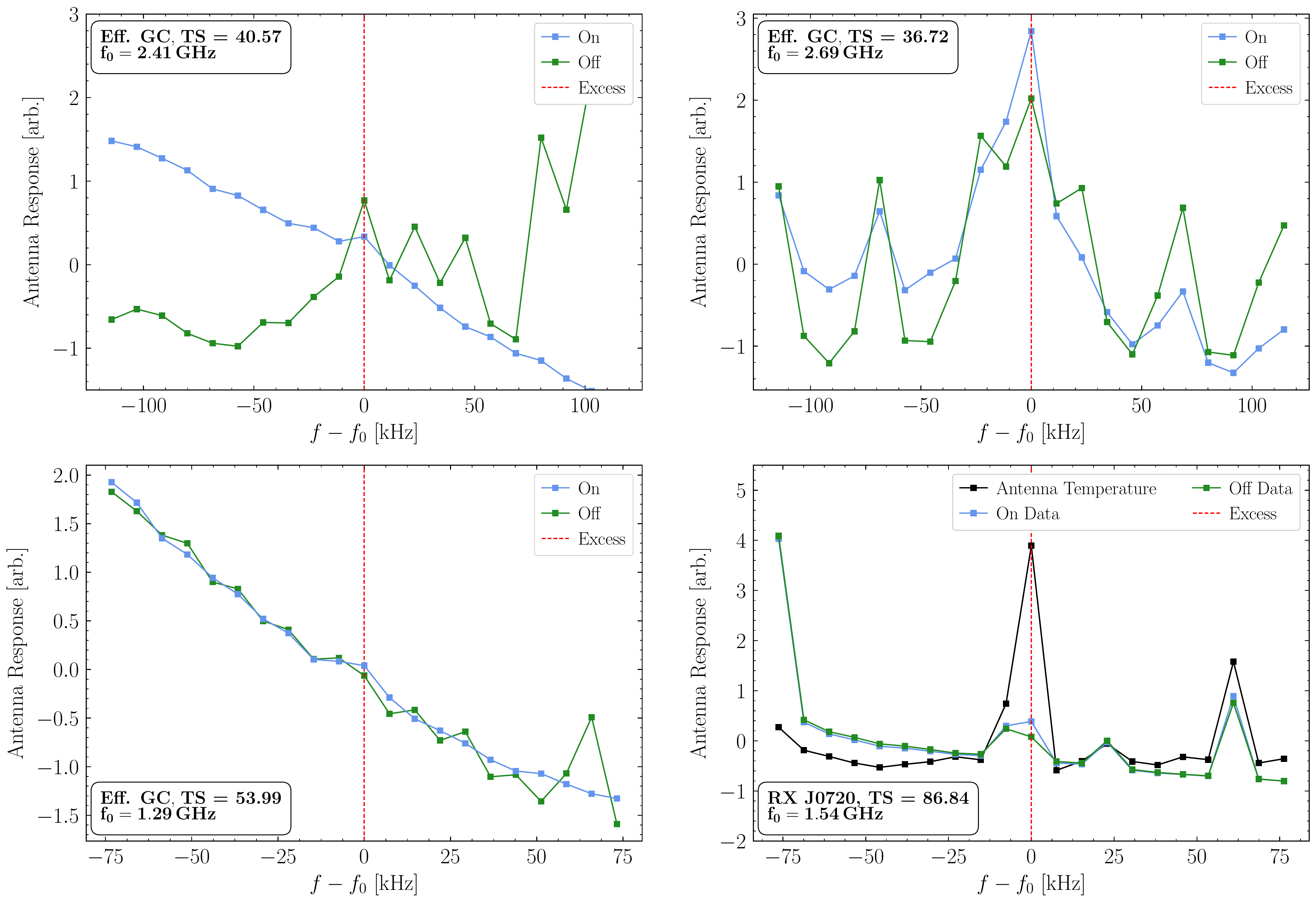} 

\caption{(\textit{Top Left}) The Effelsberg data shown in the analysis window around the excess candidate located at a central frequency of approximately 2.51 GHz in the S-band observation of the GC. Frequencies are plotted relative to the frequency corresponding to the excess channel frequency. This excess has ${\rm TS} \approx 41$.  While this excess is not vetoed by the OFF data analysis, the OFF data does should a feature at the central frequency.
(\textit{Top Right}) Similarly, the Effelsberg data shown in the analysis window around the excess candidate located at a central frequency of approximately 2.69 GHz in the S-band observation of the GC.  This excess is also not vetoed, but like the previous excess there does appear to be a corresponding feature in the OFF data. (\textit{Bottom Left}) The Effelsberg data shown in the analysis window around the excess candidate located at a central frequency of approximately 1.34 GHz in the L-band observation of the GC. It also appears that there is a similar, though not so significant, feature in the OFF data. (\textit{Bottom Right}) The GBT data shown in the analysis window around the excess candidate located at a central frequency of approximately 1.59 GHz in the observation of RX J0720.4$-$3125. As before, frequencies are plotted relative to the frequency corresponding to the excess channel frequency. The excess only appears at high significance in the antenna temperature; similar coincident features are observed in both ON and OFF data, coincident features appear in the raw ON and OFF data, although not at high enough significance in the OFF data to result in a veto of the excess. As before, this excess does not exceed our detection threshold, although it does come closer, with $\mathrm{TS} \approx 90.$}
\label{fig:ExcessCandidates}
\end{figure}

We do observe some moderate to high significance excess in our data in various observations which are depicted in \ref{fig:ExcessCandidates} with corresponding TS values and frequency locations. In the Effelsberg observation of the GC in the S-Band we observe two unvetoed excesses, both with $\mathrm{TS} \approx 40$. While these are not exceptionally large TS values, as studied in our MC signal construction, even strong axion conversion signals can result in TSs as small as 30 in the scenario with a cored DM density profile assuming the NS population is described by Model I. Similarly, we have a moderate significance excess ($\mathrm{TS} \approx 52$) in Effelsberg observation of the GC in the L-Band. A relatively higher significance excess ($\mathrm{TS} \approx 90$) is observed in the GBT observation of RX J0720.4$-$3125. We show the data corresponding to the excesses in Fig.~\ref{fig:ExcessCandidates}. We remain skeptical that these excesses requires an axion interpretation, as similar or unusual features appear in the OFF-data but at lower significances (below our veto threshold). Additionally, the location of the RX J0720.4$-$3125 excess at approximately 1.59 GHz is known to be subject to strong RFI. Follow-up observations with longer exposures and at complementary positions on the sky would be necessary to confirm or exclude the persistence of such excesses. 

Note that we also do not see any significant excesses (${\rm TS} > 100$) when shifting the frequency bins by half a bin size.  As described in the main Letter, we analyze the data shifted by half a frequency bin to account for the possibility that the axion mass falls between our frequency bins and the flux is thus split between neighboring bins.  In the \href{https://github.com/joshwfoster/RadioAxionSearch}{Supplementary Data}~\cite{SM} we present the flux and $g_{a\gamma\gamma}$ limits at each frequency point, shifted and unshifted, along with the corresponding TSs.

\begin{figure}
\includegraphics[width = .8\textwidth]{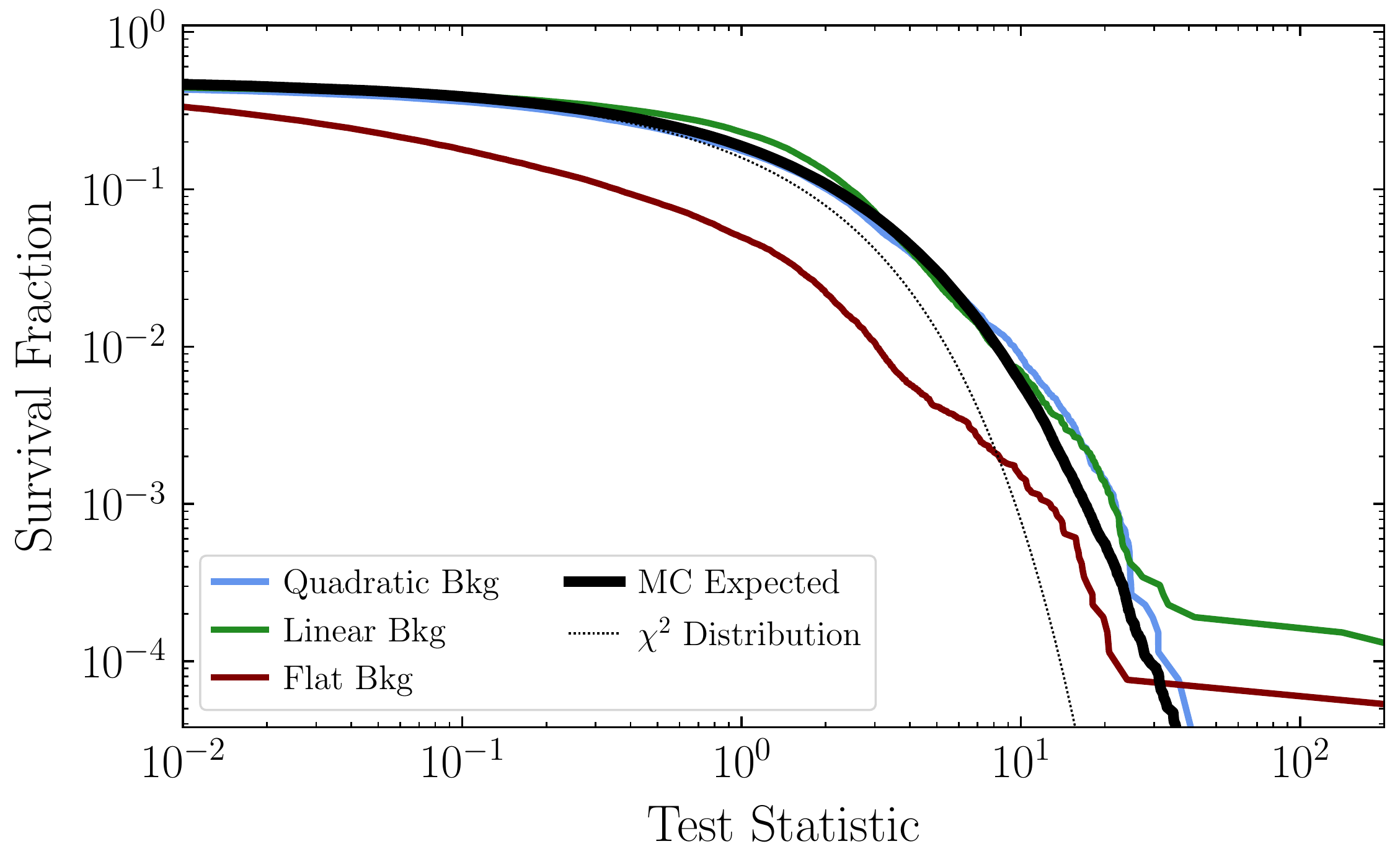} 
\caption{A comparison of survival functions for various polynomial background models for the analysis of Effelsberg S-Band data. The flat background model is unable to accurately model the null hypothesis and a significant improvement in the quality of the fits is seen by going to the linear background model.  On the other hand, there is little improvement to the quality of the fits when going from the linear to quadratic background models, except at very high TS values.  Note that we use the quadratic background model in our fiducial analyses.  Cubic background models produce similar results but are most computationally intensive to implement. 
}
\label{fig:SurvivalCompare}
\end{figure}

Additionally, our choice of a quadratic background model is somewhat arbitrary, but has limited impact on our results.  What is important is to have a background model with enough parameter freedom to describe the data under the null hypothesis but not so much freedom that the background model can be degenerate with the signal model.  The strength of axion limits set under our likelihood procedure is not appreciably affected by the background model, for small variations to the model.
In Fig.~\ref{fig:SurvivalCompare}, we show that in the Effelsberg S-Band data, the detection significances are not appreciably different across flat, linear, and quadratic background models.  The flat background model appears to be too simplistic to accurately model the form of the data, but the quadratic and linear background models give quantitatively similar results.

\begin{figure}
\includegraphics[width = .8\textwidth]{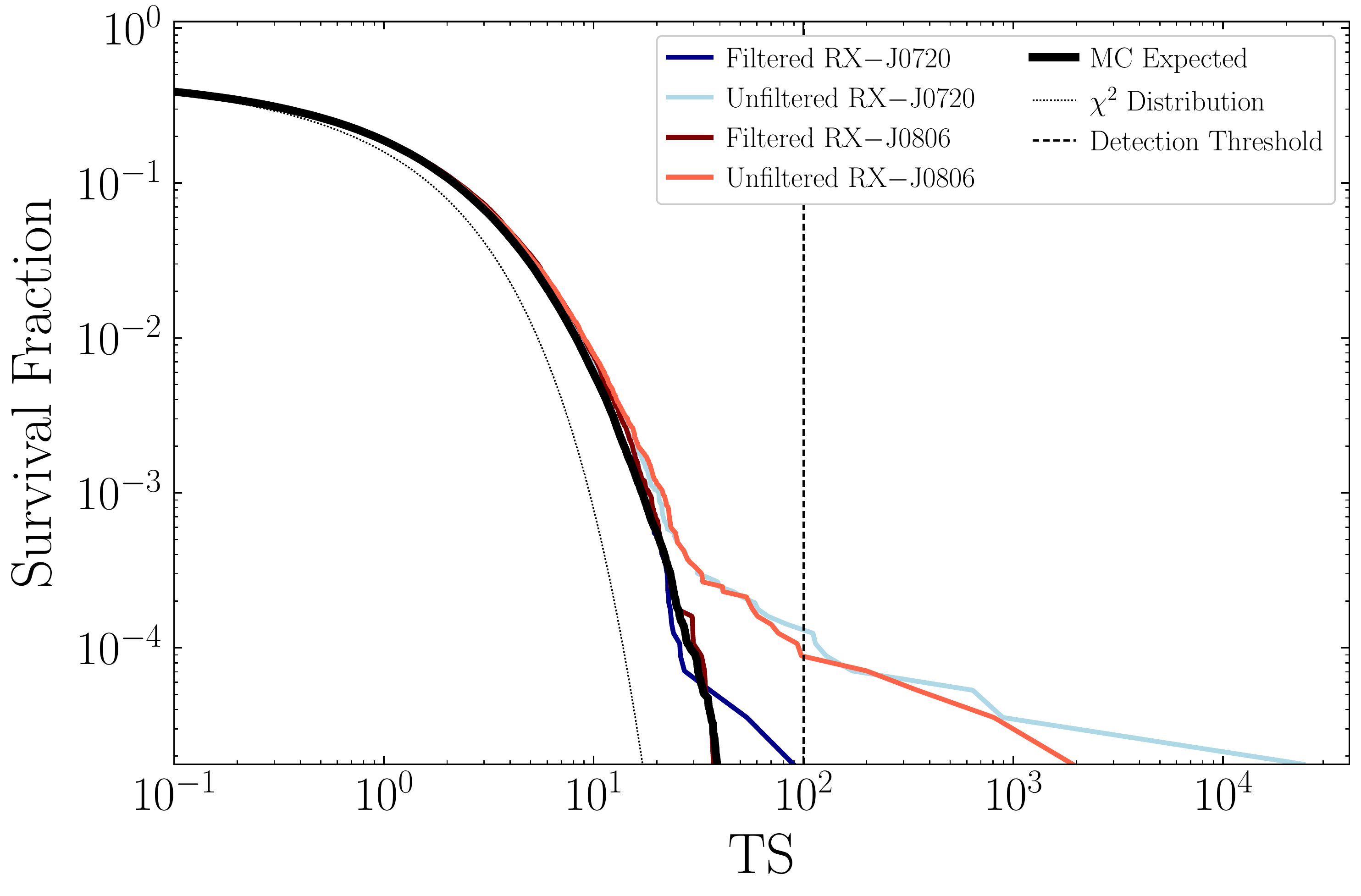} 
\caption{The discovery TS survival function for the INSs analyzed with and without time-series data filtering applied. Applying the time-series filtering eliminates a number of high significance excesses that appear due to transient noise that appears in the data.}
\label{fig:INS_Filter}
\end{figure}

In Fig.~\ref{fig:INS_Filter}, we show that in the INS GBT data the large scale features of the survival function are not appreciably changed by the application of our time-series data filtering procedure. However, the number of moderate-to-high significance excesses in RX J0720.4$-$3125 and RX J0806.4$-$4123 observations does decrease without the application of data filtering, which is consistent with the expectation that the excesses are largely sourced by transient RFI. 

\subsection{Variations to the NS and DM Density Modeling}

\begin{figure}
\includegraphics[width = .49\textwidth]{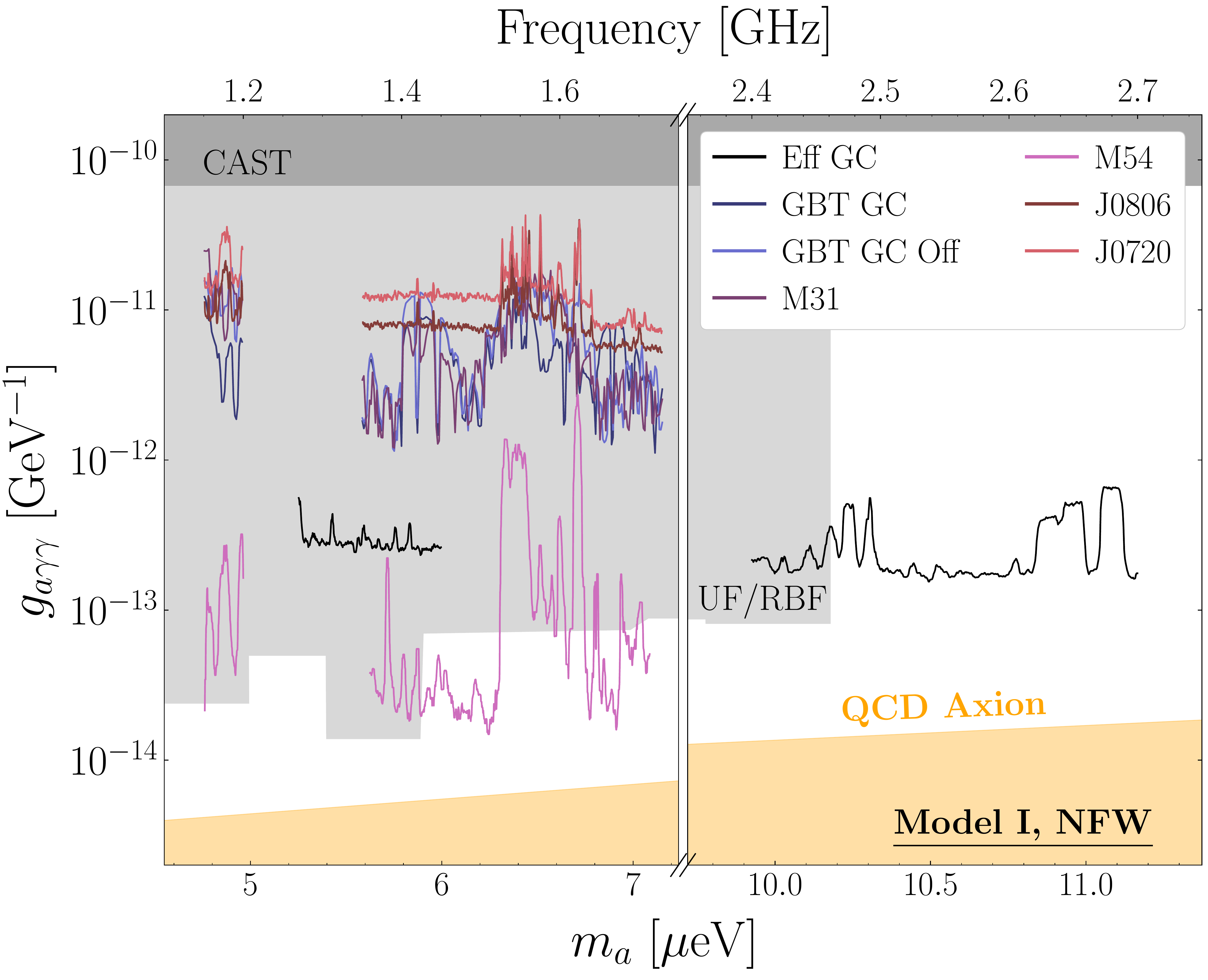} 
\includegraphics[width = .49\textwidth]{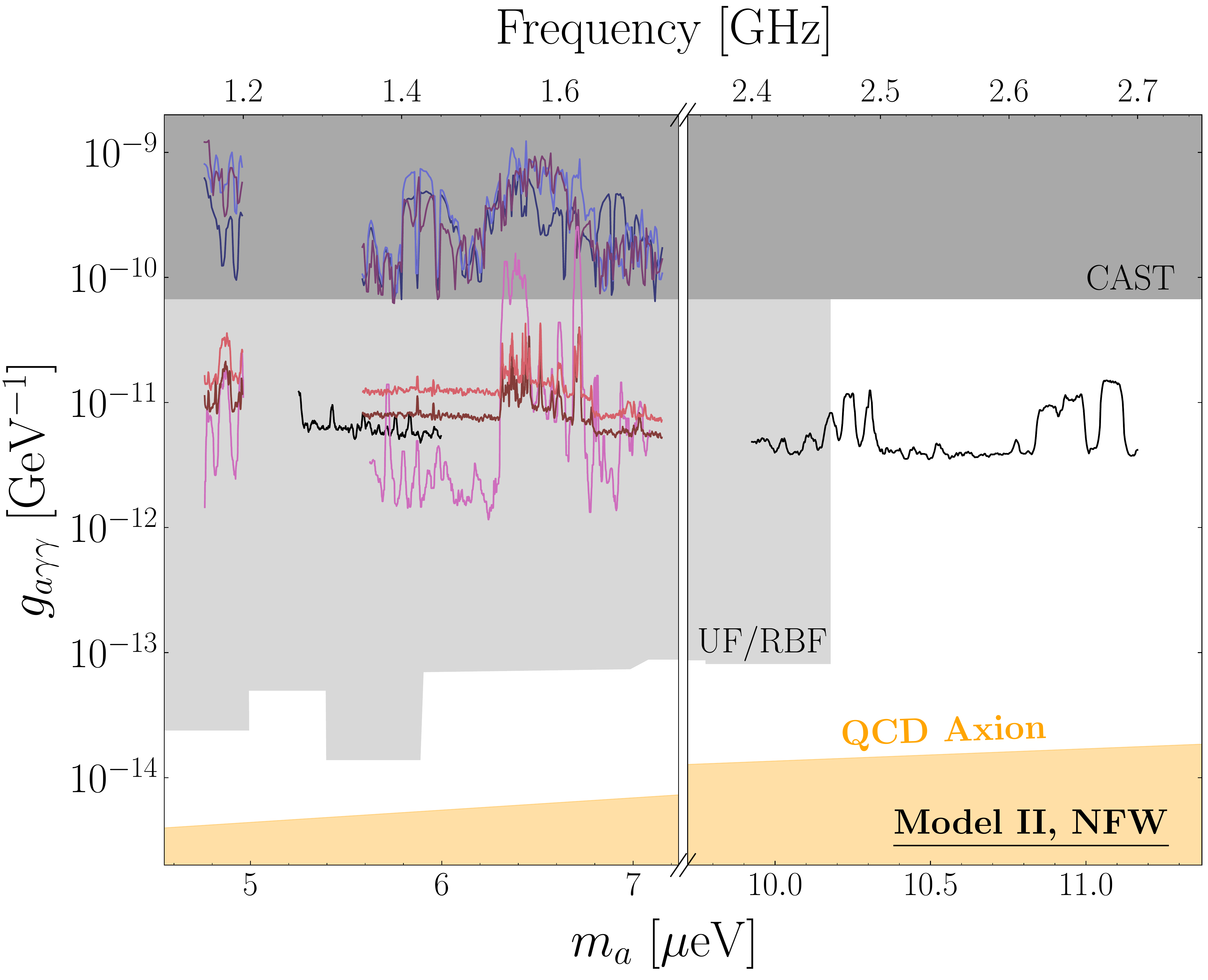}
\includegraphics[width = .49\textwidth]{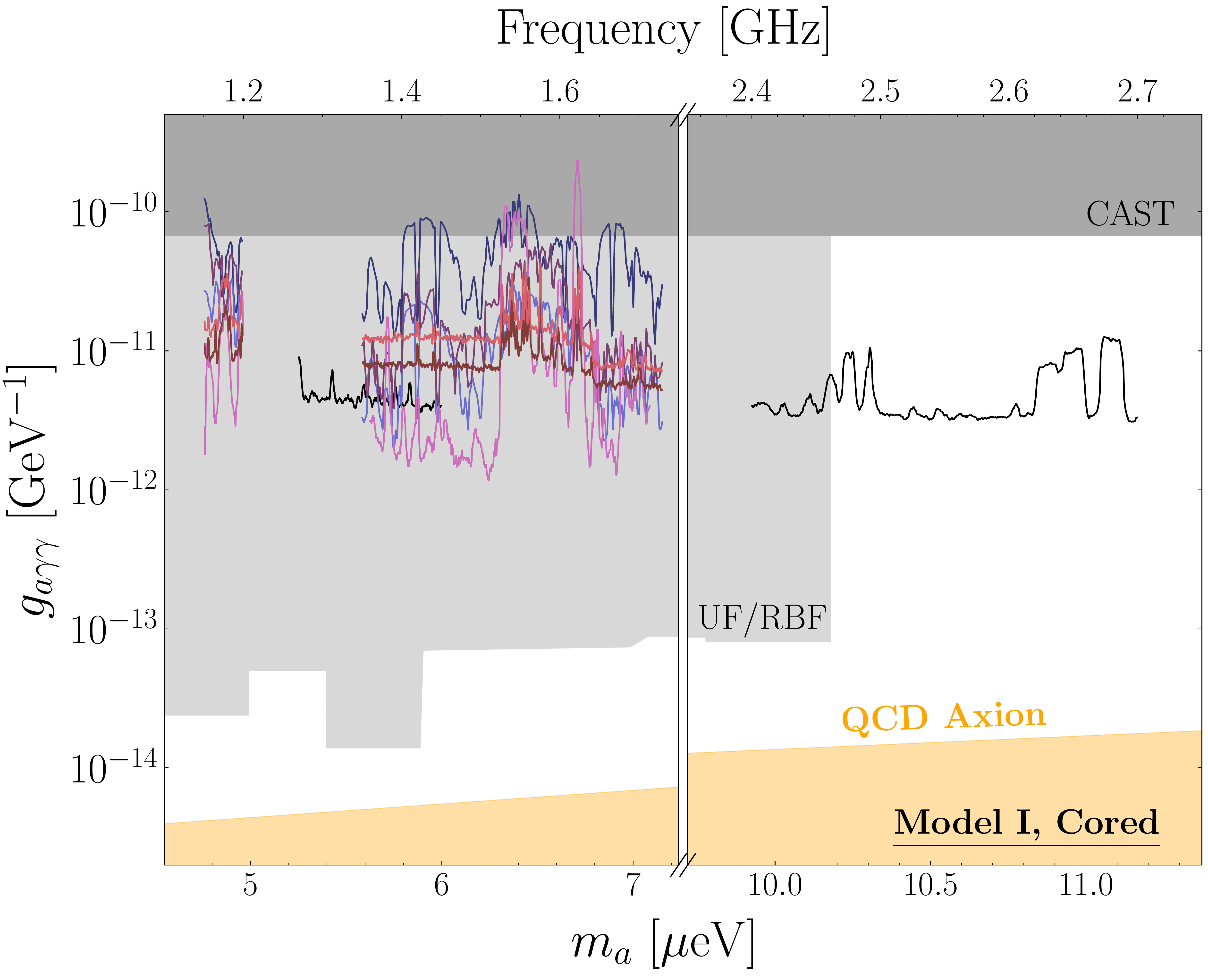}
\includegraphics[width = .49\textwidth]{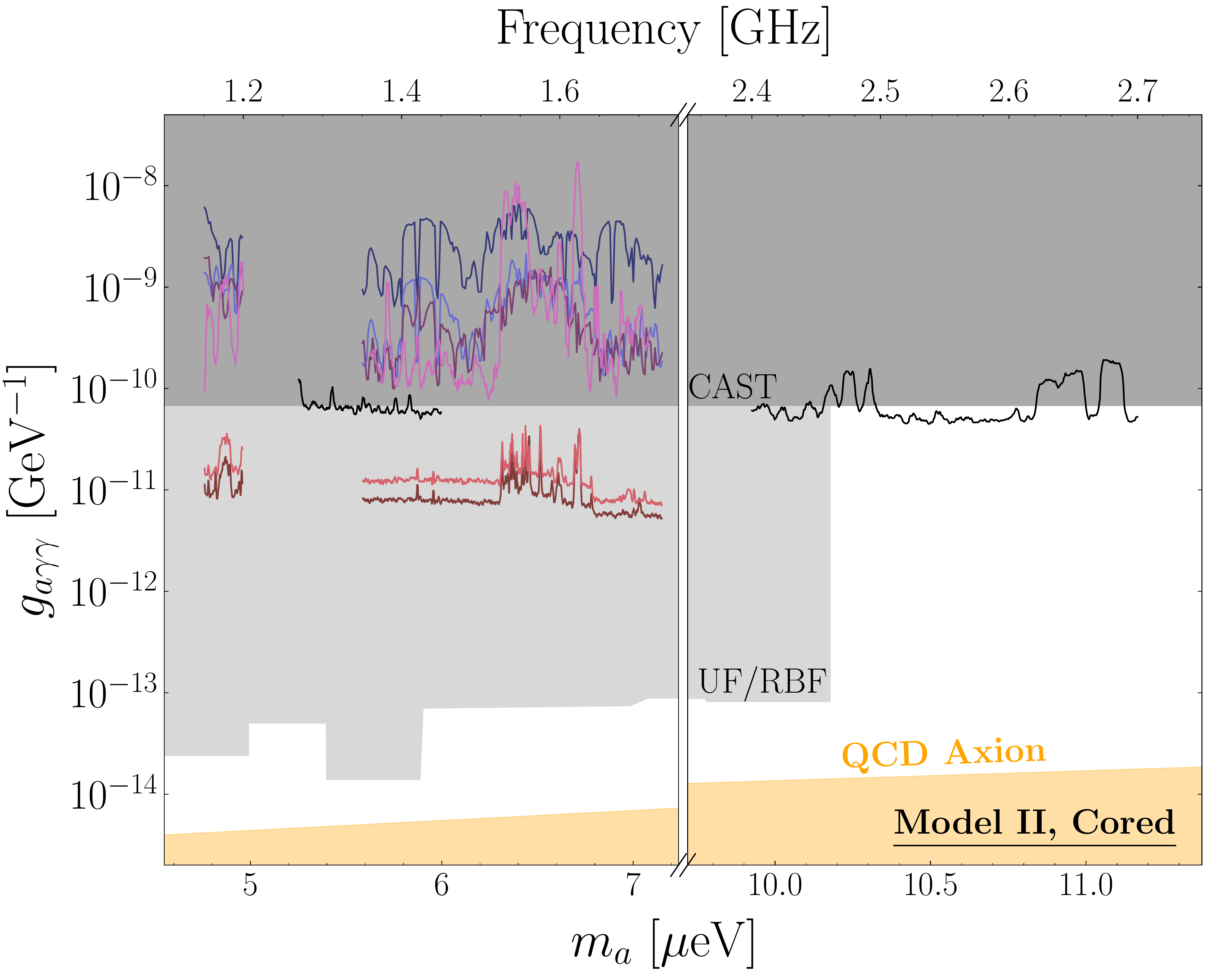} 
\caption{
Limits on the axion-photon coupling for different combinations of assumptions about the DM density profiles in the observed galaxies and the properties of the NSs within those galaxies (see~\cite{Safdi:2018oeu} and text for details).
}
\label{fig:ModelINFW}
\end{figure}

In this section, we show how variations to the NS population models and assumed DM density profiles affect the strength of the limits we are able to set. Under NS population Model I from \cite{Safdi:2018oeu}, and assuming all DM density profiles are perfectly NFW, we obtain our strongest limits, as shown in the top left panel of Fig.~\ref{fig:ModelINFW}.  Note that for M54 the NFW profile is for the host Sagittarius Dwarf Spheroidal Galaxy (see~\cite{Safdi:2018oeu} for details).  Limits are made successively weaker by the assumption of our fiducial NS population model (Model II from~\cite{Safdi:2018oeu}), as shown in Fig.~\ref{fig:ModelINFW}.  See that figure for all four combinations of NS models and DM density profile choices.  Note that the Milky Way and M31 cores are taken to be 0.6 kpc, with the DM density profile following an NFW profile outside of this radius and flat within the core radius.  
For Sagittarius, relevant for M54, under the cored DM scenario we model the DM density distribution as an isothermal sphere with scale radius of 0.2 kpc (see~\cite{Safdi:2018oeu}).

\section{Alternative Flux Density Limits}\label{subsec:alternativefluxlimits}
As a further test of the validity of our flux density limits, we recompute them using an alternative approach. The alternative 95\% C.L. flux density upper limits (``percentile limits method'') are constructed as follows. First, we smooth the calibrated spectrum with a median filter in order to remove any low-significance noise and to emphasize any potential spectral lines present in the data. The median filter assumes frequency windows containing 20 channels (for simplicity we used the same number of channels for both GBT and Effelsberg data). The outcome of this step is a data-driven estimate of the background in each respective window. Second, we single out   spectral line signals $T(f_j)$ by subtracting the smoothed background from the raw data. Third, we sort the $T(f_j)$ lines in ascending order using an ``order filter.'' Fourth, we compute the 68$^{\rm th}$ percentile band ($1\sigma$ error) at each frequency window. Finally, we obtain approximate 95\% C.L. flux density upper limits by calculating the quantity ($T(f_j)+2\sigma$). Note that this method assumes that the uncertainties are well-behaved, {\it i.e.}, going from $1\sigma$ (68\% C.L. ) to $2\sigma$ (95\% C.L.) is fairly linear. Furthermore, in order to avoid excessively strong constraints, we floor any signals that were smaller than the negative 68$^{\rm th}$ percentile error band (``power limited constraints''~\cite{Cowan:2010js}), just as we do for our profile-likelihood limit setting procedure.

\begin{figure}
\includegraphics[width = .7\textwidth]{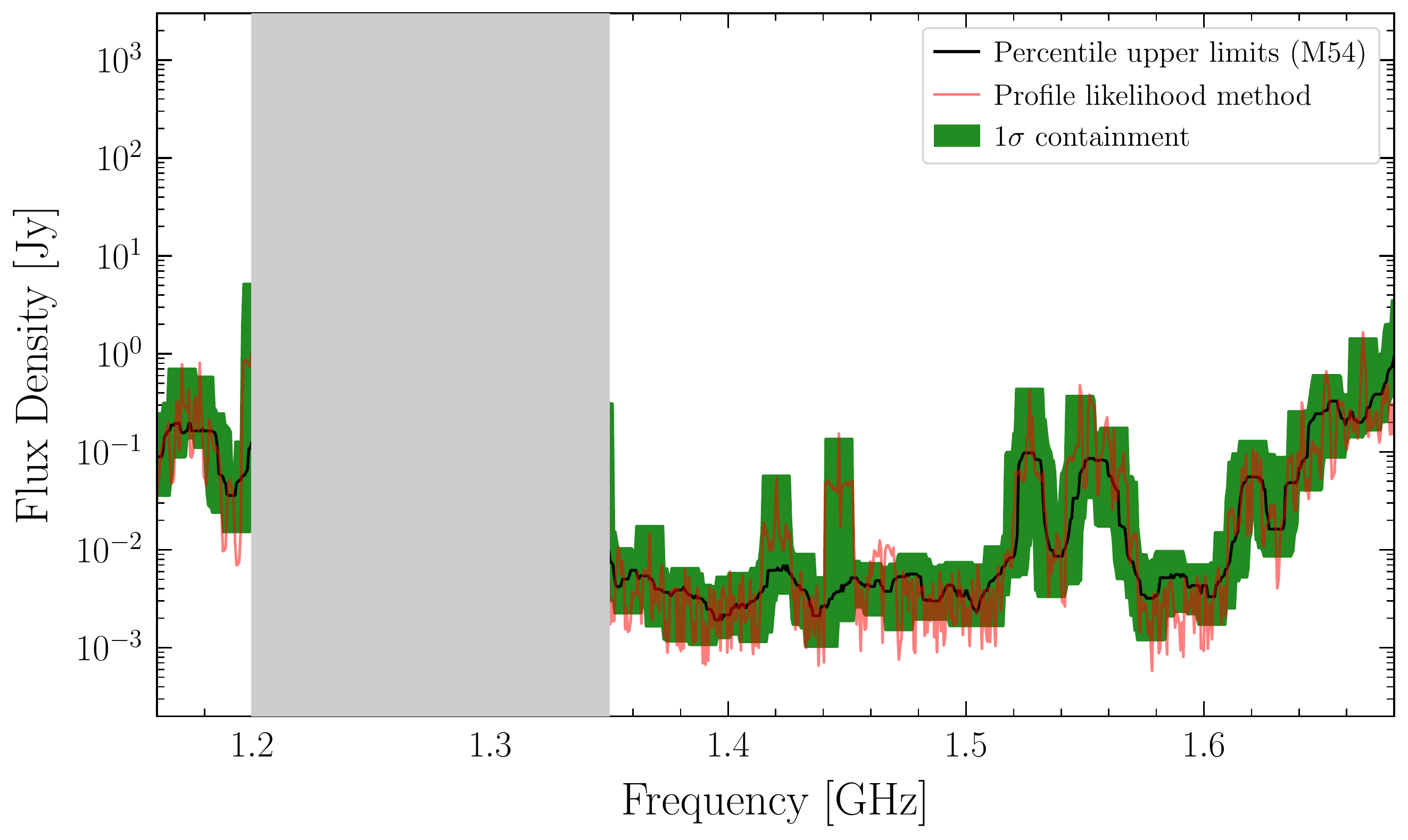} 
\caption{Comparison between the profile likelihood and percentile upper limits methods for M54 observations with GBT. The black line (green area) shows the 95\% C.L upper limits ($1\sigma$ containment band) obtained with the percentile method. The red line shows the upper limits obtained with the profile likelihood method and calibration used as default in our main pipeline; these flux limit curves are used in the main text.}
\label{fig:gbt_comparison}
\end{figure}

\begin{figure}
\includegraphics[width = .7\textwidth]{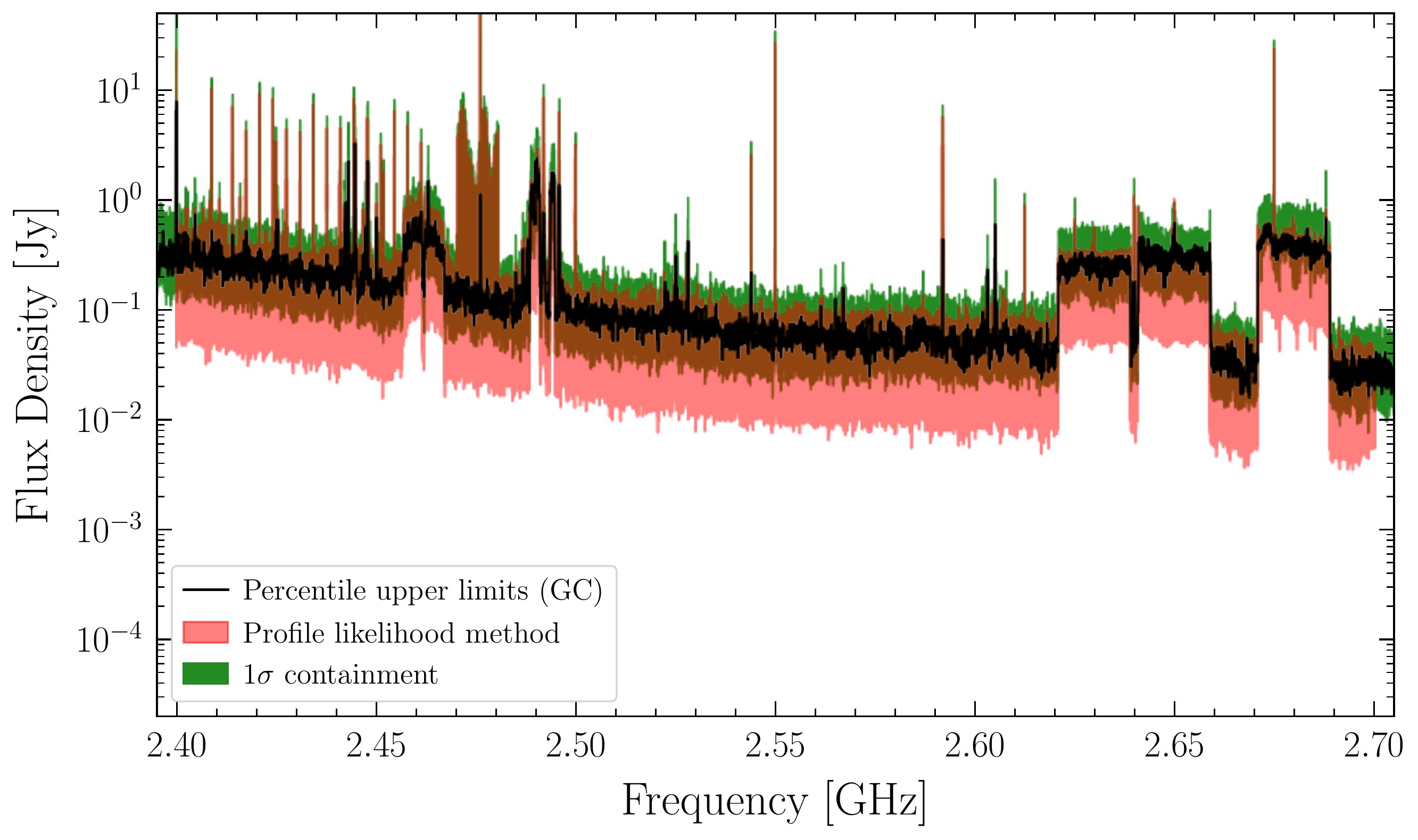} 
\caption{Same as Fig.~\ref{fig:gbt_comparison}, for Effelsberg GC observations.}
\label{fig:effelsberg_comparison}
\end{figure}

The results of our comparison for GBT observations of M54 is shown in Fig.~\ref{fig:gbt_comparison} and for Effelsberg GC observations in Fig.~\ref{fig:effelsberg_comparison}.  As can be seen, the two different methods display very good agreement. This illustrates that the profile likelihood and the percentile upper limits methods are essentially equivalent and that the upper limits obtained in this work do not depend sensitively on the limit-setting procedure. We note that the advantage of the percentile method is its efficiency: it avoids computationally expensive log-likelihood maximization computations.

\section{Bandwidth estimate from refraction in moving medium}

In Ref.~\cite{Battye:2019aco} it was pointed out that the reflected photons may acquire Doppler boosts for a rotating and misaligned magnetosphere.  In effect, one may think of the conversion surface as a mirror that reflects the incoming axions into outgoing photons.  In order for the mirror to transfer momentum to the photons, the mirror must have a component of its velocity in the normal direction.  For an aligned rotator, where the magnetic axis is aligned with the axis of rotation, the velocity vector at any point on the conversion surface is orthogonal to the normal vector from the surface and thus there is no induced Doppler broadening from reflection. However, as pointed out in~\cite{Battye:2019aco}, for misaligned rotators the conversion surface locally has a velocity that has a component in the direction of the normal vector, thus inducing a frequency shift.

Here, we consider a similar frequency shift induced by \emph{refraction} for axions that are converted to photons in the \emph{outgoing} direction.  A key point to note is that the plasma frequency profile of the magnetosphere may be interpreted as a spatially-dependent index of refraction.  For example, for the polarization component parallel to the magnetic field direction we may write the index of refraction as
\begin{equation}
  n \approx \sqrt{\frac{1 - \left({r_c \over r}\right)^3}{1 - \left({r_c \over r}\right)^3 \cos^2 \widetilde{\theta}}}   \,,
\end{equation}
for $r> r_c$, where $r_c$ is the conversion radius, $r$ is the radial direction, and $\widetilde{\theta}$ is the angle between the magnetic field and the propagation direction. The index of refraction is anisotropic, with a dependence on $\widetilde{\theta}$, because the plasma is strongly magnetized. Note that the
group velocity is always smaller than the speed of light, consistent with special relativity.

\begin{figure}
\includegraphics[width = .5\textwidth]{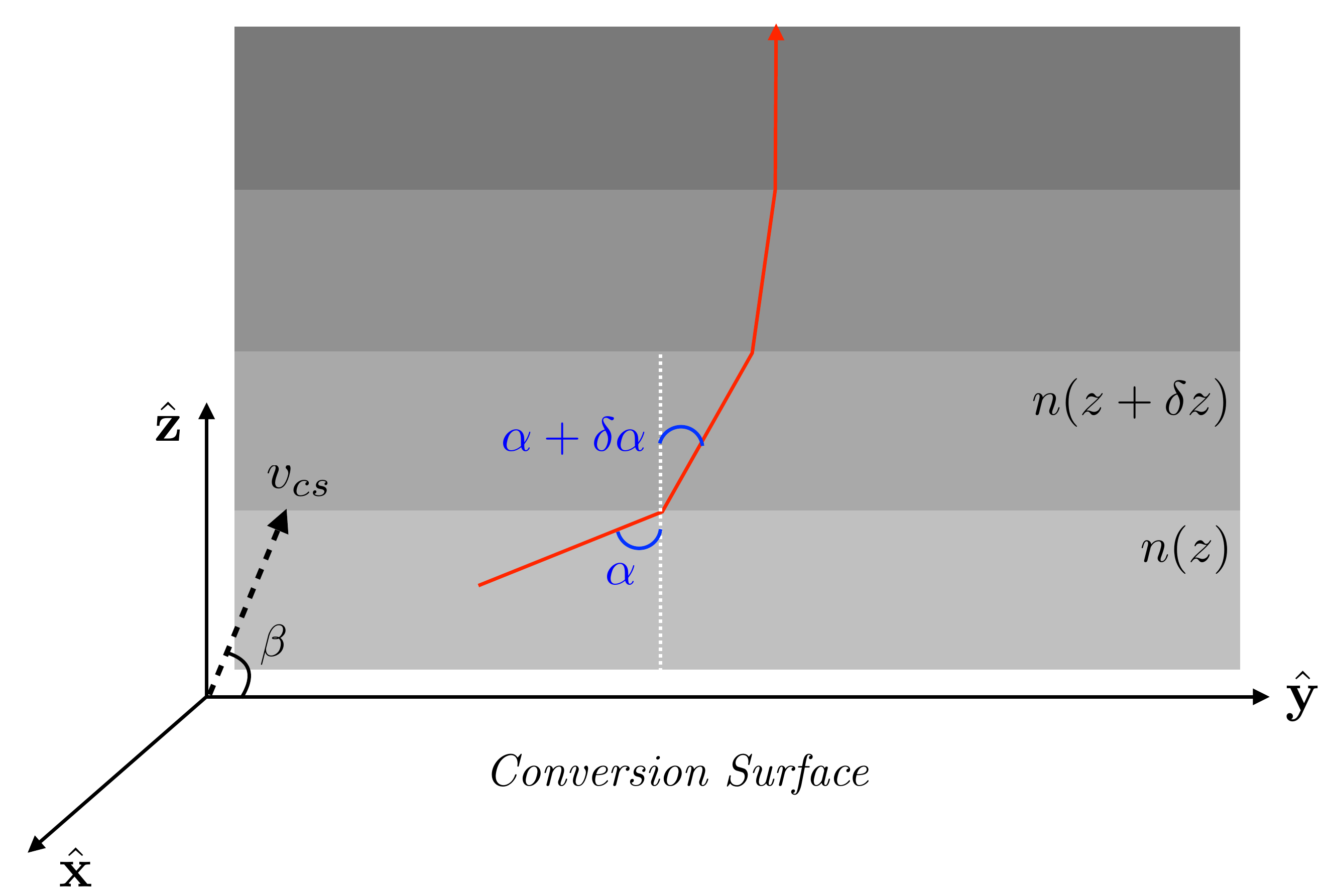} 
\caption{An illustration of how outgoing electromagnetic waves are refracted towards the normal vector to the conversion surface, labeled here by ${\bf \hat z}$.}
\label{fig:refraction}
\end{figure}

Locally around the conversion surface, we are interested in describing the scenario illustrated in Fig.~\ref{fig:refraction}.    
We choose coordinates such that the tangent plane to the conversion surface is spanned by the unit vectors ${\bf \hat x}$ and ${\bf \hat y}$, with the normal given by ${\bf \hat z}$, with ${\bf \hat y}$ chosen such that the magnetic field lies in the ${\bf \hat y}$-${\bf \hat z}$ plane. In the large-field limit the two linearly independent polarization states do not mix, so to locally describe the trajectory of outgoing electromagnetic waves, we only need to consider the waves propagating in the ${\bf \hat x}$-${\bf \hat z}$ plane.  Without loss of generality, we imagine that the local conversion surface is traveling at a speed $v_{cs}$ in the ${\bf \hat x}$-${\bf \hat z}$ plane at an angle $\beta$ from the conversion surface, as indicated in Fig.~\ref{fig:refraction}.  We will work to leading order in the speed $v_{cs}$, in natural units.  We are interested in two properties of the outgoing wave: (i) the outgoing angle with respect to the normal ${\bf \hat z}$, given an initial angle $\alpha_i$ near the conversion surface, and (ii) the frequency shift $\Delta \omega \equiv \omega(r = \infty) - \omega(r = r_c)$ between the wave asymptotically far from the conversion surface and the wave at the conversion surface. 

The outgoing wave quickly turns towards the ${\bf \hat z}$ direction, as illustrated in Fig.~\ref{fig:refraction}, because radiation refracts towards the direction of increasing index of refraction and $n$ increases with the distance from the conversion surface.  More concretely, by considering a differential form of Snell's law one may show that (assuming $\widetilde{\theta} = \pi/2$ for simplicity and since having non-trivial $\widetilde{\theta}$ does not qualitatively change these results)
\es{}{
{dz \over dy} = \sqrt{{1-z^{-3} \over 1 - z_i^{-3}}{1\over \sin(\alpha_i)^2} - 1} \,, 
}
where $z(y)$ is the trajectory of the wave that starts a distance $z_i$ from the conversion surface at an initial angle $\alpha_i$.  In the limit $z_i \to 0$, all trajectories asymptotically approach the ${\bf \hat z}$ direction, regardless of $\alpha_i$. In fact, $z_i \sim r_c v^2$ \cite{Hook:2018iia}, where $v \sim 0.1$ is the axion velocity at the conversion surface in the frame of the NS, which is close enough to zero in practice that the asymptotics hold. This is also true regardless of the magnetic field direction; the full differential equation for the trajectory is a complicated nonlinear equation because $\widetilde{\theta}$ depends on the trajectory, but such dependence is washed out because of the sharp change of index of refraction very near the conversion surface.  This itself is interesting because it says that while the initial $\alpha_i$ are isotropically distributed, since the DM phase space is isotropic, the outgoing radiation is collimated in the direction normal to the local conversion surface.

Next, we consider the frequency shift $\Delta \omega$ induced by the finite velocity $v_{cs}$ of the medium, to lowest order in $v_{cs}$. Roughly speaking, such a frequency shift results from the electromagnetic wave being created in an index of refraction which is already moving, and being measured in a stationary frame at infinity where the index of refraction is unity. To derive a differential equation for the evolution of the frequency it is useful to consider a differential step as shown in Fig.~\ref{fig:refraction} whereby we transition from a layer at distance $z$ to one at distance $z+ \delta z$, with initial angle $\alpha$ and refracted angle $\alpha + \delta \alpha$.  We perform the following set of steps.  Let the frequency of the initial state be $\omega(z)$.  First, we perform a Galilean boost by $v_{cs} \sin \beta$ in the ${\bf \hat z}$ direction so that the conversion surface is stationary in that direction.  Under this boost the material becomes birefringent, with an angle-dependent index of refraction \cite{doi:10.1119/1.2772281}.  To leading order in $v_{cs}$ the index of refraction that the incoming wave sees in the boosted frame is $\tilde n(z) = n(z) + (n^2(z) -1) v_{cs} \cos \alpha(z) \sin \beta$.  The frequency in the boosted frame (to this order in $v_{cs}$) is $\tilde \omega(z) = \omega(1 - n(z) v_{cs} \cos \alpha(z) \sin \beta)$.  We may then use Snell's law to refract the wave over the interface, where the material has index of refraction $n(z + \delta z)$.  This changes the angle $\alpha$ but does not change the frequency.  Then, we boost again by $v_{cs} \sin \beta$ but now in the negative ${\bf \hat z}$ direction. Taking the limit $\delta z \to 0$ we find the differential equation:
\es{freq_change}{
{d \log \omega(z) \over d z} = v_{cs} \sin \beta \left[ n'(z) \cos \alpha(z) - n(z) \alpha'(z) \sin \alpha(z) \right] \,.
}
In practice, since $\alpha(z)$ quickly approaches $0$ (see Fig.~\ref{fig:refraction}) the second term tends to be subdominant to the first, which remains non-zero in the limit $\alpha \to 0$. In this approximation, taking $\alpha  = 0$, the right-hand side of \eqref{freq_change} is a total derivative, and thus $\omega(z)$ only depends on the difference in index of refraction between the conversion surface (approximately zero) and infinity (approximately unity): $\delta \omega/\omega = v_{cs} \sin \beta$, to leading order in $v_{cs}$.  Again, in the limit $z_i \to 0$ this result is independent of the initial angle $\alpha_i$ and independent of the anisotropy of the index of refraction from the magnetic field direction.  

There are few interesting implications of this result.  First, when averaging the axion signal over the phase of the NS rotation there will be a frequency broadening induced by the spread in $v_{cs} \sin \beta$ across the conversion surface, appropriately averaged.  For the INSs considered in the main Letter and assuming misalignment angles $\sim$ $45^\circ$, we find that the frequency broadening is less than $\delta f / f \sim 5 \times 10^{-6}$ at 68\% containment for both NSs (not much larger than the intrinsic bandwidth $\delta f/f \sim v_0^2 \sim 10^{-6}$), justifying the bandwidths used in our fiducal analyses.  Second, this result reasserts the possibility of strong time dependence of the signal over the NS period, since the outgoing radiation is beamed normal to the conversion surface by refraction.  While Ref.~\cite{Leroy:2019ghm} claimed that the outgoing signal would not be strongly time-dependent because of the fact that the DM velocity distribution is isotropic, we have shown here that this result is modified due to the refraction of the outgoing photons.  Moreover, since the frequency shift appears to leading order to be independent of the initial angle relative to the conversion surface normal vector, it is possible that when phase-resolved, the radio signal again becomes order $v_0^2$ wide in terms of $\delta \omega / \omega$, with a central frequency that shifts by an amount $\delta \omega / \omega \sim v_{cs}$ over the period. We leave both the theoretical analysis and an investigation of this effect in the data to future work.

\end{document}